\documentclass[rmp,aps,twocolumn,nofootinbib,showkeys]{revtex4}
\usepackage{graphicx}
\usepackage{epsfig}
\usepackage{amssymb}
\usepackage{amsmath}
\usepackage{bm}
\usepackage{psfrag}
\usepackage{units}

\newcommand{\erf}{\mathop{\mathrm{erf}}\nolimits}
\newcommand{\erfc}{\mathop{\mathrm{erfc}}\nolimits}

\begin{document}

\title{Genetic Demixing and Evolutionary Forces in the One-Dimensional Stepping Stone Model}
\author{K.S.~Korolev}
\email{papers.korolev@gmail.com}
\affiliation{Department of Physics and FAS Center for Systems Biology, Harvard University, Cambridge, Massachusetts 02138, USA}
\author{Mikkel Avlund}
\affiliation{Department of Physics and FAS Center for Systems Biology, Harvard University, Cambridge, Massachusetts 02138, USA}
\affiliation{Niels Bohr Institute, University of Copenhagen, Denmark}
\author{Oskar Hallatschek}
\affiliation{Max Planck Research Group for Biological Physics and Evolutionary Dynamics, Max Planck Institute for Dynamics \& Self-Organization (MPIDS), G\"ottingen, Germany}
\author{David R. Nelson}
\email{nelson@physics.harvard.edu}
\affiliation{Department of Physics and {FAS} Center for Systems Biology, Harvard University, Cambridge, Massachusetts 02138, USA}

\begin{abstract}

We review and extend results for mutation, selection, genetic drift, and migration in a one-dimensional continuous population. The population is described by a continuous limit of the stepping stone model, which leads to the stochastic Fisher-Kolmogorov-Petrovsky-Piscounov equation with additional terms describing mutations. Although the stepping stone model was first proposed for population genetics, it is closely related to ``voter models'' of interest in nonequilibrium statistical mechanics. The stepping stone model can also be regarded as an approximation to the dynamics of a thin layer of actively growing pioneers at the frontier of a colony of microorganisms undergoing a range expansion on a Petri dish. We find that the population tends to segregate into monoallelic domains. This segregation slows down genetic drift and selection because these two evolutionary forces can only act at the boundaries between the domains; the effects of mutation, however, are not significantly affected by the segregation. Although fixation in the neutral well-mixed~(or ``zero dimensional'') model occurs exponentially in time, it occurs only algebraically fast in the one-dimensional model. We also find an unusual sublinear increase in the variance of the spatially averaged allele frequency with time. If selection is weak, selective sweeps occur exponentially fast in both well-mixed and one-dimensional populations, but the time constants are different. The relatively unexplored problem of evolutionary dynamics at the edge of an expanding circular colony is studied as well. We also briefly review how the observed patterns of genetic diversity can be used for statistical inference, and highlight the differences between the well-mixed and one-dimensional models. Although we focus on two alleles or variants, $q$-allele Potts-like models of gene segregation are considered as well. Most of our analytical results are checked with simulations, and could be tested against recent spatial experiments on range expansions off inoculations of Escherichia coli and Saccharomyces cerevisiae.

\end{abstract}

\date{\today}

\maketitle

\tableofcontents
\section{Introduction}
\label{SIntroduction}

The quantitative theory of evolution is an important open problem. The theory is necessary to determine the history of species migrations, and it could shed light on the origin and development of life. Moreover, a better understanding of the evolutionary dynamics could help control epidemics~\cite{Murray:MathematicalBiology}, fight diseases with an evolutionary character such as cancer and acquired immune deficiency syndrome~\cite{Nowak:EvolutionaryDynamics}, and guide the engineering of artificial evolution for practical applications~\cite{Poli:GeneticProgramming,MakingThingsWork}.

Most of the current understanding of evolutionary dynamics comes from population genetics, a scientific discipline that studies how evolutionary forces shape the genetic diversity of populations. The majority of theoretical models and experiments in population genetics study only one or a few well-mixed populations, i.e. populations without spatial structure, where every individual is equally likely to interact with any other individual inside the same population. Microorganisms growing and evolving in a well-mixed liquid culture provide an important example. While nonspatial models are often easier to analyze than spatial ones, they do miss what can be essential features of natural populations.

In nature, organisms often occupy areas that are much larger than the square of the dispersal distance, i.e. the distance typically traveled by an individual in one generation. This causes two main problems for well-mixed-population models. First, well-mixed-population models underestimate the role of genetic drift~(fluctuations due to the discreteness of the number of individuals). The difference arises because the organisms can only interact with their neighbors, and the number of neighbors within the dispersal distance is much smaller than the total number of organisms in the entire population. Second, well-mixed-population models neglect the spatial structure of the population that can be created by external factors or by internal dynamics. Such spatial structures often exist, and, as we show in this paper, they can significantly affect evolutionary processes in the population.

Well-mixed-population models are particularly inadequate when applied to expanding populations. Expansions are very common in biology. Species spread to new territories from the locations where they first evolved. Expansions also occur because of environmental changes such as the global warming and the glacial cycles or due to sudden long distance migrations to new habitats. Even though well-mixed-population models can account for the growing number of individuals~(population size), these models do not capture the fact that the newly settled areas are colonized by the offspring of only a small number of individuals at the expanding front. Since the ancestral population is small, the genetic drift is strong. As a result, neutral genetic diversity decreases with the distance from the origin of the expansion. This reduction in genetic diversity, which is often called ``the founder effect''~\cite{Mayr:FounderEffect}, has been observed in humans~\cite{Ramachandran:MigrationFromAfrica,Templeton:MigrationFromAfrica} and many other species. For example, the founder effect in the population waves following the receding glaciers is believed to be responsible for the reduced genetic diversity in high latitude regions compared to equatorial ones~\cite{Hewitt:SouthNorthGradient}.

\begin{figure}
\includegraphics[width=\columnwidth]{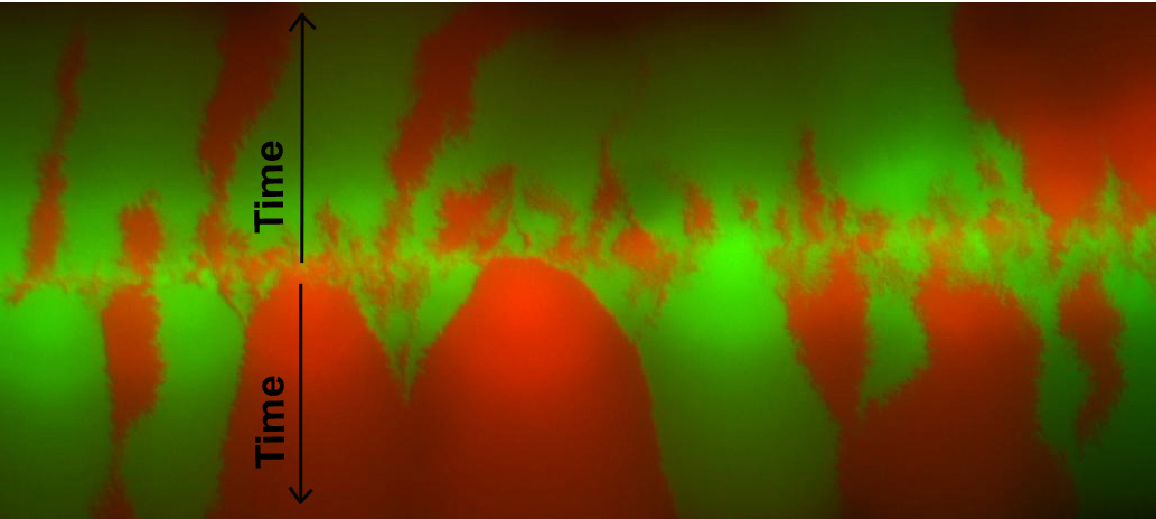}
\caption{(Color online) Spatial segregation in an expanding microbial population. Different colors label different alleles. The Petri dish was inoculated with a well-mixed population occupying a narrow horizontal linear region between the arrows, which show the direction of the growth. As this population expands, it segregates into well defined monoallelic domains. The colony is of the order $1$~cm in height. Details of the experiment are presented in~\protect{\textcite{HallatschekNelson:ExperimentalSegregation}}.}
\label{FSpatialSegregation}
\end{figure}

The spreading of Escherichia coli~(E.~coli) and Saccharomyces cerevisiae~(S. cerevisiae) on Petri dishes has been investigated in recent experiments by~\textcite{HallatschekNelson:ExperimentalSegregation}. In these experiments, microbes grown in the dark carried one of two selectively neutral alleles, differing only in a gene encoding for proteins with two distinct fluorescence spectra. Figure~\ref{FSpatialSegregation} shows the expansion of an initially well-mixed $50:50$~population of E.~coli into two unoccupied half planes initiated by a razor blade inoculation with cells grown up in liquid culture. The distinctive feature illustrated by the typical experiment in Fig.~\ref{FSpatialSegregation} is that the population does not remain well-mixed; instead, it segregates into well-defined domains. The segregation occurs because the strong genetic drift associated with reduced population size facilitates fixation of one of the two alleles at the front.

Analogous phenomena should also occur in a nonexpanding one-dimensional population because its dynamics is similar to the dynamics of the front of a growing population. The front of a population wave and a literally one-dimensional habitat are not exactly equivalent because the contour of the front undergoes undulations while a one-dimensional habitat has a fixed linear shape. Nevertheless, both are effectively one-dimensional and should deviate from the predictions of well-mixed-population models in similar ways. The advantage of a flat one-dimensional habitat is that it is easier to analyze. In addition, although most species live in effectively two dimensional habitats, a quasi one-dimensional habitat could describe a bank of a river, a sea coast, and a slope of a linear mountain range.

To study the dynamics of a population analytically, we adopt the stepping stone model proposed by Kimura and Weiss~\cite{KimuraWeiss:SSM}. This model considers many well-mixed populations, demes, located on a spatial lattice. Each deme is subject to mutation, selection, genetic drift, and short range migration between neighboring demes. In the limit of weak evolutionary forces and large number of demes, the stepping stone model is equivalent to the continuous models proposed by~\textcite{Wright:IsolationByDistance} and~\textcite{Malecot:DecreaseReltationshipDistance} and is described by the stochastic Fisher-Kolmogorov-Petrovsky-Piscounov equation~\cite{Fisher:FisherWave,Kolmogorov:FKPPEquation} with additional terms representing mutation. On the other hand, when each deme contains only one organism, the model is analogous to the Eden model~\cite{Saito:EdenModel} used to describe the growth of interfaces and the voter model~\cite{CoxGriffeath:VoterModel} discussed in Appendix~\ref{AVoter}.

\begin{figure}
\includegraphics[width=\columnwidth]{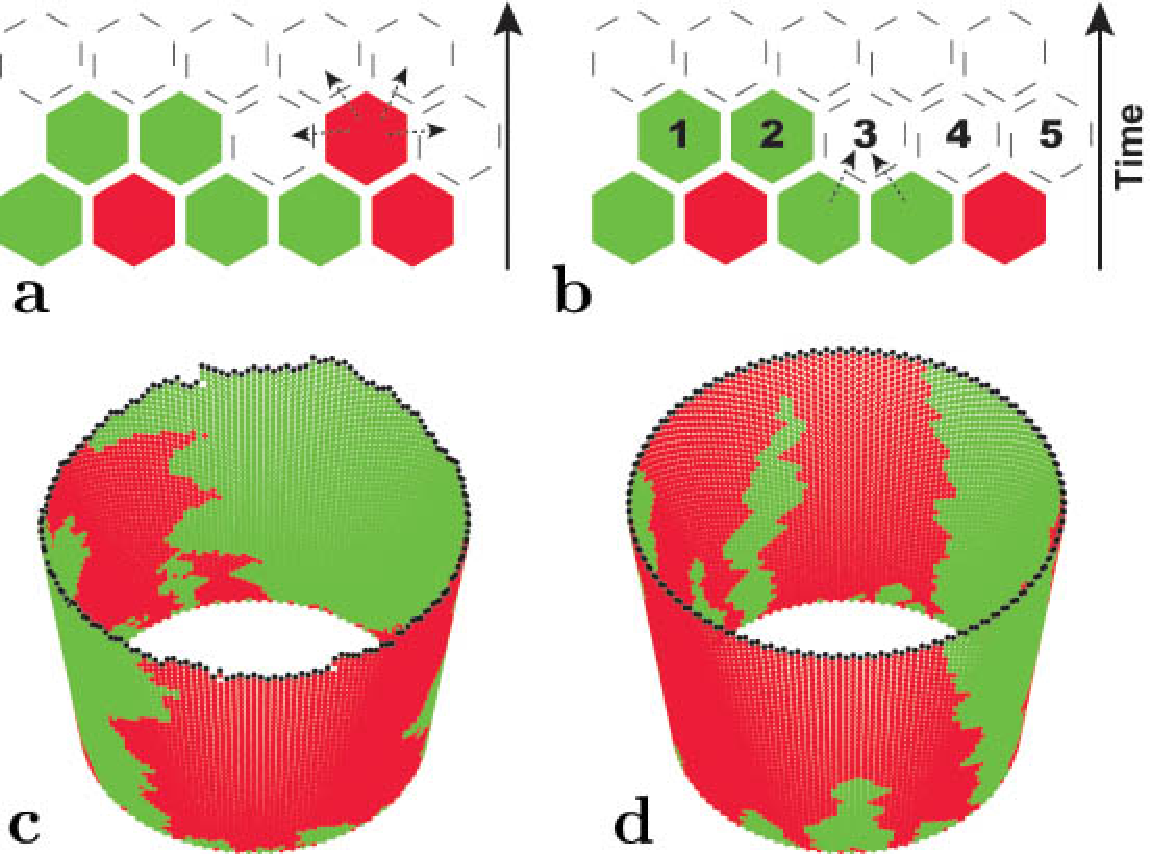}
\caption{(Color online) An illustration of the two models of a growing front. (a) and (c) illustrate the model with a rough, undulating front, which is a natural result of an unconstrained two-dimensional growth. (b) and (d) illustrate the model with a flat front, which is constrained to have no lateral undulations to simplify the analytical analysis. The blank hexagons represent empty sites, and different colors of the occupied hexagons represent different alleles. (a) The model of an undulating population front. The highlighted hexagon is a randomly chosen cell that can reproduce and deposit an identical offspring in any of its four empty nearest neighbor sites (shown with arrows) with equal probability. (b) The model of a one-dimensional habitat, where each row represents a generation. Thus, each row is completed before moving on to the next one, so an empty site can be filled only by an offspring of one of its nearest neighbors in the previous generation (shown with arrows). Both (a) and (b) show the effects of genetic drift (sampling error) when, e.g., the second from the left cell in the bottom row leaves no offspring. Such events lead to coarsening seen in (c) and (d). (c) and (d) are single simulation runs for models in (a) and (b) respectively. A population of~$100$ cells was wrapped around a cylinder to illustrate periodic boundary conditions used in this paper. Note that in (d) the front is flat whereas in (c) it is rough. This roughness affects some aspects of the shapes of the monoallelic domains shown in (c): A domain boundary followed from its lowest point to its highest point always goes up in (d), but, in (c), it sometimes goes down. As discussed in~\protect{\textcite{HallatschekNelson:ExperimentalSegregation}}, domain walls are expected to wander more vigorously in (c) than in (d). Despite the apparent differences, both models exhibit the same qualitative behavior.}
\label{FSimulationBackground}
\end{figure}

\begin{figure}
\includegraphics[width=\columnwidth]{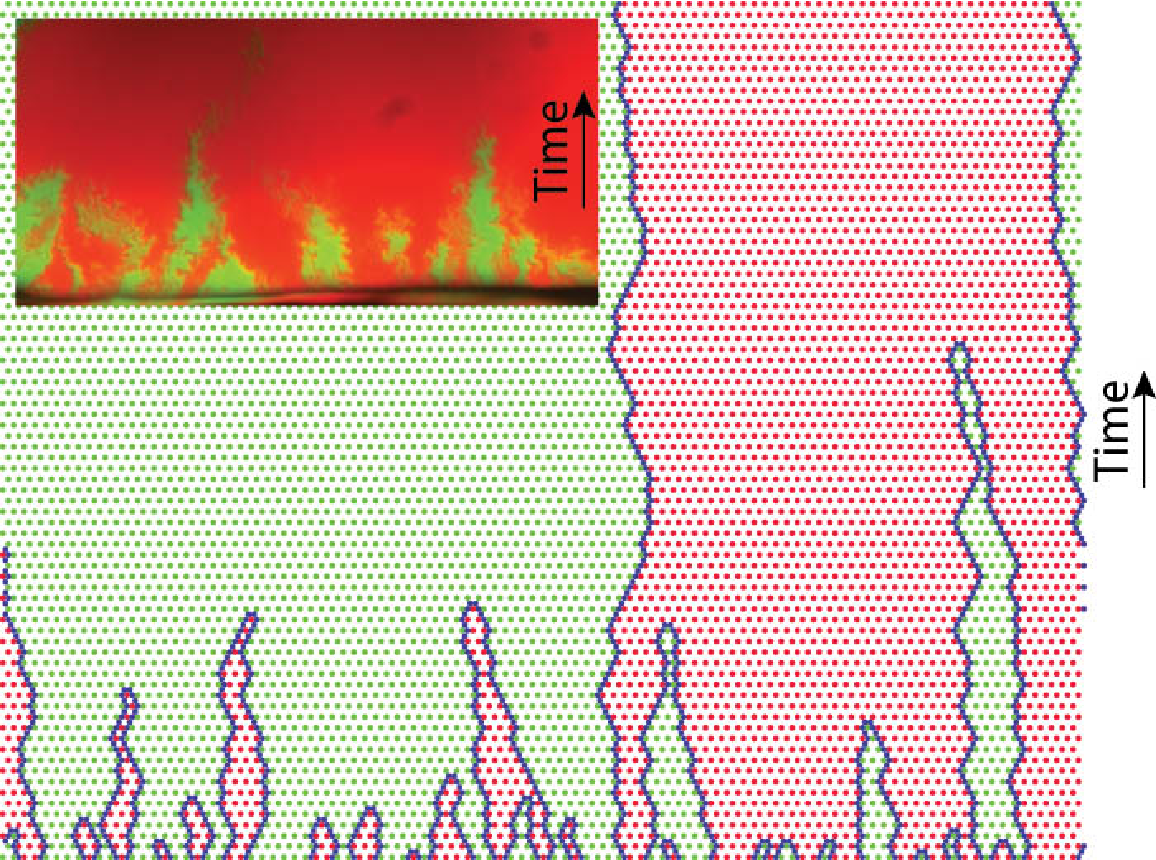}
\caption{(Color online) Qualitative comparison of a gene segregation experiment from a linear inoculation~(inset) and the simulation of a one-dimensional habitat. The experiment is analogous to the one depicted in Fig.~\protect{\ref{FSpatialSegregation}}.}
\label{FSimulationExperiment}
\end{figure}

We also performed numerical simulations to better understand the relationship between the experiments in~\textcite{HallatschekNelson:ExperimentalSegregation} and our analytical results. An illustrative simulation~(with periodic boundary conditions) is shown in Fig.~\ref{FSimulationBackground}, which also shows the difference between a growing population front with undulations and a literally one-dimensional habitat advancing uniformly in time. Figure~\ref{FSimulationExperiment} shows qualitative agreement between the experiments and the simplified row-by-row growth model that we studied analytically.

In this paper, we first focus on the spatial segregation due to genetic drift and its effect on the dynamics of a linear one-dimensional population. We find that segregation of two neutral alleles has two stages. During the first stage, distinguishable domains emerge from the well-mixed population. During the second stage, domain boundaries diffuse and annihilate upon collision. As a result, some of the domains vanish whereas others grow. We show how our calculations might be used to extract the diffusion constant and the effective population size from experiments like those in~\textcite{HallatschekNelson:ExperimentalSegregation}, and discuss how well the model describes the behavior of microbes. A \textit{detailed} comparison~(beyond the qualitative agreement we find with the main features) would require more extensive and precise experiments; we hope such experiments will be carried out in the future. The spatial segregation dramatically changes the effects of genetic drift and selection on the population compared to the predictions of well-mixed-population models. For the neutral model without mutation, we find that local diversity or ``heterozygosity'' decays as~$t^{-1/2}$, and the standard deviation of the global fraction of an allele grows subdiffusively as~$t^{1/4}$. The evolutionary dynamics during a radial expansion~(see Fig.~\ref{FCircularExperiment}) is studied as well. In this case,  migration and genetic drift slowly weaken as the circumference grows. As the result, the domains boundaries eventually stop coalescing leading to a finite number of domains in the long time limit. We find that this final number of domains grows as a square root of the initial radius of the colony. We also study the dynamics in the presence of weak selection and find that it differs markedly from that of a well-mixed population. Because of the spatial segregation into domains, selection acts only near domain boundaries, which constitute only a small fraction of the population. Hence, extinction of a deleterious allele proceeds much more slowly in one-dimensional populations than in well-mixed populations. Unlike genetic drift and selection, the effects of mutation in the spatial model are essentially the same as in the well-mixed-population model, but the spatial model gives a more accurate description of the population and accounts for the spatial correlations. Finally, we discuss how one can estimate important model parameters by sampling and sequencing DNA from organisms in a natural population. The differences between spatial and nonspatial models used for genetic inference are highlighted.

A substantial fraction of our results for the neutral dynamics in a one-dimensional habitat has been derived previously in population genetics~\cite{Malecot:Dynamics,KimuraWeiss:SSM,Barton:NeutralEvolution}, ecology~\cite{Houchmandzadeh:NeutralEcology}, and nonequilibrium statistical mechanics~\cite{CoxGriffeath:VoterModel,Bramson:FiniteSize}. Here, we present a single self-contained derivation of these earlier results in a novel context of expanding populations in two dimensions and in a language familiar to physicists, with future microbial tests of the theory in mind. Our new results are primarily confined to the analysis of radial expansions and natural selection. 

This paper is organized as follows. First, we review classical results for well-mixed populations in Sec.~\ref{SWellMixed}. We then introduce the one-dimensional stepping stone model in Sec.~\ref{S1dSSM} and derive the equations of motion for spatial correlation functions. In Sec.~\ref{SNeutralNoMutation} and Sec.~\ref{SNeutralMutation} we solve these equations for zero and nonzero mutation rates respectively. While the neutral stepping stone model has been treated before, we derive some new results and use a different technique that can be easily extended to radially expanding populations.  The effects of selection are considered in Sec.~\ref{SSelection}, and in Sec.~\ref{SSimulations} we test our analytical results with simulations. In Sec.~\ref{SInflation}, evolutionary dynamics during a radial range expansion is analyzed, and Sec.~\ref{SGeneticInference} deals with genetic inference. Various details are relegated to Appendices~\ref{AIto}--\ref{AVoter}. In Appendix~\ref{AMultipleColors}, we indicate how some of the $2$-state~(i.e., ``$2$-allele'') results can be generalized for the Potts-model-like nonequilibrium dynamics of $q$-alleles with~$q\ge3$.


\section{Population Genetics In Well-Mixed Populations}
\label{SWellMixed}

Well-mixed-population models are relevant to microorganisms vigorously shaken in a test tube, but they do not describe spatial phenomena. Indeed, if cells visit all parts of the test tube during a cell division time, they live in an effectively zero-dimensional habitat. Nevertheless, well-mixed-population models can serve as a useful reference point to which spatial models can be compared. Nonspatial populations also provide a simple context to introduce genetic drift, mutation, and selection; and the stepping stone model presented in Sec.~\ref{S1dSSM} uses a well-mixed-population model to describe the dynamics of allele frequencies within the demes. This section summarizes the classical results of nonspatial population genetics, which are primarily due to Wright, Fisher, Haldane, and Kimura; the books by~\textcite{Hartl:PopulationGenetics}, and~\textcite{Crow:PopulationGenetics} provide a good introduction to the subject and refer to the original literature, which is too extensive to be discussed here; see also~\textcite{Blythe:Review} for a recent review written for physicists.

To simplify the discussion and to make a direct connection with the experiments in~\textcite{HallatschekNelson:ExperimentalSegregation}, we consider two alleles in a population of~$N$ haploid organisms, i.e. organisms with a single set of chromosomes.~\footnote{The theory of haploid organisms also describes the dynamics of genes in cellular organelles like mitochondria and chloroplast and on certain sex chromosomes like Y-chromosome in Homo sapiens. For~$N$ diploid organisms, the theory is essentially the same under certain assumptions, provided one focuses on the dynamics of~$2N$ gene copies in each generation; see~\protect{\textcite{Hartl:PopulationGenetics}}.} The two-allele approximation may seem very restrictive, but many of our results can be generalized to an arbitrary integer number of~$q\ge3$ alleles. In addition, a two-allele model can be used to describe the dynamics of an allele of interest~(with or without a selective advantage) when all other alleles have the same fitness. We assume that each of the individuals in the population can die, give birth~(divide), and mutate. The details of this birth and death process are species dependent, but the dynamics on time scales larger than the generation time~$\tau_{g}$ is believed to be universal provided~$N$~is large. This universal dynamics is often referred to as the diffusion or continuous approximation. Two simple models are commonly used to illustrate the continuous approximation: the Wright-Fisher model and the Moran model. Here, we use the latter because it more closely resembles microbes with overlapping generations.

First, we consider the Moran model without selection and mutation. During a time step, two individuals are randomly selected with replacement from the population. The first individual is chosen to reproduce, and the second one to die; thus, the total number of the organisms is conserved. If the ``frequency'' of allele one~(i.e., the fractional number of individuals with genotype one) at time step~$\tilde{t}$ is~$f(\tilde{t})$, then, at the next time step, it is~$f+1/N$ with probability~$f(1-f)$, $f-1/N$ with probability~$f(1-f)$, and $f$ with probability~$f^{2}+(1-f)^{2}$. The expectation value and variance of~$f(\tilde{t}+1)$ are then given by,

\begin{equation}
\label{EMoranGeneticDriftFirst}
\langle f(\tilde{t}+1) \rangle=f(\tilde{t}),
\end{equation}

\begin{equation}
\label{EMoranGeneticDriftSecond}
\langle [f(\tilde{t}+1)-\langle f(\tilde{t}+1)\rangle]^{2}\rangle=\frac{2f(\tilde{t})[1-f(\tilde{t})]}{N^{2}},
\end{equation}

\noindent where angular brackets represent average with respect to the random choice of individuals for reproduction and death. Because only one of~$N$ organisms gives birth in a Moran time step,~$\tilde{t}$ measures time in fractional generation time,~$\tau_{g}/N$.

Equations~(\ref{EMoranGeneticDriftFirst}) and~(\ref{EMoranGeneticDriftSecond}) imply that~$f(\tilde{t})$ performs an unbiased random walk in the space of allele frequencies. In the continuum limit, this random walk can be described by the following Fokker-Planck equation with a frequency dependent diffusion coefficient~\cite{Crow:PopulationGenetics,Hartl:PopulationGenetics}

\begin{equation}
\label{EMoranGeneticDriftFokkerPlank}
\frac{\partial P(t,f)}{\partial t}=\frac{\mathfrak{D}_{g}}{2}\frac{\partial^{2}}{\partial f^{2}}\left[f(1-f)P(t,f)\right],
\end{equation}

\noindent where~$P(t,f)$ is the probability density function for~$f$ at time~$t$ measured in generations, and $\mathfrak{D}_{g}$ is the genetic diffusion constant. Here,~$t$ is the time measured in generations; as discussed above,~$N$ Moran time steps constitute a generation time~$\tau_{g}$. Thus, in the Moran model, we have

\begin{equation}
\label{EGeneticDiffusionCoefficient}
\mathfrak{D}_{g}=\frac{2}{N\tau_{g}}.
\end{equation}

\noindent Alternative reproduction schemes, such as Wright-Fisher sampling,~\cite{Crow:PopulationGenetics,Hartl:PopulationGenetics} lead to an equation identical to Eq.~(\ref{EMoranGeneticDriftFokkerPlank}), but with a different numerical coefficient in Eq.~(\ref{EGeneticDiffusionCoefficient}).

Equation~(\ref{EMoranGeneticDriftFokkerPlank}) is subject to absorbing boundary conditions~\footnote{Since Eq.~\protect{\ref{EMoranGeneticDriftFokkerPlank}} is singular at the boundaries, we require~$\lim_{f\rightarrow 0,1}f(1-f)P(t,f)=0$. See~\cite{Risken:FPE} and~\cite{Kimura:GeneticDrift} for a more detailed discussion.} at~$f=0$~and~$f=1$ because, if one of the alleles is lost, it cannot appear again in the absence of mutation. Therefore the population  eventually becomes fixed at one of the absorbing states. We calculate the rate of the fixation by considering the average heterozygosity of the population

\begin{equation}
\label{EDefinitionH}
H(t)\equiv \langle h(t) \rangle=\langle 2f(t)[1-f(t)] \rangle,
\end{equation}

\noindent which is the~(averaged over realizations) probability that two randomly selected individuals have different alleles. When the population is close to the fixation~($f\approx 0$ or $f\approx 1$), the heterozygosity is close to zero. The equations of motion for~$F(t) \equiv \langle f(t) \rangle$ and~$H(t)$ follow from Eq.~(\ref{EMoranGeneticDriftFokkerPlank}) by multiplying both sides with~$f$ or~$h$, integrating over~$f$, and eliminating the derivatives with respect to~$f$ via integration by parts. The results are

\begin{equation}
\label{EMoranGeneticDriftF}
\frac{dF(t)}{dt}=0,
\end{equation}

\begin{equation}
\label{EMoranGeneticDriftH}
\frac{dH(t)}{dt}=-\mathfrak{D}_{g}H(t).
\end{equation}

\noindent Equations~(\ref{EMoranGeneticDriftF})~and~(\ref{EMoranGeneticDriftH}) imply that, while the average frequencies of these neutral alleles do not change~$F=\langle f \rangle = f(t=0)\equiv F_{0}$, the population reaches fixation exponentially fast,~$H(t)=H(0)e^{-\mathfrak{D}_{g}t}=F_{0}(1-F_{0})e^{-\mathfrak{D}_{g}t}$.

The average heterozygosity is closely related to the variance of~$f(t)$, the fraction of the first allele,

\begin{equation}
\label{EHVarianceWellMixed}
V(t)=\langle(f(t)-\langle f(t)\rangle)^{2}\rangle=F(t)[1-F(t)]-\frac{1}{2}H(t).
\end{equation}

\noindent Thus, even if a population starts with zero variance, the fluctuations grow until the variance reaches its maximum value of~$F_{0}(1-F_{0})$, which corresponds to a population fixed to allele one with probability~$F_{0}$ and to allele two with probability~$1-F_{0}$. Note that, for small~$t$, $V(t)$~grows linearly with time, but, at large times, the variance approaches its limiting value exponentially fast. The linear growth of variance at small times also follows from the Fokker-Planck equation because, at small times, Eq.~(\ref{EMoranGeneticDriftFokkerPlank}) can be approximated by a diffusion equation with a constant diffusivity. 

Next, we generalize Eq.~(\ref{EMoranGeneticDriftFokkerPlank}) to account for mutations. In the Moran model, mutation is included at the end of a time step by allowing the offspring to mutate with probability~$\tilde{\mu}_{12}$ from allele one to allele two and with probability~$\tilde{\mu}_{21}$ from allele two to allele one. If the frequency of allele one at time step~$\tilde{t}$ is~$f(\tilde{t})$, then, at the next time step, the expectation value of~$f(\tilde{t}+1)$ is given by

\begin{equation}
\label{EMoranMutation}
\langle f(\tilde{t}+1)\rangle=f(\tilde{t})+\frac{\tilde{\mu}_{21}[1-f(\tilde t)]-\tilde{\mu}_{12}f(\tilde{t})}{N},
\end{equation}

\noindent and the variance of~$f(\tilde{t}+1)$ is given by Eq.~(\ref{EMoranGeneticDriftSecond}) to the leading order in the mutation rates and the inverse population size.

Since the expectation value of~$f(\tilde{t})$ changes with time, mutation leads to an~$f$-dependent drift term in the Fokker-Planck equation. Upon recalling that~$N$ Moran time steps equal one generation time, we have

\begin{equation}
\label{EMoranGeneticDriftMutationFokkerPlank}
\begin{split}
\frac{\partial P(t,f)}{\partial t}=&-\frac{\partial}{\partial f}\left\{ [\mu_{21}-(\mu_{12}+\mu_{21})f]P(t,f) \right\} \\&
+\frac{\mathfrak{D}_{g}}{2}\frac{\partial^{2}}{\partial f^{2}}\left[f(1-f)P(t,f)\right],
\end{split}
\end{equation}

\noindent where~$\mu_{12}\equiv\tilde{\mu}_{12}\tau_{g}^{-1}$~and~$\mu_{12}\equiv\tilde{\mu}_{21}\tau_{g}^{-1}$ are the mutation rates per generation.

\begin{figure}
\psfrag{P}{\hspace{-0.9cm}$\displaystyle\bm{P(\infty,f)}$}
\psfrag{f}{\hspace{0.0cm}$\displaystyle\bm{f}$}
\includegraphics[width=\columnwidth]{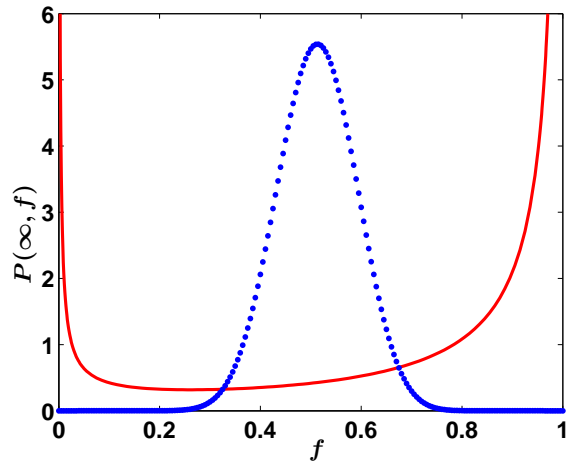}
\caption{(Color online) The stationary distribution~$P(\infty,f)$ in the presence of selection, mutation, and genetic drift, see Eqs.~(\protect{\ref{EMoranFullFokkerPlank}}) and~(\protect{\ref{EStationaryFull}}). The blue dotted line shows~$P(\infty,f)$ for~$s=\mathfrak{D}_{g}$, and~$\mu_{12}=\mu_{21}=10\mathfrak{D}_{g}$, which corresponds to weak genetic drift,~$\mu_{12}/D_{g}\gg1$. The red solid line shows~$P(\infty,f)$ for~$s=\mathfrak{D}_{g}$, and~$\mu_{12}=\mu_{21}=0.1\mathfrak{D}_{g}$, which corresponds to strong genetic drift,~$\mu_{12}/D_{g}\ll1$. Note the difference in curvature between the two cases and the fact that the distribution is dominated by the central region in the weak genetic drift limit, but by the tails in the opposite limit. The transition between these two regimes occurs when the mutation rates equal $\mathfrak{D}_{g}/2$, and~$P(\infty,f)$ diverges at~$f=0$ and~$f=1$. Also note that the effect of natural selection is to bias the distribution toward~$f=1$. As~$s$ increases, the maximum of the distribution shifts to the right for weak genetic drift, and the right tail of the distribution becomes more prominent for strong genetic drift.}
\label{FStationaryDistribution}
\end{figure}

Because the alleles can mutate into each other, the probability flux through the boundaries must be zero, so Eq.~(\ref{EMoranGeneticDriftMutationFokkerPlank}) has reflecting boundary conditions, and a nontrivial stationary solution for~$P(t,f)$ exists. While the stationary distribution can be obtained easily, see Eq.~(\ref{EStationaryFull}), Fig.~\ref{FStationaryDistribution}, and~\textcite{Crow:PopulationGenetics}, it is sufficient to analyze the moments~$F(t)$~and~$H(t)$ introduced above. This will also allow us to make a direct comparison with the corresponding solutions of the one-dimensional stepping stone model in Sec.~\ref{SNeutralMutation}. The equations of motion for~$F(t)$~and~$H(t)$ are obtained from~Eq.~(\ref{EMoranGeneticDriftMutationFokkerPlank}) in the same way as for the absence of mutation. The results are

\begin{equation}
\label{EMoranGeneticDriftMutationF}
\frac{dF(t)}{dt}=\mu_{21}-(\mu_{12}+\mu_{21})F(t),
\end{equation}

\begin{equation}
\label{EMoranGeneticDriftMutationH}
\begin{split}
\frac{dH(t)}{dt}=&-(\mathfrak{D}_{g}+2\mu_{12}+2\mu_{21})H(t)\\&+2[\mu_{21}+(\mu_{12}-\mu_{21})F(t)].
\end{split}
\end{equation}

\noindent Since these equations are linear differential equations with constant coefficients, the equilibrium is approached exponentially fast. The stationary solutions, which are obtained in the limit~$t\rightarrow\infty$, are given below

\begin{equation}
\label{EMoranGeneticDriftMutationFSolution}
F(\infty)=\frac{\mu_{21}}{\mu_{12}+\mu_{21}},
\end{equation}

\begin{equation}
\label{EMoranGeneticDriftMutationHSolution}
H(\infty)=\frac{2F(\infty)[1-F(\infty)]}{1+\frac{\mathfrak{D}_{g}}{2(\mu_{12}+\mu_{21})}}.
\end{equation}

From~Eqs.~(\ref{EMoranGeneticDriftMutationHSolution}) and~(\ref{EHVarianceWellMixed}), we see that, when the population size is large enough, i.e.~$\mathfrak{D}_{g}\ll(\mu_{12}+\mu_{21})$,~$H(\infty)\approx 2F(\infty)[1-F(\infty)]$, the stationary value of the heterozygosity is consistent with~$f(t)\approx F(\infty)$. Thus~$V(\infty)\approx0$, and the fluctuations of~$f(t)$ are negligible. In the opposite limit,~$H(\infty)\propto\frac{\mu_{12}+\mu_{21}}{\mathfrak{D}_{g}}$ is significantly smaller, which suggests that most of the time the population is fixed to one of the alleles, and mutations lead to rare transitions between states with~$f=0$~and~$f=1$. Consequently, the stationary distribution is dominated by the regions around~$f=0$ and~$f=1$, as one can see in Fig.~\ref{FStationaryDistribution}. Our interpretation of~Eq.~(\ref{EMoranGeneticDriftMutationHSolution}) is consistent with a more rigorous analytical and numerical analysis by Duty~\cite{Duty:Thesis}.

Finally, we introduce Darwinian natural selection, which is usually related to the difference in the reproduction or survival probability of the organisms. In the continuous time limit considered here, both mechanisms of selection lead to the same dynamics; therefore, we only consider selection due to different growth rates. In the Moran model, a growth rate difference is embodied in modified probabilities of reproduction: the individual with allele one is chosen to reproduce not with probability~$f$ but with probability~$\frac{w_{1}f}{w_{1}f+w_{2}(1-f)}$, where~$w_{1}$~and~$w_{2}$ are the fitnesses~(i.e., growth rates) of alleles one and two respectively. In the absence of mutations, this modification results in

\begin{equation}
\label{EMoranFitness}
\langle f(\tilde{t}+1)\rangle=f(\tilde{t})+\frac{f(\tilde{t})[1-f(\tilde{t})](w_{1}-w_{2})}{N\{w_{1}f(\tilde{t})+w_{2}[1-f(\tilde{t})]\}}.
\end{equation}

\noindent When selection is weak, that is~$|w_{1}-w_{2}|\ll w_{1}+w_{2}$, Eq.~(\ref{EMoranFitness}) reduces to

\begin{equation}
\label{EMoranSelection}
\langle f(\tilde{t}+1)\rangle=f(\tilde{t})+\frac{\tilde{s}}{N}f(\tilde{t})[1-f(\tilde{t})],
\end{equation}

\noindent where~$\tilde{s}=2(w_{1}-w_{2})/(w_{1}+w_{2})$ is the selective advantage of allele one, which has to be much smaller than one for the approximation to hold. When~$\tilde{s}>0$, allele one is advantageous; for~$\tilde{s}<0$, it is deleterious. In the following, we assume that allele one is advantageous because one can always relabel the alleles to satisfy this condition.

Similar to the case of mutations without selection, the variance of~$f(\tilde{t}+1)$ is given by Eq.~(\ref{EMoranGeneticDriftSecond}) to the leading order in~$\tilde{s}$~and~$N^{-1}$, and the corresponding Fokker-Planck equation acquires an $f$-dependent drift term due to selection:

\begin{equation}
\label{EMoranGeneticDriftSelectionFokkerPlank}
\begin{split}
\frac{\partial P(t,f)}{\partial t}=&-s\frac{\partial}{\partial f}\left[f(1-f)P(t,f)\right] \\&
+\frac{\mathfrak{D}_{g}}{2}\frac{\partial^{2}}{\partial f^{2}}\left[f(1-f)P(t,f)\right],
\end{split}
\end{equation}

\noindent where~$s=\tilde{s}\tau_{g}^{-1}$. The equations for~$F$~and~$H$ are not as useful as before because the hierarchy of the moment equations does not close. Nevertheless, Eq.~(\ref{EMoranGeneticDriftSelectionFokkerPlank}) can be easily analyzed in two limits. When the population size is large~($\mathfrak{D}_{g} \ll s$), fluctuations are not important, and~$\frac{dF(t)}{dt}\approx sF(t)[1-F(t)]$. Upon setting~$F_{0} \equiv F(0)$, we have

\begin{equation}
\label{EDeterministicSecetiveSweepWellMixed}
F(t)\approx\frac{1}{1+\frac{1-F_{0}}{F_{0}}e^{-st}},
\end{equation}

\noindent so the selective sweep is exponentially fast. When the fluctuations dominate the dynamics, the selection slightly increases the odds of fixation of the advantageous allele, but does not significantly affect the rate of fixation. For a detailed analysis of Eq.~(\ref{EMoranGeneticDriftSelectionFokkerPlank}) see~\textcite{Crow:PopulationGenetics}.

In the continuous limit, the population genetics of a well-mixed effectively zero-dimensional population with genetic drift, selection, and mutation is summarized by the following Fokker-Planck~(or forward Kolmogorov) equation:

\begin{equation}
\label{EMoranFullFokkerPlank}
\begin{split}
\frac{\partial P(t,f)}{\partial t}=&-s\frac{\partial}{\partial f}\left[f(1-f)P(t,f)\right] \\&
-\frac{\partial}{\partial f}\left\{ [\mu_{21}-(\mu_{12}+\mu_{21})f]P(t,f) \right\} \\&
+\frac{\mathfrak{D}_{g}}{2}\frac{\partial^{2}}{\partial f^{2}}\left[f(1-f)P(t,f)\right].
\end{split}
\end{equation}

\noindent The stationary distribution for Eq.~(\ref{EMoranFullFokkerPlank}) is reached exponentially fast and takes the following form~\cite{Crow:PopulationGenetics,Duty:Thesis}

\begin{equation}
\label{EStationaryFull}
P(\infty,f)=C e^{2sf/\mathfrak{D}_{g}}f^{2\mu_{21}/D_{g}-1}(1-f)^{2\mu_{12}/D_{g}-1},
\end{equation}

\noindent where~$C$ is the normalization constant chosen to set~$\int_{0}^{1}P(\infty,f)df=1$. This stationary distribution is plotted in Fig.~\ref{FStationaryDistribution} for both strong and weak genetic drift. 

Although the formulation in terms of a Fokker-Plank equation is appropriate for nonspatial models, an alternative formulation via a stochastic differential equation can be generalized to spatial models more easily. Equation~(\ref{EMoranFullFokkerPlank}) is equivalent to

\begin{equation}
\label{EMoranStochastic}
\begin{split}
\frac{df(t)}{dt}=&sf(t)[1-f(t)]+\mu_{21}-(\mu_{12}+\mu_{21})f(t) \\ &+\sqrt{\mathfrak{D}_{g}f(t)[1-f(t)]}\Gamma(t) \quad \mbox{(It\^{o})},
\end{split}
\end{equation}

\begin{equation}
\label{EMoranStochasticGamma}
\langle\Gamma(t_{1})\Gamma(t_{2})\rangle=\delta(t_{1}-t_{2}),
\end{equation}

\noindent where~$\Gamma(t)$ is a white, zero mean Gaussian noise, and~$\delta(t)$ is Dirac's delta-function; to get the correct Fokker-Planck Equation~(\ref{EMoranFullFokkerPlank}), one must use It\^o's prescription to define how Eq.~(\ref{EMoranStochastic}) steps the dynamics forward in time. This interpretation of the noise term ensures that~$f(t)$ depends only on~$\Gamma(t')$ with~$t'<t$ as it is appropriate for population genetics. It\^o's prescription is adopted throughout the paper, and a brief introduction to the It\^o calculus is given in Appendix~\ref{AIto}~[see also~\cite{Risken:FPE,Gardiner:Handbook,Duty:Thesis}]. In Sec.~\ref{S1dSSM}, we use Eq.~(\ref{EMoranStochastic}) to formulate the stepping stone model in one dimension.

Well-mixed-population-models do not describe migration and subdivision of natural populations~\cite{Hartl:PopulationGenetics}. To remedy this deficiency, two common approaches exist: to assume a uniformly populated spatial habitat with free diffusion or to assume a patchy habitat with a prescribed pattern of limited migration between the patches. The former is the subject of this paper, and can be regarded as the continuum limit of the stepping stone model~\cite{KimuraWeiss:SSM}, see Sec.~\ref{S1dSSM}. The simplest variant of the latter approach is known as the island model~\cite{Wright:IslandModel}. The island model assumes that all patches or islands have the same number of organisms and populations in every patch obey well-mixed-population dynamics. The migration occurs between \textit{any} two patches with equal probability, so, in some sense, this is a mean field or infinite-dimensional model. The island model successfully predicts that the organisms are more likely to be related locally than globally, but most of its predictions are similar to those of well-mixed-population models because the migration does not account for spatial structure. In the limit of an infinitely large number of islands, the effect of migration in and out of any patch is equivalent to an effective mutation rate; however, this is not the case in a one-dimensional model considered below. 


\section{One-Dimensional Stepping Stone Model}
\label{S1dSSM}

In Sec.~\ref{SWellMixed}, we formulated a model to describe genetic drift, mutation, and selection in an effectively zero-dimensional habitat. For a one-dimensional population considered in this section, we extend the model to account for short range migrations during every generation. Migration is usually modeled either as exchange of individuals between neighboring island populations~(demes)~\cite{Wright:IslandModel,KimuraWeiss:SSM} or as dispersal of offspring or adults within a continuous population~\cite{Wright:IsolationByDistance,Nagylaki:DecayGeneticVariability,Malecot:Dynamics}. Although the first approach was developed to model patchy populations, it can be used to describe continuous populations if the deme sizes are much smaller than the whole population, and spatial variations are gradual. In this limit, both migration models should give essentially the same results. Here, we adopt the first approach because it is conceptually simpler.

To specify the one-dimensional stepping stone model, we consider an infinite set of demes arranged on a line. Neighboring demes are separated by distance~$a$ and indexed by an integer~$l=-\infty,...,-1,0,1,...,\infty$. Each deme has~$N$ organisms~(but the total population size is infinite), and the frequency of allele one in deme~$l$ is~$f_{l}(t)$. Migration occurs only between nearest neighbors, and, every generation, a deme exchanges~$\tilde{m}N/2$ individuals with its right neighbor and~$\tilde{m}N/2$ individuals with its left neighbor. We assume that the exchange fraction~$\tilde{m}$ is much smaller than one, and that the individuals of both allelic types are equally likely to be exchanged. Thus, in one generation,~$\langle f_{l}\rangle$ changes by~$\tilde{m}(f_{l-1}+f_{l+1}-2f_{l})/2$ due to migration. The variance of~$f_{l}$ grows due to randomness in the exchange process, but this increase is negligible compared to the genetic drift within an island. In the continuous time limit, $f_{l}(t)$~obeys the following generalization of Eq.~(\ref{EMoranStochastic}):

\begin{equation}
\label{EDiscrete1dSSM}
\begin{split}
\frac{df_{l}}{dt}=&\frac{m}{2}(f_{l-1}+f_{l+1}-2f_{l})+sf_{l}(1-f_{l})+\mu_{21}\\ &-(\mu_{12}+\mu_{21})f_{l}  +\sqrt{\mathfrak{D}_{g}f_{l}(1-f_{l})}\Gamma_{l} \quad \mbox{(It\^{o})},
\end{split}
\end{equation}

\begin{equation}
\label{EDiscrete1dSSMGamma}
\langle\Gamma_{l_{1}}(t_{1})\Gamma_{l_{2}}(t_{2})\rangle=\delta_{l_{1}l_{2}}\delta(t_{1}-t_{2}),
\end{equation}

\noindent where~$m=\tilde{m}\tau^{-1}_{g}$ and $\delta_{l_{1}l_{2}}$~is Kronecker's delta. We can also write Eq.~(\ref{EDiscrete1dSSM}) in the continuous space limit by introducing a spatial coordinate~$x=la$,

\begin{equation}
\label{EContinuous1dSSM}
\begin{split}
\frac{\partial f}{\partial t}=&D_{s}\frac{\partial^{2}f}{\partial x^{2}}+sf(1-f)+\mu_{21}-(\mu_{12}+\mu_{21})f \\ &+\sqrt{D_{g}f(1-f)}\Gamma \quad \mbox{(It\^{o})},
\end{split}
\end{equation}

\begin{equation}
\label{EContinuous1dSSMGamma}
\langle\Gamma(t_{1},x_{1})\Gamma(t_{2},x_{2})\rangle=\delta(t_{1}-t_{2})\delta(x_{1}-x_{2}),
\end{equation}

\noindent where the spatial and genetic diffusion constants are~$D_{s}=ma^{2}/2$ and $D_{g}=a\mathfrak{D}_{g}=(2a)/(\tau_{g}N)$ respectively. Thus the continuous time and space limit the stepping stone model is described by the stochastic Fisher-Kolmogorov-Petrovsky-Piscounov equation~\cite{Fisher:FisherWave,Kolmogorov:FKPPEquation} with additional terms describing mutation.

Similar to the analysis of the well-mixed-population model discussed in Sec.~\ref{SWellMixed}, we use equal-time correlation functions of~$f(t,x)$ to characterize the dynamics of the stepping stone model. The spatial versions of the average frequency and heterozygosity are defined as follows:

\begin{equation}
\label{ESpatialF}
F(t,x)=\langle f(t,x) \rangle,
\end{equation}

\begin{equation}
\label{ESpatialH}
H(t,x_{1},x_{2})=\langle f(t,x_{1})[1-f(t,x_{2})] \rangle + \langle f(t,x_{2})[1-f(t,x_{1})] \rangle.
\end{equation}

\noindent The equation of motion for~$F(t,x)$ depends on~$H(t,x,x)$, and is readily derived by averaging Eq.~(\ref{EContinuous1dSSM}), which gives

\begin{equation}
\label{EContinuous1dSSMF}
\frac{\partial F}{\partial t}=D_{s}\frac{\partial^{2}F}{\partial x^{2}}+\mu_{21}-(\mu_{12}+\mu_{21})F+\frac{s}{2}H(t,x,x).
\end{equation}

\noindent The dynamics of~$H(t,x_{1},x_{2})$ is obtained by differentiating Eq.~(\ref{ESpatialH}) with respect to~$t$ and then eliminating~$\frac{\partial f}{\partial t}$ with the help of Eq.~(\ref{EContinuous1dSSM}). Note, It\^o's formula~(see Appendix~\ref{AIto}) must be used to differentiate Eq.~(\ref{ESpatialH}) correctly. The result is

\begin{equation}
\label{EContinuous1dSSMH}
\begin{split}
\frac{\partial}{\partial t}H(t,x_{1},x_{2})=&D_{s}\left(\frac{\partial^{2}}{\partial x_{1}^{2}}+\frac{\partial^{2}}{\partial x_{2}^{2}}\right)H(t,x_{1},x_{2})\\ &-D_{g}H(t,x_{1},x_{2})\delta(x_{1}-x_{2})\\ &-2(\mu_{12}+\mu_{21})H(t,x_{1},x_{2})\\&+ 2\mu_{21}+(\mu_{12}-\mu_{21})[F(t,x_{1})+F(t,x_{2})]\\ &+\frac{s}{2}[H(t,x_{1},x_{1})+H(t,x_{2},x_{2})]\\&-s\langle2f(t,x_{1})[1-f(t,x_{1})]f(t,x_{2})\rangle \\ &-s\langle2f(t,x_{2})[1-f(t,x_{2})]f(t,x_{1})\rangle.
\end{split}
\end{equation}

\noindent Equations~(\ref{EContinuous1dSSMF}) and~(\ref{EContinuous1dSSMH}) agree with the ones derived in~\textcite{Nagylaki:Hierarchy} in the limit of no mutations considered there.

From Eq.~(\ref{EContinuous1dSSMH}), one can see that the hierarchy of the moment equations does not close unless selection is absent. Similar to the well-mixed case, the correlation functions for neutral models with and without mutations can be found analytically, see Secs.~\ref{SNeutralMutation}~and~\ref{SNeutralNoMutation}, but different methods are required to analyze the dynamics in the presence of selection, see Sec.~\ref{SSelection}. To simplify the analysis, we consider well-mixed, spatially homogeneous initial conditions. Then~$F$ is only a function of~$t$, and~$H$ is a function of~$t$ and~$x=x_{1}-x_{2}$. With these simplifying assumptions, the equations of motion for~$F(t)$ and~$H(t,x)$ take the following form:

\begin{equation}
\label{EContinuous1dSSMFHomogeneous}
\frac{dF(t)}{dt}=\mu_{21}-(\mu_{12}+\mu_{21})F(t)+\frac{s}{2}H(t,0),
\end{equation}

\begin{equation}
\label{EContinuous1dSSMHHomogeneous}
\begin{split}
\frac{\partial}{\partial t}H(t,x)=&2D_{s}\frac{\partial^{2}}{\partial x^{2}}H(t,x)-D_{g}H(t,0)\delta(x) \\ &-2(\mu_{12}+\mu_{21})H(t,x)\\&+ 2\mu_{21}+2(\mu_{12}-\mu_{21})F(t,x)+sH(t,0)\\ &-2s\langle2f(t,0)[1-f(t,0)]f(t,x)\rangle.
\end{split}
\end{equation}


\section{Neutral Model Without Mutations}
\label{SNeutralNoMutation}

We start the analysis of the one-dimensional stepping stone model by considering neutral alleles that do not mutate. In practice, this means~$N^{2}\tilde{\mu}_{12},N^{2}\tilde{\mu}_{21}\ll1$ and~$N^{2}\tilde{s}\ll1$~(as we show below). Although these assumptions are not always realistic, they help to clarify the role of genetic drift in a spatial context. In addition, neglecting mutations is a good approximation on time scales shorter than the waiting times for the mutations~$\mu_{12}^{-1}$ and~$\mu_{21}^{-1}$. Under these assumptions, $F$~does not change,~$F(t)=F_{0}$, and Eq.~(\ref{EContinuous1dSSMHHomogeneous}) reads

\begin{equation}
\label{EHNoMutationEquation}
\frac{\partial}{\partial t}H(t,x)=2D_{s}\frac{\partial^{2}}{\partial x^{2}}H(t,x)-D_{g}H(t,0)\delta(x).
\end{equation}

Equation~(\ref{EHNoMutationEquation}) can also be derived by tracing the ancestral lineages of organisms backward in time. The average spatial heterozygosity~$H(t,x)$ is the average probability of sampling two different individuals chosen at time~$t$ from demes separated by distance~$x$. As we trace the lineages of the two sampled organisms backward in time, the lineages diffuse in space due to migration and, when they are at the same point, they have a chance to coalesce, in which case the sampled organisms must be identical because they have a common ancestor. Such a coalescence event changes the probability of being different from~$H(t,0)$ to~$0$, and acts like a sink at~$x=0$. The first term on the right hand side of Eq.~(\ref{EHNoMutationEquation}) describes the diffusion, and the second term describes the coalescence. Since this argument is valid for an arbitrary number of alleles, Eq.~(\ref{EHNoMutationEquation}) is valid for an arbitrary number of spatially diffusing neutral alleles. See Appendix~\ref{AMultipleColors} for a more detailed discussion of the $q$-allele problem, with~$q\ge3$.

To better understand the microbiology experiments on neutral alleles by~\textcite{HallatschekNelson:ExperimentalSegregation}, we consider uncorrelated initial conditions~$F(0)=F_{0}$~and~$H(0,x)=H_{0}$, where~$F_{0}$ is the fraction of allele one and~$H_{0}=2F_{0}(1-F_{0})$, which is the heterozygosity of a well-mixed population with the frequency of allele one equal to~$F_{0}$. For these initial conditions, Eq.~(\ref{EHNoMutationEquation}) is solved in Appendix~\ref{ANeutral}. The results are

\begin{equation}
\label{EHSolutionNoMutation}
\begin{split}
H(t,x)=&H_{0}-D_{g}\int_{0}^{t}dt'e^{-\frac{x^{2}}{8D_{s}(t-t')}}\frac{H(t',0)}{\sqrt{8\pi D_{s}(t-t')}},
\end{split}
\end{equation}

\begin{equation}
\label{EH0SolutionNoMutation}
H(t,0)=H_{0}\erfc\left(\sqrt{\frac{D_{g}^{2}t}{8D_{s}}}\right)e^{\frac{D_{g}^{2}t}{8D_{s}}},
\end{equation}

\noindent where~$\erfc(y)$ is the complementary error function.

The spatial heterozygosity at vanishing separation,~$H(t,0)$, is particularly interesting because it indicates the degree of spatial segregation: if~$H(t,0)\ll1$, then, locally, the demes are fixed to one of the two alleles. From Eq.~(\ref{EH0SolutionNoMutation}), one can see that, for~$t\gg 8D_{s}/D_{g}^2$,

\begin{equation}
\label{EH0LimitNoMutation}
H(t,0)=H_{0}\left(\frac{\pi D_{g}^2 t}{8D_{s}}\right)^{-1/2}+O(t^{-3/2}),
\end{equation}

\noindent which means that at long times one of the alleles reaches fixation locally. Therefore we see that the spatial model we are considering is consistent with the experiments by~\textcite{HallatschekNelson:ExperimentalSegregation}~(see Fig.~\ref{FSpatialSegregation}) because it predicts the formation of domains~(regions of local fixation). Thus, similar to the well-mixed model considered in Sec.~\ref{SWellMixed}, one of the alleles reaches fixation locally with fixation time~$\tau_{f}=8D_{s}/(\pi D_{g}^{2})\sim N^{2}$. But not only is this fixation time proportional to~$N^{2}$, instead of~$N$, the functional form of heterozygosity decay is different: instead of a rapid exponential decay, the spatial model shows a slow algebraic decay of local heterozygosity. These results agree with the previous works on population genetics by Mal\'ecot~\cite{Malecot:Dynamics} and Nagylaki~\cite{Nagylaki:DecayGeneticVariability}. Local fixation and~$t^{-1/2}$ decay of local heterozygosity have also been found in the voter model~\cite{CoxGriffeath:VoterModel}, which corresponds to the stepping stone model with~$N=1$ (see Appendix~\ref{AVoter}).  

The characteristic demixing time can also be estimated by the following scaling argument: the characteristic population size at time~$t$ in the coarsening process is~$N_{ch}(t)\sim n_{0}\sqrt{tD_{s}}$, where the population density~$n_{0}\sim N/a$. Upon recalling that the fixation time in zero dimensions is~$\tau_{f}\sim N_{ch}(\tau_{f})\tau_{g}$, and solving self-consistently for~$\tau_{f}$, we have~$\tau_{f}\sim D_{s}N^{2}\tau_{g}^{2}/a^{2}\sim D_{s}/D_{g}^{2}\sim N^{2}\tau_{g}$.

\begin{figure}
\includegraphics[width=\columnwidth]{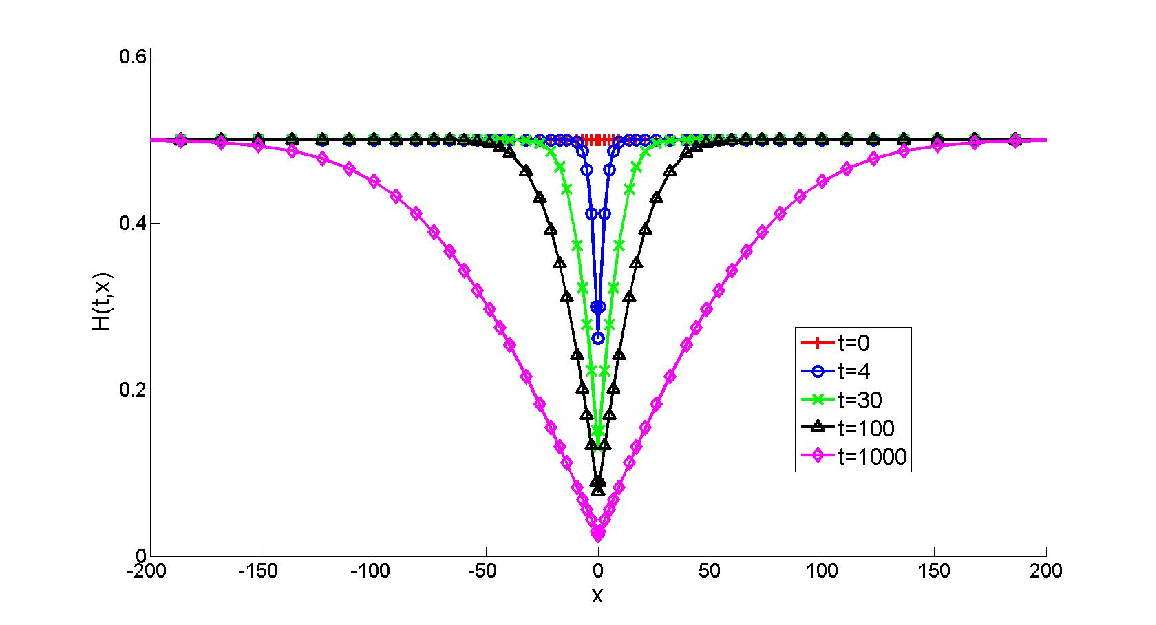}
\caption{(Color online) Solutions of Eq.~(\protect{\ref{EHNoMutationEquation}}) at various times, given~$H(0,x)=\frac{1}{2}$. Time and distance are measured in units such that~$D_{s}=1$ and~$D_{g}=1$. Time increases from the top curves to the bottom curves. Note the statistical reflection symmetry,~$H(t,x)=H(t,-x)$.}
\label{FHNoMutationTheory}
\end{figure}

Another important characteristic of~$H(t,x)$ is the length scale over which~$H(t,x)$ changes from its minimum to its maximum values. Figure~\ref{FHNoMutationTheory} plots Eq.~(\ref{EHSolutionNoMutation}) and shows the spatial variation of~$H(t,x)$ at different times. One can see that the spatial heterozygosity is reduced near the origin due to the local fixation, but~$H(t,x)$ rises to its initial value~$H_{0}$ at large~$x$, where the alleles remain uncorrelated. After the domains form, i.e. for~$t\gg 8D_{s}/D_{g}^2$, this change from~$H(t,0)$ to~$H(0,x)$ happens on a length scale that is set by the average size of the domains~$\ell$, which is proportional to the diffusion length~$\sqrt{2D_{s}t}$, as follows from Eq.~(\ref{EHSolutionNoMutation}). Since this characteristic length scale changes with time, it is convenient to rescale distances:~$\bar{x}=x/\sqrt{2D_{s}t}$. Upon using Eq.~(\ref{EH0LimitNoMutation}) to simplify Eq.~(\ref{EHSolutionNoMutation}), we see that~$H(t,x)$ approaches a nontrivial limit in terms of~$\bar{x}$ as time goes to infinity:

\begin{equation}
\label{ENoMutationHLimitingShape}
H(t,\bar{x})\stackrel{t\rightarrow\infty}{\longrightarrow}H_{0}\left(1-\frac{1}{\pi}\int_{0}^{1}\frac{d\zeta}{\sqrt{\zeta(1-\zeta)}}e^{-\frac{|\bar{x}|^{2}}{4\zeta}}\right),
\end{equation}

\noindent which agrees with the known results for the voter model~\cite{CoxGriffeath:VoterModel}.

A more precise evaluation of the domain density and hence an average domain size~$\ell(t)$ can be obtained from~$H(t,x)$, as shown in Appendix~\ref{ADomainSize}. From Eq.~(\ref{EDomainSizeH0}), we know that~$\ell(t)=\frac{4D_{s}}{D_{g}H(t,0)}$, so using Eq.~(\ref{EH0LimitNoMutation}) we see that

\begin{equation}
\label{EDomainSizeNoMutation}
\ell=\frac{\sqrt{2\pi D_{s}t}}{2f_{0}(1-f_{0})},
\end{equation}

\noindent which is consistent with the analysis of~\textcite{Hallatschek:LifeFront}. Note that the genetic diffusion constant~$D_{g}\sim1/N$ drops out because, at large times, the only dynamics left is the diffusive motion of the domain walls. With neutral alleles, these boundaries behave as annihilating random walks, and the average domain size can be easily calculated~\cite{Hallatschek:LifeFront}.

Equations~(\ref{EH0SolutionNoMutation}) and (\ref{EDomainSizeNoMutation}) suggest that the processes driven by the genetic drift slow down with time because the logarithmic time derivatives of~$H(t,0)$ and~$\ell$ tend to zero as time goes to infinity. In the annihilating random walk picture of~\textcite{Hallatschek:LifeFront}, annihilations become rarer and rarer as the coarsening progresses. A more direct measure of genetic drift, which is also interesting from the biological point of view, is the fluctuations of the total fraction of, say, the first allele~$\mathfrak{f}(t)$ in a finite population of length~$L$. We define~$\mathfrak{f}(t)$~as

\begin{equation}
\label{ETotalFraction}
\mathfrak{f}(t)=\frac{1}{L}\int_{0}^{L}f(t,x)dx,
\end{equation}
 
\noindent and compute its variance~$\nu(t)$ to characterize its fluctuations.

Upon integrating Eq.~(\ref{EContinuous1dSSM})over~$x$ with~$s=\mu_{12}=\mu_{21}=0$, we obtain the equation of motion for~$\mathfrak{f}$:

\begin{equation}
\label{ETotalFractionEquationMotion}
\frac{d\mathfrak{f}}{dt}=\frac{1}{L}\int_{0}^{L}\sqrt{D_{g}f(t,x)[1-f(t,x)]}\Gamma(t,x)dx,
\end{equation}

\noindent where the spatial diffusion term vanishes after integration by parts provided periodic or Newman boundary conditions are imposed. Upon noting that~$\langle \mathfrak{f}\rangle=F=\rm{const}$~(the It\^{o} interpretation of the noise~$\Gamma(t,x)$ is crucial here) and defining

\begin{equation}
\nu=\langle\mathfrak{f}^{2}\rangle-\langle\mathfrak{f}\rangle^{2},
\label{ENu}
\end{equation}

\noindent one finds immediately that $\frac{d\nu}{dt}=\frac{d}{dt}\langle\mathfrak{f}^{2}\rangle$. To evaluate the time derivative, we use the rules of the It\^{o} calculus, sketched in Appendix~\ref{AIto}, and find

\begin{equation}
\label{ETotalFractionVarianceEquationMotion}
\begin{split}
\frac{d\nu(t)}{dt}=&\frac{1}{L^{2}}\int_{0}^{L}\int_{0}^{L}\sqrt{D_{g}f(t,x_{1})[1-f(t,x_{1})]}\\ &\times\sqrt{D_{g}f(t,x_{2})[1-f(t,x_{2})]}\delta(x_{1}-x_{2})dx_{1}dx_{2},
\end{split}
\end{equation}

\noindent where the delta function comes from averaging over the noise and using Eq.~(\ref{EContinuous1dSSMGamma}). From Eq.~\ref{ETotalFractionVarianceEquationMotion}, it follows that

\begin{equation}
\label{ETotalFractionVarianceSolution}
\nu(t)=\frac{D_{g}}{2L}\int_{0}^{t}H(t',0)dt',
\end{equation}

\noindent where we assume~$\nu(0)=0$. Hence, we know~$\nu(t)$ exactly because~$H(t,0)$ is given by Eq.~(\ref{EH0SolutionNoMutation}). 

For small times,~$t\ll8D_{s}/D_{g}^{2}$, the variance grows linearly with time. For large times, we can use the asymptotic expansion of~$H(t,0)$ given by Eq.~(\ref{EH0LimitNoMutation}) to calculate~$\nu(t)$. The result is

\begin{equation}
\label{ETotalFractionVarianceSolutionLimit}
\nu(t)=\frac{H_{0}\sqrt{8D_{s}t}}{\sqrt{\pi}L}+O\left(\frac{D_{s}}{D_{g}L}\right).
\end{equation}

\noindent Equation~(\ref{ETotalFractionVarianceSolutionLimit}) is consistent with~\textcite{Bramson:FiniteSize}, and we immediately conclude that the standard deviation~$\Delta(t)=\sqrt{\nu(t)}$ grows as~$t^{1/4}$ for large times! This important result is generalized for the flat-front and undulating-front models with $q$-alleles in Appendix~\ref{AMultipleColors} by approximating the dynamics of the domain boundaries by annihilating random walks. Thus,~$\mathfrak{f}$ performs a subdiffusive random walk, and genetic drift of the global frequency~$\mathfrak{f}(t)$ becomes weaker with time. Equation~(\ref{ETotalFractionVarianceSolutionLimit}) is valid only for~$t\ll\frac{L^{2}}{D_{s}}$ because it relies on Eq.~(\ref{EH0LimitNoMutation}), which is valid for an infinite population, and should break down at times that are long enough for a domain boundary to diffuse from one end of the population to the other. Using Equations~(\ref{EHVarianceWellMixed}) and~(\ref{ETotalFractionVarianceSolution}), one can also calculate the behavior of the global heterozygosity~$\mathcal{H}(t)=L^{-1}\int_{0}^{L}H(t,x)dx$, i.e. the probability to sample two different alleles from the population regardless of their spatial locations:

\begin{equation}
\label{EGlobalHSolution}
\begin{split}
2F_{0}(1-F_{0})-\mathcal{H}(t)=&\frac{D_{g}}{L}\int_{0}^{t}H(t',0)dt'\\ =&\frac{H_{0}\sqrt{32D_{s}t}}{\sqrt{\pi}L}+O\left(\frac{D_{s}}{D_{g}L}\right),
\end{split}
\end{equation}

\noindent where the second equality requires~$8D_{s}/D_{g}^{2}\gg t\ll\frac{L^{2}}{D_{s}}$ for the reasons mentioned above. In the opposite limit~$t\gg\frac{L^{2}}{D_{s}}$, the global heterozygosity~$\mathcal{H}(t,x)$ obeys zero-dimensional dynamics of a well-mixed population with an effective~$\mathfrak{D}_{g}=D_{g}/L$ as shown in~\textcite{Nagylaki:DecayGeneticVariability}.

The local heterozygosity and average domain size can be obtained from experiments on microbial spreading like the one shown in Fig.~\ref{FSpatialSegregation}. If the data are sufficiently precise, Eqs.~(\ref{EH0LimitNoMutation}) and~(\ref{EDomainSizeNoMutation}) could be used to extract~$D_{s}$ and~$D_{g}$ from the experiments. Since~$D_{g}\sim1/N$, extracting~$D_{g}$ from experimental data determines the effective deme size for the equivalent stepping stone model. $D_{s}$ can be obtained from the diffusion of individual domain boundaries or~$\nu(t)$. These two parameters completely determine the neutral dynamics without mutation and play an important role when selection or mutation is present.


\section{Neutral Model With Mutations}
\label{SNeutralMutation}

While on short time scales mutation can be neglected, it is the long time scales and the patterns of genetic diversity created by mutations that are of particular interest in population genetics. Noticeable mutations also arise in microbiology experiments like those in Fig.~\ref{FSpatialSegregation}, especially if mutation rates are enhanced by DNA damaging chemicals or radiation. In this section, we extend the results of Sec.~\ref{SNeutralNoMutation} by allowing for nonzero mutation rates between the two alleles. We assume, as before, statistically homogeneous initial conditions and note that the dynamics of the one and two-point correlation functions is then given by

\begin{equation}
\label{EMutationFEquation}
\frac{dF(t)}{dt}=\mu_{21}-(\mu_{12}+\mu_{21})F(t),
\end{equation}

\begin{equation}
\label{EMutationHEquation}
\begin{split}
\frac{\partial}{\partial t}H(t,x)=&2D_{s}\frac{\partial^{2}}{\partial x^{2}}H(t,x)-D_{g}H(t,x)\delta(x)\\ &-2(\mu_{12}+\mu_{21})H(t,x)\\&+ 2\mu_{21}+2(\mu_{12}-\mu_{21})F(t),
\end{split}
\end{equation}

\noindent where~$F(t)\equiv\langle f(t,x)\rangle$ is independent of~$x$. The equation of motion for~$F$ in the spatial model is exactly the same as~Eq.~(\ref{EMoranGeneticDriftMutationF}), which describes the well-mixed-population model. Therefore $F$ relaxes to its equilibrium value,~$F(\infty)=\frac{\mu_{21}}{\mu_{12}+\mu_{21}}$,~[see Eq.~(\ref{EMoranGeneticDriftMutationFSolution})] exponentially fast with time constant~$(\mu_{12}+\mu_{21})^{-1}\gg\tau_{g}$. The similarity to the nonspatial model is not surprising because neutral mutations occur equally likely at any point within the population, regardless of its spatial structure. The dynamics of~$H(t,x)$ is, however, more complicated because both mutation and genetic drift determine the behavior of the spatial heterozygosity.

The stationary solution of Eq.~(\ref{EMutationHEquation}) reads

\begin{equation}
\label{EMutationHStationarySolution}
H(\infty,x)=\frac{2\mu_{12}\mu_{21}}{(\mu_{12}+\mu_{21})^{2}}\left(1-\frac{e^{-\sqrt{\frac{\mu_{12}+\mu_{21}}{D_{s}}}|x|}}{1+4\sqrt{\frac{D_{s}(\mu_{12}+\mu_{21})}{D_{g}^{2}}}}\right).
\end{equation}

\noindent Equation~(\ref{EMutationHStationarySolution}) agrees with the solution by Kimura and Weiss~\cite{KimuraWeiss:SSM}, which was obtained in the discrete space and time limit. One can see that, for~$x\gg\sqrt{\frac{D_{s}}{\mu_{12}+\mu_{21}}}$, the spatial heterozygosity approaches~$2F(\infty)[1-F(\infty)]$. Thus mutations cause the frequencies of allele one to eventually become uncorrelated at large separations. At shorter distances, however, there are correlations, and~$H(\infty,x)<H(\infty,\infty)$ for all~$x<\infty$. Note, in particular, that

\begin{equation}
\label{EMutationH0StationarySolution}
H(\infty,0)=\frac{2F(\infty)[1-F(\infty)]}{1+\frac{1}{4}\frac{D_{g}}{\sqrt{D_{s}(\mu_{12}+\mu_{21})}}}<H(\infty,\infty).
\end{equation}

\noindent Note also that, for small mutation rates, the heterozygosity is proportional to~$\mu_{12}+\mu_{21}$~[see Eq.~(\ref{EMoranGeneticDriftMutationHSolution})] in a well-mixed population, but the local heterozygosity in a one-dimensional population is proportional to~$\sqrt{\mu_{12}+\mu_{21}}$ whenever~$\tau_{g}(\mu_{12}+\mu_{21})\ll a^2/(N^2D_{s}\tau_{g})$, which is a consequence of weaker genetic drift in one dimension.~\footnote{In population genetics, population structure and spatial correlations are often reported via~$F_{\rm{st}}=H(\infty,0)/H(\infty,\infty)$, which can readily be obtained from Eq.~(\protect{\ref{EMutationH0StationarySolution}}).}

When~$H(\infty,0)\ll H(\infty,\infty)$, the population is segregated into domains of different allelic types. Upon invoking Eq.~(\ref{EDomainSizeH0}), we obtain the following average domain size:

\begin{equation}
\label{EMutationDomainSize}
\begin{split}
\ell=&\frac{2D_{s}(\mu_{12}+\mu_{21})^{2}}{D_{g}\mu_{12}\mu_{21}}\left(1+\frac{1}{4}\frac{D_{g}}{\sqrt{D_{s}(\mu_{12}+\mu_{21})}}\right)\\ \approx&\frac{D_{s}^{\frac{1}{2}}(\mu_{12}+\mu_{21})^{\frac{3}{2}}}{2\mu_{12}\mu_{21}}.
\end{split}
\end{equation}

\noindent This result, together with Eq.~(\ref{EMoranGeneticDriftMutationFSolution}), can be used to extract the mutation rates from experimental data.

We can also determine how fast~$H(t,x)$ reaches its stationary value. Since the heterozygosity cannot be in equilibrium unless the frequency of the alleles has equilibrated, we assume, for simplicity, that~$F(0)$ equals its stationary value. Then, the deviation of the spatial heterozygosity from its long time equilibrium value~$\tilde{H}(t,x)=H(t,x)-H(\infty,x)$ obeys the following equation:

\begin{equation}
\label{EMutationTildeH}
\frac{\partial}{\partial t}\tilde{H}=2D_{s}\frac{\partial^{2}}{\partial x^{2}}\tilde{H}-D_{g}\tilde{H}\delta(x)-2(\mu_{12}+\mu_{21})\tilde{H}.
\end{equation}

Equation~(\ref{EMutationTildeH}) can be further simplified by the change of variables~$\tilde{H}=e^{-2(\mu_{12}+\mu_{21})t}\hat{H}$, which leads to

\begin{equation}
\label{EMutationhatH}
\frac{\partial}{\partial t}\hat{H}=2D_{s}\frac{\partial^{2}}{\partial x^{2}}\hat{H}-D_{g}\hat{H}\delta(x).
\end{equation}

\noindent Since~Eq.~(\ref{EMutationhatH}) is identical to Eq.~(\ref{EHNoMutationEquation}), we conclude that, at long times, the difference between~$H(t,x)$  and the stationary solution decays as~$\hat{C}\sqrt{\frac{D_{s}}{D_{g}^{2}t}}e^{-2(\mu_{12}+\mu_{21})t}$, where~$\hat{C}$ is a constant. Thus, apart from an algebraic prefactor~(and a nontrivial spatial dependence), the dynamics of~$H(t,x)$ is essentially the same as in the well-mixed case.

In this section, we considered a model with only two alleles; however, in many circumstances, an infinite alleles model is more appropriate. The infinite alleles model is briefly discussed in Appendix~\ref{AInfiniteAlleleModel}. Some results for$q$-alleles,~$2<q<\infty$, are discussed in Appendix~\ref{AMultipleColors}.


\section{Selection}
\label{SSelection}

Unlike the neutral models with spatial diffusion and mutation discussed above, the one-dimensional stepping stone model with selection are difficult to treat analytically because the hierarchy of moment equations does not close. We briefly examined three closure schemes:

\begin{equation}
\label{EClosureSimple}
\langle 2f(t,x_{1})[1-f(t,x_{1})]f(t,x_{2})\rangle \approx H(t,0)F(t),
\end{equation}

\begin{equation}
\label{EClosureNagylaki}
\begin{split}
&\langle 2f(t,x_{1})[1-f(t,x_{1})]f(t,x_{2})\rangle\approx \\& 2F(t)[1-F(t)][1-2F(t)]\\ &-[1-2F(t)]H(t,x)+H(t,0)F(t),
\end{split}
\end{equation}

\noindent and

\begin{equation}
\label{EClosureRD}
\begin{split}
\langle f(t,x_{1})f(t,x_{2})f(t,x_{3})\rangle\approx & \frac{\langle f(t,x_{1})f(t,x_{2})\rangle \langle f(t,x_{2})f(t,x_{3})\rangle}{F},\\ & \quad {\rm{assuming}} \quad x_{1}\leq x_{2} \leq x_{3}.
\end{split}
\end{equation}

\noindent The first scheme is a simple factorization approximation; the second scheme, which assumes small fluctuations, is due to~\textcite{Nagylaki:Hierarchy}; and the third scheme, which provides a good approximation for some diffusion limited reactions, was proposed in~\textcite{Lin:ClosureSchemes}. Unfortunately, none of the schemes describe the behavior of the system correctly. The progress can be made, however, for some initial conditions in two limiting cases of strong selection~$\frac{sD_{s}}{D_{g}^{2}}\gg1$ and weak selection~$\frac{sD_{s}}{D_{g}^{2}}\ll1$ Note that we now use the term weak selection in a different sense than in Sec.~\ref{SWellMixed}. For the rest of this section, we include spatial diffusion and genetic drift, but neglect mutations, which is justified on short time scales.

First, let us consider the initial condition~$f(0,x)=1-\theta(x)$, where~$\theta(x)$ is the Heaviside step function. This initial condition specifies just one domain boundary, which, for any positive~$s$, undergoes Brownian motion with a drift to the right. This is a good description of an expansion of a new advantageous mutant spreading through the population. In the strong selection limit~($s\gg D_{g}^{2}/D_{s}$), \textcite{Fisher:FisherWave} found that the sharp boundary above broadens to a width of order~$\sqrt{D_{s}/s}$, and the velocity of the genetic wave is given by

\begin{equation}
\label{EWaveVelocityStrongSelection}
v_{s}=2\sqrt{sD_{s}}.
\end{equation}

\noindent When, in contrast, selection is weak compared to genetic drift, it was recently found that the velocity is given by~\cite{Doering:FisherWaveWeakSelection,Hallatschek:FisherWave}

\begin{equation}
\label{EWaveVelocityWeakSelection}
v_{w}=\frac{2sD_{s}}{D_{g}}.
\end{equation}

\noindent When the population contains multiple domains, the domain walls bordering a favorable genetic variant~(``allele one'') expand to engulf the regions occupied by the more deleterious allele.

Another interesting initial condition is~$f(0,x)=F_{0}=\rm{const.}$, i.e. the population is initially well-mixed. This scenario, for example, describes the quasi-one-dimensional strip of pioneers advancing at the front of a two-dimensional population wave that originated from a well-mixed ancestral population and is propagating in the region where one of the alleles has higher fitness~(see~\textcite{Hallatschek:LifeFront}).

If selection is strong enough, then allele one~(``the preferred variant'') takes over the population before spatial correlations have time to appear. To see this, note from Eq.~(\ref{EDeterministicSecetiveSweepWellMixed}) that allele one wins locally on the time scales of~$s^{-1}$, but, from Eq.~(\ref{EH0SolutionNoMutation}), the time for spatial correlations to appear is on the order of~$D_{s}/D_{g}^{2}$, which is much larger than~$s^{-1}$ in the strong selection limit. Thus the behaviors of one-dimensional and well-mixed populations are similar when~$s\gg D_{g}^{2}/D_{s}=a^{2}/(\tau_{g}^{2}N^{2}D_{s})$. 

In the limit of weak selection, however, spatial correlations appear before allele two is eliminated. Qualitatively, we can divide the selective sweep into two stages. During the first stage, the effects of selection are negligible and spatial segregation occurs as described in Sec.~\ref{SNeutralNoMutation}. During the second stage, the domains of allele two shrink at each end with wall velocity~$v_{w}$ given by Eq.~(\ref{EWaveVelocityWeakSelection}), and the stochastic motion of the domain boundaries can be neglected. The crossover time between the stages occurs when the diffusive displacement of the walls is of the same order as their deterministic displacement, i.e. when~$v_{w}t=\sqrt{2D_{s}t}$. Thus the crossover time~$\tau^{*}$ is on the order of~$D_{s}/v_{w}^{2}$, which can be expressed as~$D_{g}^{2}/(s^{2}D_{s})$ with the help of Eq.~(\ref{EWaveVelocityWeakSelection}). Then, from Eq.~(\ref{EDomainSizeNoMutation}), the average domain size at the crossover~$\ell^{*}$ is on the order of~$D_{g}/(sH_{0})$.

The dynamics during the second stage depends on the probability distribution of domains of size~$\eta$,~$P_{d}(\eta)$ at time~$\tau^{*}$. For annihilating random walkers, which are a good approximation to domain boundaries during the first stage, \textcite{Bramson:DomainSizeDistributionARW} proved that~$P_{d}(\eta)$ has exponential tail for large~$\eta$ of the form~$e^{-\gamma' \eta/\ell^{*}}$, where~$\gamma'$ is a number of the order~$1$. Since each domain shrinks with velocity~$2v_{w}$, the fraction of allele two can be expressed as

\begin{equation}
\label{EDeleteriousAlleleFractionEquation}
1-F(t)\propto \int_{0}^{\infty}\eta e^{-\frac{\gamma (\eta+2v_{w}t)}{\ell^{*}}}d\eta\propto e^{-\frac{\lambda s^{2}D_{s}t}{D_{g}^{2}}},
\end{equation}

\noindent where~$\lambda$ is a number of order~$1$. From Eq.~(\ref{EDeleteriousAlleleFractionEquation}) it follows that, as in the well-mixed case, the selective sweep is exponentially fast, but the time constant of this process is proportional to~$s^{-2}$ rather than~$s^{-1}$.

The analysis leading to Eq.~(\ref{EDeleteriousAlleleFractionEquation}) can be generalized to an arbitrary initial probability distribution,~$P_{d}(\eta)$, provided the dynamics is dominated by selection and genetic drift. For example, if a population initially in equilibrium with respect to mutations and genetic drift~(see Sec.~\ref{SNeutralMutation}) is affected by an abrupt environmental change that makes allele one advantageous, then the shift to the new equilibrium occurs exponentially fast with a time constant proportional to~$\frac{sD_{s}^{1/2}\mu_{12}\mu_{21}}{D_{g}(\mu_{12}+\mu_{21})^{3/2}}$, assuming~$P_{d}(\eta)$ has exponential tail of the form~$e^{-\gamma''\eta/\ell}$, where~$\gamma''$ is a constant of order unity, and~$\ell$ is given by Eq.~(\ref{EMutationDomainSize}).

Finally, we address a slightly different, but equally important, question: What is the probability~$p_{\rm{surv}}$ that a few copies of the advantageous allele survive and establish a growing domain? \textcite{Doering:FisherWaveWeakSelection} solved this problem exactly:

\begin{equation}
\label{EPSurvive1d}
p_{\rm{surv}}=1-\exp\left[-\frac{2s}{D_{g}}\int_{-\infty}^{+\infty}f(0,x)dx\right].
\end{equation}

\noindent Surprisingly, the survival probability does not depend on the diffusion constant. For a small initial number of advantageous alleles, we can qualitatively explain this result in the limit of weak selection by the following argument. Initially, the dynamics is almost neutral, and the probability of survival within a small interval of length~$\Delta x$ is proportional to the relative fraction of the advantageous allele in this interval,~$\int f(0,x)dx/\Delta x$, because every organism has approximately the same probability to reach fixation. Once a domain of size~$\Delta x$ is formed, its survival probability equals the probability that the two biased random walks performed by the domain boundaries never meet, which is~$1-\exp(-v\Delta x/D_{s})\approx v\Delta x/D_{s}$, for small~$\Delta x$ and~$s$, see~\cite{Hallatschek:LifeFront,Redner:FirstPassageTime}. Then, using Eq.~(\ref{EWaveVelocityWeakSelection}), the survival probability is~$v\Delta x/D_{s}\int f(0,x)dx/\Delta x=2s\int f(0,x)dx/D_{g}\approx1-\exp\left[-2s\int f(0,x)dx\right/D_{g}]$. Note that, even though~$p_{\rm{surv}}$ does not depend on~$D_{s}$, the expression for the survival in one dimension does not reduce to its analog in well-mixed populations~\cite{Crow:PopulationGenetics,Doering:FisherWaveWeakSelection}

\begin{equation}
\label{EPSurviveWellMixed}
p_{\rm{surv}}=\frac{1-e^{-2sf(0)/\mathfrak{D}_{g}}}{1-e^{-2s/\mathfrak{D}_{g}}},
\end{equation}

\noindent unless one assumes~$s/\mathfrak{D}_{g}\gg 1$.

We make two important observations based on the results of this section. First, the temporal dynamics of the one-dimensional stepping stone model with selection can depend strongly on the initial conditions. Second, the results in the weak selection limit are sometimes related to the results in the strong selection limit or in well-mixed-population models by a parameter substitution, e.g.~$s\rightarrow s^{2}D_{s}/D_{g}^{2}$, at least up to a numerical factor. The second observation suggests that, while data can be naively fitted to a well-mixed-population model, the fit in fact gives the ``renormalized'' value of~$s$ instead of the ``bare'' one.

Although the one-dimensional stepping stone model provides a reasonable approximation to neutral genetic demixing at a linear front of an expanding two-dimensional microbial colony~(see Fig.\ref{FSimulationExperiment}), there are additional subtleties associated with the dimensional reduction from two to one dimensions when one of the alleles is more fit. Apart form the undulations of the front mentioned in the caption of Fig.~\ref{FSimulationBackground}, the front develops additional structure because favorable sectors bulge out ahead of their less fit neighbors. In the limit of very small genetic drift, there are kink singularities where favorable and unfavorable domains meet. Nevertheless, the basic picture of domain boundaries engulfing unfavorable sectors at a constant velocity is still valid. See~\textcite{Hallatschek:LifeFront} for further details.


\section{Simulations}
\label{SSimulations}

In Secs.~\ref{S1dSSM},~\ref{SNeutralNoMutation},~\ref{SNeutralMutation}, and~\ref{SSelection}, we reviewed and extended the theoretical analysis of the one-dimensional stepping stone model. This model, while of great theoretical interest, relies on a restrictive set of assumptions including large deme sizes and slow diffusive migration. The recent experiments by~\textcite{HallatschekNelson:ExperimentalSegregation}, on the other hand, were carried out with bacterial fronts that were only a monolayer thick; therefore, demes consisted of only a few microbes. Moreover, depending on the microorganism, nearby demes can exchange a significant fraction of cells in each generation. In this section, we discuss numerical simulations not subject to these restrictions and compare them with the theoretical predictions.

We simulate~$L$ organisms arranged on a line and labeled by an integer~$l$, $l=1,2,...,L$. Each organism can be either of allelic type~$1$ or allelic type~$2$. During even generations, the offspring at site~$l$ comes from an organism at either site~$l-1$ or site~$l$, whereas, during odd generations it comes from either site~$l$ or~$l+1$. The simulations embody the process illustrated in Fig.~\ref{FSimulationBackground}, laid out on a triangular lattice in space and time. Periodic boundary conditions are imposed at the left and right ends of the population. Let~$12\rightarrow2$ refer to the event that the offspring has allelic type~$2$, while one of its possible parents has allelic type~$1$ and the other has allelic type~$2$, etc. The transition probabilities, which depend on the states of the possible ancestors, are then given by

\begin{equation}
\label{ETransitionProbabilitiesSimulation}
\begin{split}
&11\rightarrow1:\qquad 1-\tilde{\mu}_{12},\\
&11\rightarrow2:\qquad \tilde{\mu}_{12},\\
&22\rightarrow1:\qquad \tilde{\mu}_{21},\\
&22\rightarrow2:\qquad 1-\tilde{\mu}_{21},\\
&12\rightarrow1:\qquad \left(\frac{1}{2}+\frac{\tilde{s}}{4}\right)(1-\tilde{\mu}_{12})+\left(\frac{1}{2}-\frac{\tilde{s}}{4}\right)\tilde{\mu}_{21},\\
&12\rightarrow2:\qquad \left(\frac{1}{2}+\frac{\tilde{s}}{4}\right)\tilde{\mu}_{12}+\left(\frac{1}{2}-\frac{\tilde{s}}{4}\right)(1-\tilde{\mu}_{21}),\\
&21\rightarrow1:\qquad \left(\frac{1}{2}+\frac{\tilde{s}}{4}\right)(1-\tilde{\mu}_{12})+\left(\frac{1}{2}-\frac{\tilde{s}}{4}\right)\tilde{\mu}_{21},\\
&21\rightarrow2:\qquad \left(\frac{1}{2}+\frac{\tilde{s}}{4}\right)\tilde{\mu}_{12}+\left(\frac{1}{2}-\frac{\tilde{s}}{4}\right)(1-\tilde{\mu}_{21}).
\end{split}
\end{equation}

\noindent  The event~$12\rightarrow2$ can happen either if allele one was selected for reproduction~(probability~$1/2+\tilde{s}/4$) and mutated~(probability~$\tilde{\mu}_{12}$), or if allele two was selected for reproduction~(probability~$1/2-\tilde{s}/4$) and did not mutate~(probability~$1-\tilde{\mu}_{21}$). Other transition probabilities are obtained analogously. Thus, the system we simulate is very similar to the voter model~\cite{CoxGriffeath:VoterModel}, which equivalent to population genetics models with~$N=1$ (see Appendix~\ref{AVoter}). However, to make the calculation faster, we use discrete generations rather than exponentially distributed waiting times until reproduction. We found no significant differences in dynamics between the voter model and the model used here.

\begin{figure}
\includegraphics[width=\columnwidth]{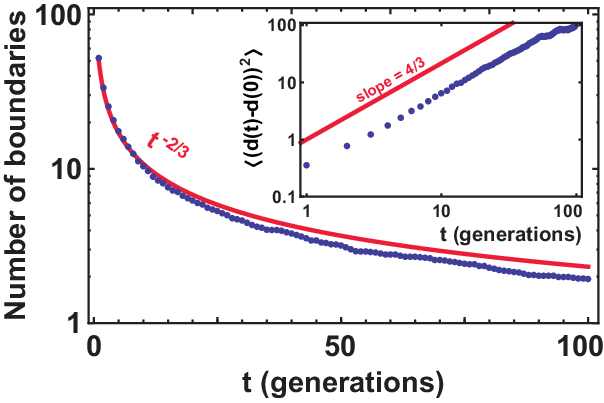}
\caption{(Color online) The number of monoallelic domain boundaries as a function of time in the undulating-front model. The simulation of~$100$ demes averaged over~$100$ runs is plotted in blue~(dots), and the theoretically expected decay of the number of boundaries as~$t^{-2/3}$ [see~\textcite{Saito:EdenModel}] is plotted in red~(solid line). Note that the agreement between the theory and the simulations is not expected during the transitory regime at early times. At~$t=0$, each site is assigned either allele one or allele two with equal probability. Inset: Log-log plot of the mean square displacement of the domain boundaries as a function of time in the same set of simulations as the main plot. The blue dots are the simulation data, and the solid red line is the expected slope according to~\textcite{Saito:EdenModel}.}
\label{FBoundariesFront}
\end{figure}

\begin{figure}
\includegraphics[width=\columnwidth]{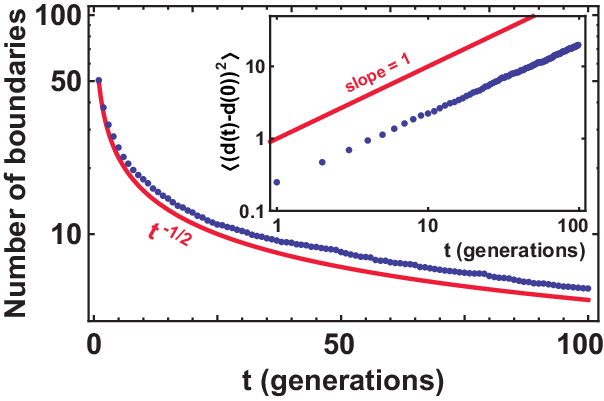}
\caption{(Color online) The number of monoallelic domain boundaries as a function of time in the flat-front model. The simulation of~$100$ demes averaged over~$100$ runs is plotted in blue~(dots), and the theoretically expected decay of the number of boundaries as~$t^{-1/2}$~(see~\textcite{Hallatschek:LifeFront}) is plotted in red~(solid line). Note that the agreement between the theory and the simulations is not expected during the transitory regime at early times. At~$t=0$, each site is assigned either allele one or allele two with equal probability. Inset: Log-log plot of the mean square displacement of the domain boundaries as a function of time in the same set of simulations as the main plot. The blue dots are the simulation data, and the solid red line is the expected slope according to~\textcite{Hallatschek:LifeFront}, \textcite{Bramson:FiniteSize}, and Eq.~(\ref{EDomainSizeNoMutation}).}
\label{FBoundaries1d}
\end{figure}

First, we simulate the neutral model without mutations. To illustrate the similarities and differences between the stepping stone model and the undulating-front model, we also simulate a linear population wave in a two-dimensional habitat; both models are displayed in Fig.~\ref{FSimulationBackground}. Our model with an undulating front is the same as in~\textcite{Saito:EdenModel}, but we use a triangular grid instead of a square one. Figures~\ref{FBoundariesFront} and~\ref{FBoundaries1d} show how the average number of domain boundaries decreases with time; the insets show the mean square displacements of the boundaries as a function of time. The simulations confirm that the domain move diffusively for the one-dimensional model with a flat front and superdiffusively for the undulating front~(Eden) model, in agreement with~\textcite{Saito:EdenModel}. Therefore, the number of domain boundaries decays faster for the undulating-front model compared to the flat-front model, which, as we show below, most closely tracks the prediction of the one-dimensional stepping stone model.

\begin{figure}
\includegraphics[width=\columnwidth]{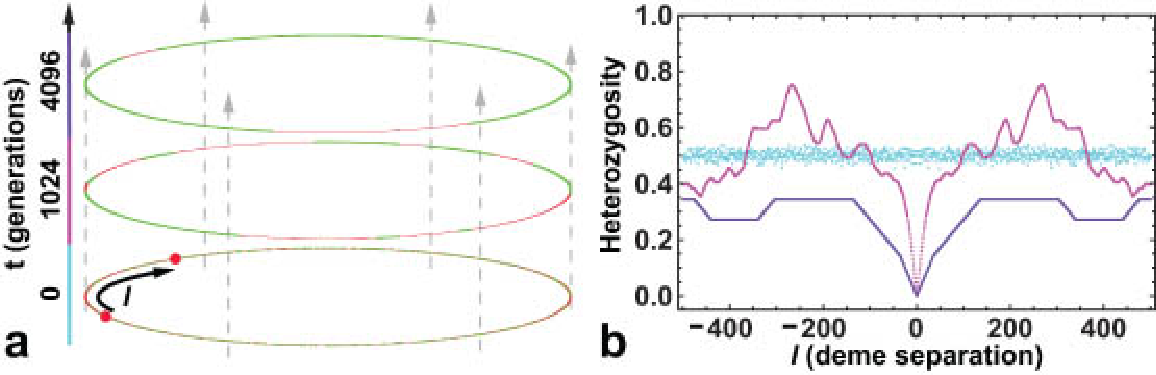}
\caption{(Color online) A single run of the flat-front model with~$1000$ organisms. At~$t=0$, each site is assigned either allele one or allele two with equal probability. The spatially averaged heterozygosity~[defined in the sense of Eq.~(\protect{\ref{ESpaceAverageH}}) but without taking the limit~$L\rightarrow\infty$] for three times measured in generations:~$t=0$ shown in cyan,~$t=1024$ shown in red, and~$t=4096$ shown in dark blue. The separation~$l$ is the shortest distance between two points around the cylinder, and we take the clockwise direction to be positive. At inoculation, the heterozygosity fluctuates around~$1/2$ since the two alleles have equal probabilities of occupying any site. The only exception is the site~$l=0$, where the heterozygosity is zero automatically because we only allow a single microorganism per site. After~$1024$ generations, short range correlations are clearly visible, and, after~$4096$ generations, one can relate the abrupt changes in the slope of~$H(t,l)$ to the sizes of the sectors in the population~(not shown). The wiggles are eliminated when averaged over many realizations, as shown in Fig.~\protect{\ref{FAverageHeterozygosity}}. Note that the curve for~$t=4096\tau_{g}$ lies completely below~$1/2$ because, at this time, the relative fraction of the alleles deviates significantly from the initial $50:50$~ratio due to genetic drift.}
\label{FSingleSimulationRun}
\end{figure}

\begin{figure}
\includegraphics[width=\columnwidth]{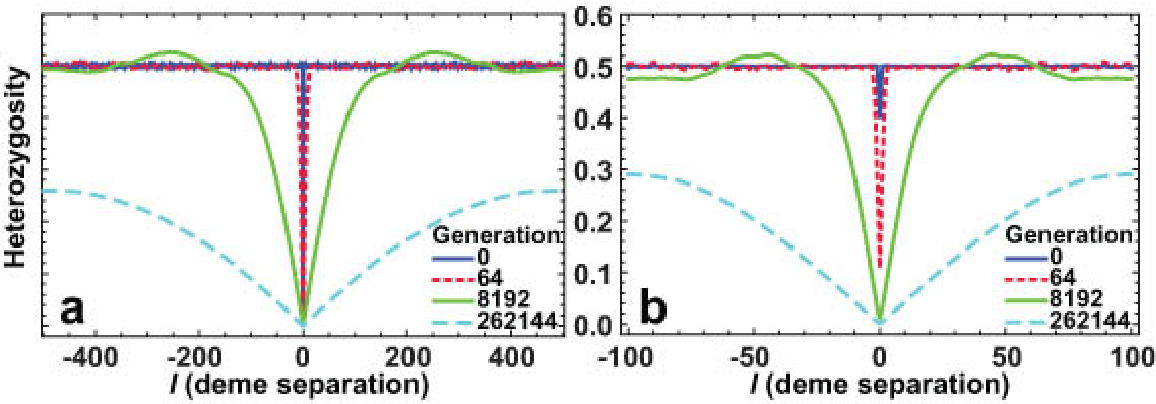}
\caption{(Color online) The effects of coarse-graining on the time evolution of the spatial heterozygosity~$H(t,l)$ averaged over~$100$ realizations of the flat-front model with a thousand organisms. At~$t=0$, each site is assigned either allele one or allele two with equal probability. (a) Each deme hosts only one organism. Consequently, the heterozygosity at~$l=0$ is zero at all times. (b) The same set of simulations, but the organisms have now been grouped into demes of size~$5$ for the purpose of calculating~$H(t,l)$. Part b shows qualitative agreement with the solution of stepping stone model displayed in Fig.~\ref{FHNoMutationTheory}, as does part a outside the region around~$l=0$. Note that, unlike the calculation presented in Fig.~\ref{FHNoMutationTheory}, this simulation was run sufficiently long to show the effects of the boundary conditions.}
\label{FAverageHeterozygosity}
\end{figure}

\begin{figure}
\psfrag{y}{\hspace{-0.3cm}$\bm{H(t,x)}$}
\psfrag{x}{\hspace{-1.2cm}$\mbox{rescaled distance,}\;\bm{\bar{x}}$}
\includegraphics[width=\columnwidth]{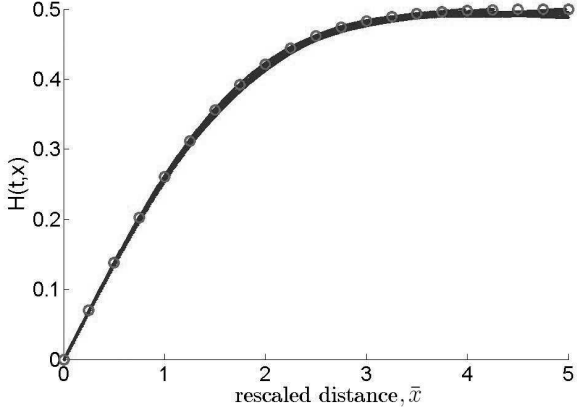}
\caption{(Color online) Comparison between the analytical prediction for the limiting shape of~$H(\infty,\bar{x})$ and the simulations of the flat-front model. The continuous curve~(black) is formed by the data points representing~$H(t,x)$ for several times between $t=2\cdot10^{4}$ and~$t=4\cdot10^5$ plotted in the rescaled coordinates~$\bar{x}$, and the circles~(red) represent the theoretical prediction of the limiting shape of the average spatial heterozygosity, see Eq.~(\protect{\ref{ENoMutationHLimitingShape}}). The data are obtained in a simulation of~$3200$ individuals for~$4\cdot10^5$ generations with averaging over~$500$ realizations. At~$t=0$, each site is assigned either allele one or allele two with equal probability.}
\label{FFitLimitinShape}
\end{figure}

A single run of a simulation is shown in Fig.~\ref{FSingleSimulationRun}, and the spatial heterozygosity averaged over many realizations is shown in Fig.~\ref{FAverageHeterozygosity}. Figure~\ref{FFitLimitinShape} shows that~$H(t,x)$ for large~$t$ is described well by the limiting shape of spatial heterozygosity given by Eq.~(\ref{ENoMutationHLimitingShape}).

To properly represent~$H(t,0)$, we artificially define demes of a larger size by grouping~$M$ neighboring individuals into one deme. From a theoretical point of view, this procedure is similar to the formation of Kadanoff block spins as in renormalization group methods~\cite{Goldenfeld:Lectures,Wilson:Renormalization} whereas, from the point of view of population genetics, this procedure is similar to the methods of collecting data from a dispersed natural population. In field studies, scientists do not typically sample every single individual; instead, they often divide the habitat into patches and sample a representative number of individuals from those patches. To summarize, we keep the dynamics of the simulation exactly the same, but define the spatial heterozygosity on the demes of size~$M$ rather than on single individuals, see Fig.~\ref{FAverageHeterozygosity}. We found that the local heterozygosity has the form~$H(t,0)=\beta(M)t^{-1/2}$, as predicted by our analysis of the stepping stone model, for all~$M$ studied ($1\leq M\leq64$). From Eq.~(\ref{EH0LimitNoMutation}), we expect that~$\beta\propto M^{-2}$; this expectation is also confirmed by our simulations.

\begin{figure}

\includegraphics[width=\columnwidth]{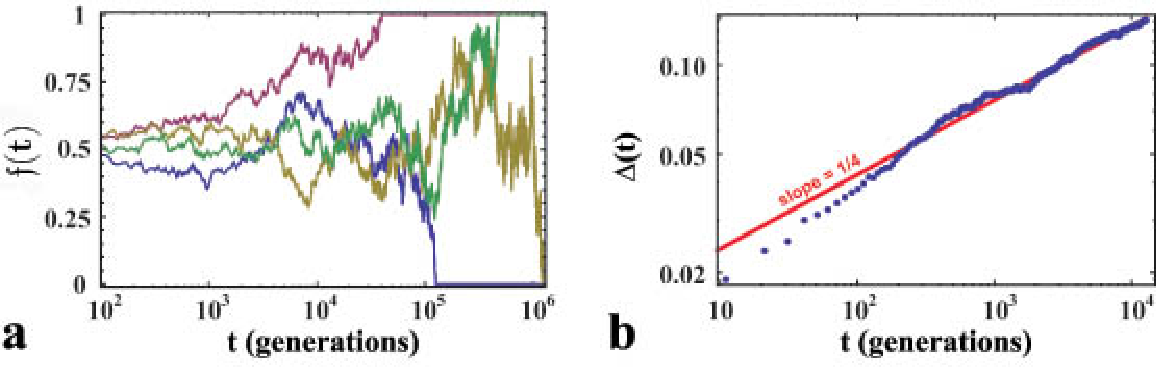}

\caption{(Color online)  Genetic drift in a finite population. At~$t=0$, each site is assigned either allele one or allele two with equal probability. (a)~The total fraction of allele one~$\mathfrak{f}(t)$ versus time in four single runs of the neutral model with a flat front. Here,~$L=1000$, and there are no mutations. 
(b)~The average standard deviation of the frequency of allele one~$\Delta(t)$, shown in blue, is obtained from~$200$ realizations of the simulations described in~a. The red solid line shows the best power-law fit, and the slope is close to the exponent expected from~Eq.~(\protect{\ref{ETotalFractionVarianceSolutionLimit}}). The gray area encloses the points within one standard deviation from the mean.}
\label{FTotalFractionRW}
\end{figure}

\begin{figure}
\psfrag{y}{\hspace{-0.8cm}$\bm{H(\infty,x)}$}
\psfrag{x}{\hspace{-0cm}$$\mbox{\fontsize{12}{14}\selectfont $\bm{x}$}}
\includegraphics[width=\columnwidth]{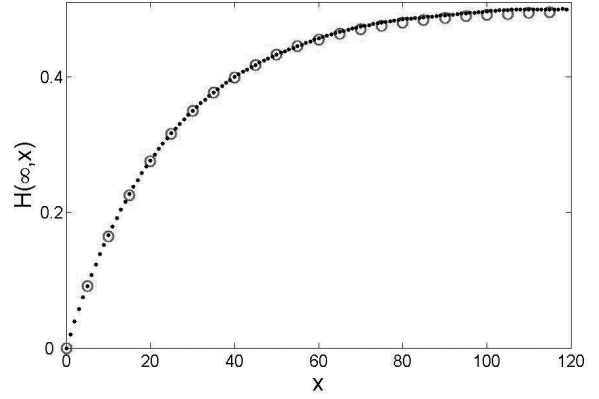}
\caption{(Color online)Equilibrium between mutation and genetic drift in the absence of selection. Comparison between the analytical prediction for the steady state heterozygosity~$H(\infty,x)$ and the simulations with~$\tilde{\mu}_{12}=\tilde{\mu}_{21}=10^{-4}$. The black dots represent the results of the simulation, and the red circles represent the best fit of theoretical result given by Eq.~(\protect{\ref{EMutationHStationarySolution}}) to the data. Here, only~$D_{g}$ is a fitting parameter; the values of~$D_{s}$,~$\mu_{12}$, and~$\mu_{21}$ follow from the correspondence between the discrete and continuum models. The data are obtained in a simulation of~$3200$ individuals for~$2\cdot10^5$ generations with averaging over~$100$ realizations. At~$t=0$, each site is assigned either allele one or allele two with equal probability.}
\label{FMutation}
\end{figure}

\begin{figure}
\psfrag{y}{\hspace{0.0cm}$$\mbox{\fontsize{12}{14}\selectfont $\bm{\alpha}$}}
\psfrag{x}{\hspace{-0cm}$$\mbox{\fontsize{12}{14}\selectfont $\bm{\tilde{s}}$}}
\psfrag{f}{\hspace{-0.6cm}$$\mbox{\fontsize{7}{8}\selectfont $\bm{\ln(1-F)}$}}
\psfrag{t}{\hspace{-0cm}$\bm{t}$}
\includegraphics[width=\columnwidth]{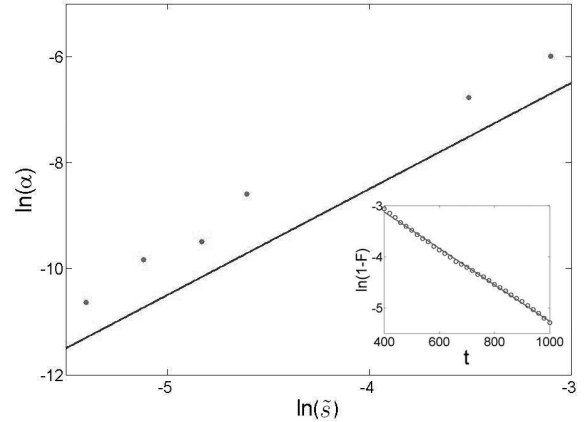}
\caption{ The effective extinction rate $\alpha$ versus~$\tilde{s}$ in the limit of weak selection. The red dots are the data from the simulation, and the black line has the slope equal to~$2$, which is the expected slope from Eq.~(\protect{\ref{EDeleteriousAlleleFractionEquation}}). The data supports~$\alpha \propto \tilde{s}^{2}. $The values of~$\alpha$ are obtained from graphs like the one shown in the inset. Inset: $\ln(1-F)$~versus~$t$ for~$\tilde{s}=0.12$. The green circles are the actual data points, and the blue line is the best least squares linear fit. The simulation confirms exponentially fast fixation. The data are obtained in a simulation of~$1600$ individuals for~$6000$ generations with averaging over~$100$ realizations. At~$t=0$, each site is assigned either allele one or allele two with equal probability.}
\label{FSelectionAlpha}
\end{figure}

As discussed in Sec.~\ref{SNeutralNoMutation}, the~\textit{total} fraction of allele one~$\mathfrak{f}(t)$ fluctuates in an unusual way with time. Figure~\ref{FTotalFractionRW}a shows examples of these remarkable variable-step-length random walks. The fluctuations of~$\mathfrak{f}(t)$ obey Eq.~(\ref{ETotalFractionVarianceSolutionLimit}) and grow subdiffusively, as shown in Fig.~\ref{FTotalFractionRW}b. We also find good agreement between the theory and the simulations in the presence of mutation for all values of~$M$ studied. The stationary average heterozygosity for~$M=1$ is shown in Fig.~\ref{FMutation}. Finally, we studied selective sweeps in an initially well-mixed population. Figure~\ref{FSelectionAlpha}a confirms the prediction from Eq.~(\ref{EDeleteriousAlleleFractionEquation}) that~$F\propto(1-e^{-\alpha t})$, and Fig.~\ref{FSelectionAlpha}b confirms the result of Sec.~\ref{SSelection} that, for strong genetic drift, the effective extinction rate~$\alpha$ is proportional to~$s^{2}$.

Numerical results for \textit{three} neutral alleles are presented in Appendix~\ref{AMultipleColors}.

\section{Inflation}
\label{SInflation}

\begin{figure}
\includegraphics[width=\columnwidth]{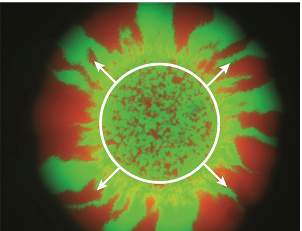}
\caption{(Color online) Spatial segregation in an expanding circular bacterial colony of E. coli. Different colors label different alleles. The Petri dish was inoculated with a well-mixed population occupying the circled region of the colony, leading to many small domains in the central ``homeland.'' As this population expands~(shown with arrows), it segregates into well defined monoallelic domains, which coalesce at early times, but seem to stop coalescing in the final stages of the experiment. Note that the boundaries between the domains are biased to move away from each other due to inflation, in addition to their diffusive random-walk-like motion. Details of the experiment are presented in~\protect{\textcite{HallatschekNelson:ExperimentalSegregation}}.}
\label{FCircularExperiment}
\end{figure}

Throughout this review we have focused on the evolutionary forces acting at a linear~(flat or undulating) front, whose total length~(averaged over the undulations) does not change in time. In this section, we explore the changes in the evolutionary dynamics caused by a constant increase of the total front length, for example, at the edge of an expanding circular colony; see Fig.~\ref{FCircularExperiment}. We now show that this increase, which we term ``inflation,'' in an analogy with cosmology~\cite{UniverseInflation}, slows down genetic drift and natural selection at the front. 

Models of both linear and circular fronts are relevant to biology. Linear fronts describe the essential features of the dynamics when the effects of curvature and changes in the front length are negligible, or when the spreading is limited by some geographical barriers, say, a receding glacier between two parallel rivers. If one focuses on genetic markers in Homo sapiens~(e.g. in mitochondria DNA), dynamics of a linear front also resembles the abrupt settlement by pioneers via a ``land run'' in 1889 across the border of Oklahoma~\cite{Gibson:Oklahoma}.  Circular fronts are more appropriate for modeling an initial colonization by a small number of pioneers arriving in the interior of a large, spatially homogeneous habitat. Semicircular fronts are relevant to colonization after landing on a coast line. The circular scenario is often realized in microbiological experiments when a Petri dish is inoculated with microorganisms. A radial range expansion of E.~coli is illustrated in Fig.~\ref{FCircularExperiment}, which highlights the effects of genetic drift at the front.

The growth of a circular colony lengthens the front, thereby increasing characteristic local length scales. This inflation is specified by the dependence of the radius of the colony~$R$ on time~$t$. Here, we assume~$R(t)=R_{0}+\mathsf{v}t$, which corresponds to a colony expanding with a constant velocity~$\mathsf{v}$ from an initial radius~$R_{0}$. The velocity of the expansion has been found constant in the experiments by~\textcite{HallatschekNelson:ExperimentalSegregation} and in the theoretical studies of the two dimensional Fisher equation~\cite{Murray:MathematicalBiology}, provided the width of the front is much smaller than its length, a condition necessary for a one-dimensional model to hold.

To highlight the effects of inflation, we consider the simplest version of the one-dimensional stepping stone model without mutations and natural selection. In a circular geometry, it is convenient to use the angle~$\varphi=x/R(t)$ instead of~$x$ to reference positions along the front. Then, the equation of motion for~$H(t,x)$ takes a form analogous to  Eq.~(\ref{EHNoMutationEquation}):

\begin{equation}
\label{EHCircular}
\frac{\partial}{\partial t}H(t,\varphi)=\frac{2D_{s}}{(R_{0}+\mathsf{v}t)^{2}}\frac{\partial^{2}}{\partial \varphi^{2}}H(t,\varphi)-\frac{D_{g}}{R_{0}+\mathsf{v}t}H(t,0)\delta(\varphi),
\end{equation}

\noindent where the factors of~$R_{0}+\mathsf{v}t$ have been introduced to account for the inflation. Like Eq.~(\ref{EHNoMutationEquation}), Eq.~(\ref{EHCircular}) is valid for an arbitrary number of neutral alleles, and can be understood by tracing two lineages backward in time. The time dependence of the coefficients in front of the diffusion and coalescence terms accounts for the fact that, as the colony grows, the same sizes in the~$\varphi$-space correspond to different sizes in the~$x$-space, where the diffusion and coalescence terms have their familiar, time independent form as in Eq.~(\ref{EHNoMutationEquation}). When reexpressed in terms of~$t$ and~$x=(R_{0}+\mathsf{v}t)\varphi$, Eq.~(\ref{EHCircular}) contains an advection term describing the deterministic decrease of the separation between the lineages as they go back to the initial radius~$R_{0}$. 

Equation~(\ref{EHCircular}) is defined on a bounded domain~$\varphi\in[-\pi,\pi]$ with periodic boundary conditions. Nevertheless, we can approximate the problem well by considering an unbounded domain~$\varphi\in(-\infty,\infty)$, provided two diametrically opposite lineages are sufficiently unlikely to coalesce. From Eq.~(\ref{EHCircular}), we see that diffusion effectively stops after a characteristic time~$R_{0}/\mathsf{v}$, so our approximation of an unbounded domain should be valid if the distance traveled by the lineages during this time is small compared to the radius of the colony:~$\sqrt{R_{0}D_{s}/\mathsf{v}}\ll R_{0}$ or $D_{s}/\mathsf{v}\ll R_{0}$; this corresponds to a regime with many sectors as we show later. One can also test the goodness of the approximation by evaluating~$H(0,\pi)-H(t,\pi)$, which is expected to be small if the approximation is valid.

To simplify the analysis, we make Eq.~(\ref{EHCircular}) dimensionless in terms of the new variables~$\mathsf{t}$ and~$\phi$ such that~$t=\mathsf{t}D_{s}/D_{g}^{2}$ and~$\varphi=\phi D_{s}/(D_{g}R_{0})$. The equation of motion for~$H(\mathsf{t},\phi)$ then reads

\begin{equation}
\label{EHCircularDimensionless}
\frac{\partial}{\partial \mathsf{t}}H(\mathsf{t},\phi)=\frac{2}{(1+\sigma\mathsf{t})^{2}}\frac{\partial^{2}}{\partial \phi^{2}}H(\mathsf{t},\phi)-\frac{1}{1+\sigma\mathsf{t}}H(\mathsf{t},0)\delta(\phi),
\end{equation}
 
\noindent where the dimensionless parameter~$\sigma=\mathsf{v}D_{s}/(R_{0}D_{g}^{2})$ is proportional to the ratio of two characteristic time scales in the problem: the local fixation time~$\tau_{f}\sim D_{s}/D^{2}_{g}$ in the model of a linear front and the time in which the colony doubles its initial radius. 

Upon assuming~$\phi\in(-\infty,\infty)$, we obtain the exact solution of Eq.~(\ref{EHCircularDimensionless}) for the initial condition~$H(0,\varphi)=H_{0}$ by a generalization of the method presented in Appendix~\ref{ANeutral}:

\begin{equation}
\begin{split}
H(\mathsf{t},\phi)=&H_{0}-\int_{0}^{\mathsf{t}}d\mathsf{t}'\frac{H(\mathsf{t}',0)}{1+\sigma\mathsf{t}'}\sqrt{\frac{(1+\sigma\mathsf{t})(1+\sigma\mathsf{t}')}{8\pi(\mathsf{t}-\mathsf{t}')}}\\& \times\exp\left[-\frac{\phi^{2}(1+\sigma\mathsf{t})(1+\sigma\mathsf{t}')}{8(\mathsf{t}-\mathsf{t}')}\right],
\end{split}
\label{EHCircularSolution}
\end{equation}

\noindent where

\begin{equation}
\begin{split}
H(\mathsf{t},0)=&H_{0}-\frac{H_{0}\sqrt{1+\sigma\mathsf{t}}}{\sqrt{2\pi\sigma}}\left[\frac{\pi}{2}-\arcsin\left(\frac{1}{\sqrt{1+\sigma\mathsf{t}}}\right)\right]\\&+H_{0}\sqrt{\frac{\pi(1+\sigma\mathsf{t})}{8\sigma}}\exp\left(\frac{1+\sigma\mathsf{t}}{8\sigma}\right)\\&\times\left[\erf\left(\sqrt{\frac{1+\sigma\mathsf{t}}{8\sigma}}\right)-\erf\left(\frac{1}{\sqrt{8\sigma}}\right)\right]\\&-H_{0}\sqrt{\frac{\pi(1+\sigma\mathsf{t})}{8\sigma}}\left(e^{\mathsf{t}/8}-1\right)\\&+H_{0}\sqrt{\frac{1+\sigma\mathsf{t}}{128\pi\sigma}}e^{\mathsf{t}/8}\int_{0}^{\mathsf{t}}d\mathsf{t}'e^{-\mathsf{t}'/8}\arcsin\left(\frac{1}{\sqrt{1+\sigma\mathsf{t}'}}\right).
\end{split}
\label{EH0CircularSolution}
\end{equation}

\begin{figure}
\psfrag{y}{\hspace{-0.9cm}$\bm{H(\mathsf{t},\phi)}$}
\psfrag{x}{\hspace{0.1cm}$$\mbox{\fontsize{12}{14}\selectfont $\bm{\phi}$}}
\includegraphics[width=\columnwidth]{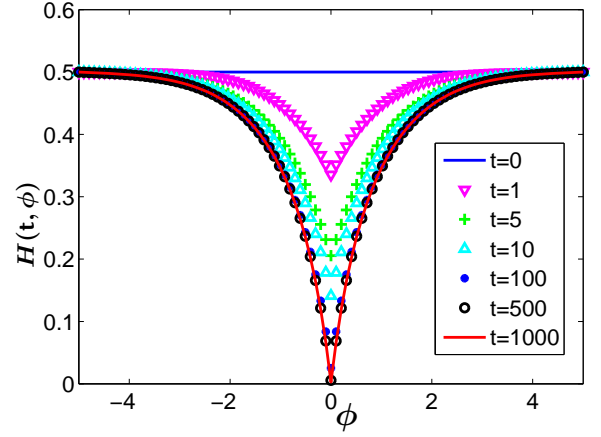}
\caption{(Color online) Solutions of Eq.~(\protect{\ref{EHCircularDimensionless}}) with~$\sigma=1$ at various rescaled times~$\mathsf{t}$, given random initial conditions on the circle bounding the homeland,~$H(0,\phi)=\frac{1}{2}$. Time increases from the top curves to the bottom curves. Note that there is no observable difference between~$H(500,\phi)$ and~$H(1000,\phi)$ because~$H(\mathsf{t},\varphi)$ reaches a nontrivial limit-shape as~$\mathsf{t}\rightarrow\infty$.}
\label{FHCircularSolution}
\end{figure}

\noindent The behavior of~$H(\mathsf{t},\phi)$ is shown in Fig.~\ref{FHCircularSolution}. Similar to a linear front, the local heterozygosity~$H(t,0)$ vanishes for large times, and the characteristic angular length scale over which~$H(t,\varphi)$ changes from~$0$ to~$H_{0}$ increases. Thus, Eq.~(\ref{EHCircular}) predicts the formation and growth of the domains shown in Fig.~\ref{FCircularExperiment}. However, there are two important differences that distinguish radial expansions from linear ones. First,~$H(t,0)$ tends to zero as~$t^{-1}$ rather than as~$t^{-1/2}$. Second, the curve~$H(t,\varphi)$ approaches a nontrivial limit-shape, unlike the case with a linear front, where the analogous curve widens indefinitely. 

Following ~\textcite{HallatschekNelson:ExperimentalSegregation}, we can qualitatively understand this behavior by noticing that the diffusion and lineage coalescence effectively stop after the characteristic time~$R_{0}/\mathsf{v}$. After this point, the number of domain boundaries and the angular width of the domains remain approximately constant. Therefore,~$H(t,0)$, which is proportional to the fraction of the circumference occupied by the boundaries between the domains, should decay as~$t^{-1}$, and the shape of~$H(t,\varphi)$ should approach a nontrivial limit.

From the exact solution~(\ref{EHCircularSolution}),~(\ref{EH0CircularSolution}), we compute the average angular size of the domains~$\ell_{\varphi}(t)$ and the average number of the domains~$\mathcal{N}(t)=2\pi/\ell_{\varphi}(t)$. Similar quantities were calculated by \textcite{Hallatschek:LifeFront} in the approximation of random walking domain boundaries, which appropriate when~$t>\tau_{f}$. Although we cannot use Eq.~(\ref{EDomainSizeH0}) because of the inflation, Eq.~(\ref{EDomainSizedHdx}) remains valid and takes the following form

\begin{equation}
\ell_{\varphi}(t)=\left(\frac{\partial H(t,+0)}{\partial \varphi}\right)^{-1}.
\label{EDomainSizedHdphi}
\end{equation}

\noindent By integrating Eq.~(\ref{EHCircular}) over~$\varphi$ in the neighborhood of~$0$, we express~$\frac{\partial H(t,+0)}{\partial \varphi}$ in terms of~$H(t,0)$ and obtain

\begin{equation}
\label{EDomainSizeH0Circular}
\ell_{\varphi}(t)=\frac{4D_{s}}{D_{g}(R_{0}+\mathsf{v}t)H(t,0)},
\end{equation}

\noindent which approaches a constant at long times. This limit can be computed analytically, with the results

\begin{equation}
\ell_{\phi}(\infty)=\left(\frac{ H_{0}\mathsf{v}}{D_{g}}+H_{0}\sqrt{\frac{R_{0}\mathsf{v}}{2\pi D_{s}}}\right)^{-1},
\label{ElCircular}
\end{equation}

\begin{equation}
\mathcal{N}(\infty)=\frac{2\pi H_{0}\mathsf{v}}{D_{g}}+H_{0}\sqrt{\frac{2\pi R_{0}\mathsf{v}}{D_{s}}}.
\label{ENR}
\end{equation}

Equation~(\ref{ENR}) implies two things. First, by measuring~$\mathcal{N}(\infty)$ as a function of the initial homeland radius~$R_{0}$, one can estimate both~$D_{s}$ and~$D_{g}$ for a microbial population, which could potentially be easier than the experiments with linear fronts that we proposed in Sec.~\ref{SNeutralNoMutation}. Second, if all individuals in the founding population are distinguishable~($H_{0}=1$), then each of the final sectors must originate from a single ancestor. Hence,~$\mathcal{N}(\infty)$ for~$H_{0}=1$ gives the average number of ancestors of the genetically segregated population at the periphery, which contains most of the organisms. This number is remarkably small. Figure~\ref{FCircularExperiment}, where~$H_{0}=1/2$, has about~$20$ domains, so, since~$\mathcal{N}(\infty)\propto H_{0}$, the segregated part of the population descended from only about~$40$ ancestors, a tiny fraction of about~$20,000$ founding cells. Although some of these cells are trapped in the interior of the homeland, a large number of them are piled in a ring at the edge of the homeland within minutes of inoculation, as the carrier fluid dries out~\cite{HallatschekNelson:ExperimentalSegregation}.

We can further quantify the amount of genetic drift in the population by the variance~$\nu(t)$ of the total fraction of allele one~$\mathfrak{f}(t)$. For simplicity, we assume a population with only two alleles. For several alleles, the global heterozygosity~$\mathcal{H}(t)$ is more appropriate and can be easily obtained from our expressions for~$\nu(t)$ because~$\mathcal{H}(t)=H_{0}-2\nu(t)$. We compute~$\mathcal{H}(t)$ and thus~$\nu(t)$ by integrating Eq.~(\ref{EHCircular}) over~$\varphi$; the result is

\begin{equation}
\label{ETotalFractionVarianceSolutionCirlular}
\nu(t)=\frac{D_{g}}{4\pi}\int_{0}^{t}\frac{H(t',0)}{R_{0}+\mathsf{v}t}dt'.
\end{equation}

\noindent We are mostly interested in the long time limit~$\nu(\infty)$, which is approached asymptotically as~$t^{-1}$. This limit can be expressed as

\begin{equation}
\nu(\infty)=\frac{D_{s}H_{0}}{4\pi R_{0}D_{g}}K(\sigma),
\label{ENuK}
\end{equation}

\noindent where~$K(\sigma)=\int_{0}^{\infty}H^{*}(\mathsf{t}',0)/(1+\sigma\mathsf{t}')d\mathsf{t}'$, and~$H^{*}(\mathsf{t},\phi)$ is the solution of Eq.~(\ref{EHCircularDimensionless}) for~$H_{0}=1$. The dependence of~$K$ on~$\sigma$ is shown in Fig.~\ref{FK}. In the limit of large~$D_{g}$~(approximated by the voter model, see Appendix~\ref{AVoter}), an analytical expression for~$\nu(\infty)$ is given by Eq.~(\ref{EANuRVoter}).

\begin{figure}
\psfrag{y}{\hspace{-0.9cm}$\bm{K(\sigma)}$}
\psfrag{x}{\hspace{0.1cm}$$\mbox{\fontsize{12}{14}\selectfont $\bm{\sigma}$}}
\includegraphics[width=\columnwidth]{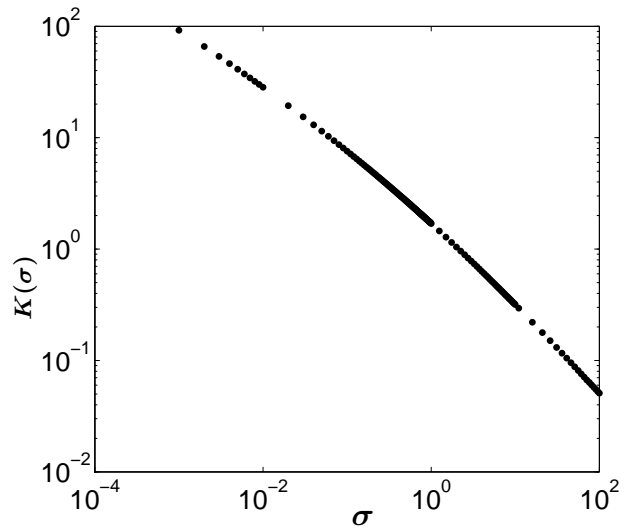}
\caption{A plot of~$K(\sigma)$ for Eq.~(\protect{\ref{ENuK}}) from a numerical solution of Eq.~(\protect{\ref{EHCircularDimensionless}}).}
\label{FK}
\end{figure}

Even though inflation slows down lineage diffusion and coalescence, genetic drift can still cause large fluctuations in the relative frequency of the alleles. These fluctuations are particularly important for any organism that undergoes spatial colonizations followed by almost complete extinctions~(such life cycles are common both in nature and in the laboratory). For such organisms,~$\nu(\infty)$ or~$\mathcal{H}(\infty)$ characterizes the effective genetic drift, which is much larger than the one predicted by well-mixed-population models. We also note that the effects of genetic drift could be more pronounced at a circular front because natural selection is less efficient in the presence of inflation: Domains of deleterious alleles persist longer because the contraction of the domain due to Darwinian selection must also be able to overcome its natural expansion due to inflation. 


\section{Genetic Inference}
\label{SGeneticInference}

So far, we have focused on forward-in-time dynamics, while trying to calculate the patterns of genetic diversity from simple models of evolutionary dynamics. However, it is often necessary to reverse the question: Given the observed genetic diversity, how do we infer the recent history of the population and estimate important parameters like the mutation rates and the effective population size? This question is particularly important because the current state of genetic diversity is often the only clue to the past. Fortunately, genetic inference can be very powerful because the differences among the genomes of individuals contain valuable information about the evolution of the population, and these differences can now be easily measured via DNA sequencing. For example, genetic inference has been used to determine the time and origin of the recent expansion of Homo sapiens~\cite{Ramachandran:MigrationFromAfrica} and to test whether Homo sapiens and Homo neanderthalensis used to interbreed~\cite{Nordborg:Neanderthal}.

Genetic inference is a well-developed subject, which becomes rather technical when one wants to incorporate biological details and use advanced statistical tools. In this section, we address some of the basic questions in genetic inference and highlight the differences between the spatial and nonspatial models. The results for well-mixed-population models presented here are usually attributed to~\textcite{Kingman:Coalescent}; we refer the interested reader to the book by~\textcite{Wakeley:Coalescent} for a lucid introduction.
 
\begin{figure}
\psfrag{t}{\hspace{-0.0cm}$$\mbox{\fontsize{12}{14}\selectfont $\bm{t}$}}
\psfrag{p}{\hspace{0.0cm}$$\mbox{\fontsize{12}{14}\selectfont $\bm{\tau}$}}
\psfrag{x}{\hspace{0.0cm}$$\mbox{\fontsize{12}{14}\selectfont $\bm{x}$}}
\includegraphics[width=\columnwidth]{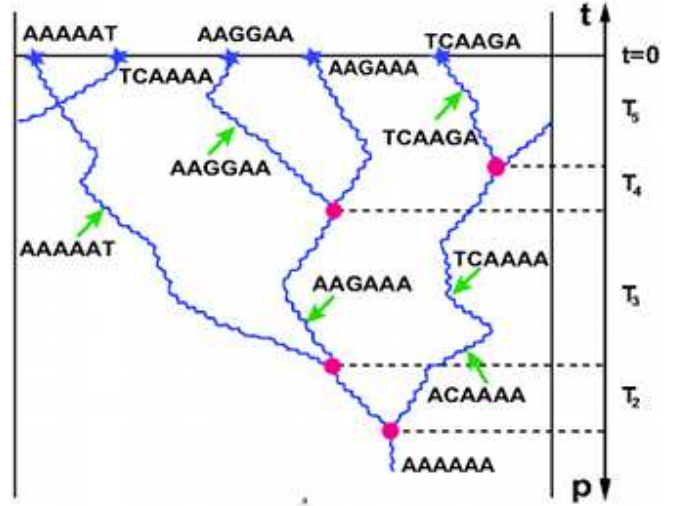}
\caption{(Color online) An illustration of the backward-in-time dynamics of the ancestral lineages in a one-dimensional habitat with periodic boundary conditions. Five organisms~(i.e.~$n=5$) are sampled from the much larger population at~$t=0$ and their DNA is sequenced. We do not display the sites that are identical for all organisms, which are usually the majority of the sequenced sites, i.e. only the segregating sites are shown. For illustration purposes, we also assumed that all samples differ in at least one nucleotide, but in experiments one often finds organisms that have identical sequences. We trace the spatial diffusion and coalescence of the lineages backward in time until they merge into a single lineage of the common ancestor of the whole sample. The coalescence events are denoted by red circles, and the mutations are denoted by arrows and the resulting mutated sequences. Note that lineages may cross without coalescing as shown in the top left corner of the figure. The ancestral process shown here satisfies the infinite sites model and illustrates the fact that the more genetically similar the lineages are the more likely they are to have a common ancestor in the recent past. Most genetic inference methods rely on this relationship as we show in this section.}
\label{FCoalescent}
\end{figure}

In a typical study,~$n$ organisms are sampled from the population, and parts of their genomes are sequenced (see Fig.~\ref{FCoalescent}). A genetic sequence, say \dots{ACTGAA}\dots, is an ordered string of letters taken from a four-letter alphabet: A, T, C, and G, where the letter stand for the nucleotides: adenine, thymine, cytosine, and guanine respectively. For haploid organisms considered here, an offspring inherits its sequences from the parent with possibly a few mutations (but no recombination). While a wide range of mutations is possible, we consider only point mutations, i.e. substitutions of one letter for another. Moreover, we assume that every new point mutation occurs at a new site~(position) in the genome~\cite{Kimura:InfiniteSiteModel}. Because the mutation rate per site~$\mu$ is very small and the total number of sites~$\mathcal{L}_{g}$~(i.e. the length of the sequenced section of the genome) is large, most of the mutations occur at different positions along the sequence, so this infinite sites approximation is reasonable on time scales shorter than~$\mu^{-1}$. For simplicity, we neglect the dependence of the mutation rates on the position within the genome as well as on the type of substitution, i.e. all~$12$ possible substitutions are assumed to occur at the same rate~$\mu$. We further assume that all genetic variation is neutral~\cite{Kimura:NeutralEvolution}.

In principle, the complete data set of~$n$ sequences of length~$\mathcal{L}_{g}$ can be and often is used to estimate parameters in the model. However, we can understand the basic principles of genetic inference by considering two simple summary statistics: the average number of pairwise differences~${\varPi}$ and the number of segregating sites~$S$, i.e the number of sites that do not have identical nucleotides in at least two sequences in the sample. The former is intimately related to the average heterozygosity~$H$ and the latter illustrates the use of genealogical trees in genetic inference, see Fig.~\ref{FCoalescent}. 

We first consider~$\varPi$, which is defined as the expected number of different sites in two randomly selected sequences. In a finite population, any two sequences have a common ancestor, so, as we trace them backward in time, their lineages must coalesce. Let us denote the average time it takes two lineages to coalesce by~$T_{2}$. Then the expected number of pairwise differences is given by

\begin{equation}
\label{EPiT}
\varPi=2\mu\mathcal{L}_{g}T_{2},
\end{equation}

\noindent where the factor of~$2$ accounts for the fact that mutations occur in both lineages. 

Since genetic inference deals with backward-in-time dynamics, it is convenient to use reverse time~$\tau=-t$. To calculate~$T_{2}$, we introduce the persistence probability~$U_{2}(\tau)$, the probability that two lineages sampled at~$t=0$ have not coalesced between~$t=0$ and~$t=-\tau$. Because~$U_{2}(\tau)$ is the cumulative probability distribution function for the coalescence times, the desired probability density function is~$-dU_{2}(\tau)/d\tau$, and~$T_{2}$ can be calculated from the following equation

\begin{equation}
\label{EUT}
T_{2}=\int_{0}^{\infty}\tau\left[-\frac{dU_{2}(\tau)}{d\tau}\right]d\tau=\int_{0}^{\infty}U_{2}(\tau)d\tau.
\end{equation}

For a one-dimensional population, we have to take into account the positions where the organisms are sampled. Therefore, we introduce~$U_{2}(\tau,x_{1},x_{2})$ as the probability that two lineages have not coalesced and are at positions~$x_{1}$ and~$x_{2}$ respectively at reverse time~$\tau$. Then,~$U_{2}(\tau)$ is given by

\begin{equation}
\label{EUU}
U_{2}(\tau)=\int_{0}^{L}\int_{0}^{L}U_{2}(\tau,x_{1},x_{2})dx_{1}dx_{2},
\end{equation}

\noindent where~$L$ is the length of the habitat.

The time evolution of the persistence probability and the average heterozygosity are intimately related both in well-mixed and spatial populations because both quantities describe the fate of two lineages traced backward in time. In fact, the equation of motion for~$H(t)$ is identical to that of~$U_{2}(\tau)$, and the same is true for~$H(t,x_{1},x_{2})$ and~$U_{2}(\tau,x_{1},x_{2})$. For example, in the well-mixed-population model considered in Sec.~\ref{SWellMixed}, $H(t)$~and~$U_{2}(\tau)$ change only due to coalescence events, and each coalescent event changes both quantities from their current values to zero. Thus, analogously to Eq.~(\ref{EMoranGeneticDriftH}) the equation of motion for~$U_{2}(\tau)$ reads

\begin{equation}
\label{EOMWellMixedU}
\frac{d}{d\tau}U_{2}(\tau)=-\mathfrak{D}_{g}U_{2}(\tau),
\end{equation}

\noindent with the initial conditions~$U_{2}(0)=1$. For the one-dimensional stepping stone model, we obtain that

\begin{equation}
\label{EOM1dU}
\begin{split}
\frac{\partial}{\partial \tau}U_{2}(\tau,x_{1},x_{2})=&D_{s}\left(\frac{\partial^{2}}{\partial x_{1}^{2}}+\frac{\partial^{2}}{\partial x_{2}^{2}}\right)U_{2}(\tau,x_{1},x_{2})\\ &-D_{g}U_{2}(\tau,x_{1},x_{2})\delta(x_{1}-x_{2})
\end{split}
\end{equation}

\noindent in analogy with Eq.~(\ref{EContinuous1dSSMH}) for~$\mu_{12}=\mu_{21}=s=0$. The initial condition is~$U_{2}(0,x_{1},x_{2})=\delta(x_{1}-x_{1}^{0})\delta(x_{2}-x_{2}^{0})$, where~$x_{1}^{0}$ and~$x_{2}^{0}$ are the positions of the first and second lineages at the time of sampling.

For the well-mixed case, we integrate both sides of Eq.~(\ref{EOMWellMixedU}) with respect to~$\tau$ from zero~($U_{2}(0)=1$) to infinity~($U_{2}(\infty)=0$) and use Eq.~(\ref{EUT}) to find that~$T_{2}=\mathfrak{D}_{g}^{-1}$. Then, from Eq.~(\ref{EPiT}), we obtain the average number of pairwise differences

\begin{equation}
\label{EPiWellMixed}
\varPi_{\rm{well-mixed}}=\frac{2\mu\mathcal{L}_{g}}{\mathfrak{D}_{g}}.
\end{equation}

\noindent The mutation rate~$\mu$ can often be measured experimentally~\cite{Drake:MutationRates,Araten:MutationRates}, so Eq.~(\ref{EPiWellMixed}) and the knowledge of~$\varPi_{\rm{well-mixed}}$ can be used to estimate the effective population size encoded in~$\mathfrak{D}_{g}$~[see Eq.~(\ref{EGeneticDiffusionCoefficient})].

For the one-dimensional stepping stone model, one has to specify the spatial boundary conditions for Eq.~(\ref{EOM1dU}). Since a lineage can neither go outside the habitat nor disappear at its edge, reflecting boundary conditions should be used. With these boundary conditions, Eq.~(\ref{EOM1dU}) has been analyzed by~\textcite{Wilkins:Linear}. Here, we assume periodic boundary conditions, which are appropriate for a population living on a coast line of an island. These boundary conditions are simpler because they ensure translational invariance: the average coalescence time for two lineages sampled at~$x_{1}^{0}$ and~$x_{2}^{0}$ can only depend on~$|x_{1}^{0}-x_{2}^{0}|$, but not on~$x_{1}^{0}$ and~$x_{2}^{0}$ separately. 

Following~\cite{Wilkinson:1dCoalescent}, we solve Eq.~(\ref{EOM1dU}) with periodic boundary conditions by the Fourier transform in the positions and the Laplace transform in reverse time. The result is

\begin{equation}
\label{ET1d}
T_{2}(x_{1}^{0},x_{2}^{0})=\frac{L}{D_{g}}+\frac{|x_{1}^{0}-x_{2}^{0}|(L-|x_{1}^{0}-x_{2}^{0}|)}{4D_{s}},
\end{equation}

\noindent where the first term on the right hand side is the average coalescence time for two lineage sampled at the same point, and the second term is the average time for two lineages to meet for the first time.\footnote{The average time to the first encounter of two random walks on an interval with periodic boundary conditions is equal to the average survival time of a single random walk with twice the diffusion constant on the same interval, but with absorbing boundary conditions. This equivalent problem can be solved by a standard method; see, e.g.,~\cite{Redner:FirstPassageTime}.} Note that~$T_{2}(x_{1},x_{1})$ is identical to $T_{2}$ in a well-mixed population, provided we take the effective population size to be the total size of the spatially extended population:~$L/D_{g}=\mathfrak{D}_{g}^{-1}L/a=(N\tau_{g}/2)L/a$, where~$L/a$ is the number of demes~($a$ is the distance between neighboring demes). Note that the distribution of the coalescence times is highly skewed, and the average coalescence time does not characterize the distribution well: most of the time coalescence occurs very fast compared to~$T_{2}(x_{1},x_{1})$, but in rare cases lineages persist for times much longer than~$T_{2}(x_{1},x_{1})$~\cite{Charlesworth:NeutralVariation}.

The average number of pairwise differences for the whole data set is obtained by averaging over the spatial positions of the samples~$x_{j},\; j=1,2,\dots,n$:

\begin{equation}
\label{EPi1d}
\varPi_{\rm{1d}}=\frac{4\mu\mathcal{L}_{g}}{n(n-1)}\sum_{j_{1}=2}^{n}\sum_{j_{2}=1}^{j_{1}-1}T_{2}(x_{j_{1}},x_{j_{2}}),
	\end{equation}

	\noindent where the factor~$n(n-1)/2$ accounts for the total number of different ways to pair up the sequences. Given the mutation rate~$\mu$ and a sufficiently large sample size~$n$, one can use Eqs.~(\ref{EPi1d}) and~(\ref{ET1d}) to estimate~$D_{g}$ and~$D_{s}$. Both parameters can be estimated because~$\varPi_{\rm{1d}}$, unlike~$\varPi_{\rm{well-mixed}}$, depends on the spatial positions of the samples as well as on the properties of the population. Thus, one can generate independent equations to estimate~$D_{g}$ and~$D_{s}$ by using different subsets of the samples; for example,~$D_{g}$ can be estimated from the samples taken from the same point, and~$D_{s}$ can be estimated from the remaining samples. Note, however, that~$\varPi_{1d}$ depends only on~$\mu/D_{g}$ and~$\mu/D_{s}$, so at most two parameters can be estimated from the data; similar considerations hold for the well-mixed case as well. 

	The average number of pairwise differences is relatively easy to compute in both spatial and nonspatial models because it depends on the history of only two lineages. For the same reason,~$\varPi$ does not illuminate the underlying tree-like genealogy of the sample~(see Fig.\ref{FCoalescent}), and a different statistic is needed for that purpose. Under the infinite site assumption, a given site is either monomorphic, i.e. all samples have the same nucleotide at this site, or polymorphic, i.e. two different nucleotides are found: one is ancestral and the other is due to a mutation. Only the polymorphic sites contain information about the underlying genealogy, and the frequencies of mutations at each site are often used for genetic inference. Here, we consider a simpler summary statistic, the number of segregating sites~$S$, i.e. the expected number of polymorphic sites in the sample. 

	As we go backward in time, the number of lineages decreases due to coalescence events from~$n$ to~$n-1$, to~$n-2$, etc. until it eventually reaches~$1$; we consider only pairwise coalescence events assuming the population size is sufficiently large so that the coalescence of more than two lineages at one time is unlikely. Let~$T_{j}$ be the average time when~$j$ lineages are present. Then the expected number of polymorphic sites is given by

	\begin{equation}
	\label{ESTj}
	S=\mu\mathcal{L}_{g}\sum_{j_{1}=2}^{n}jT_{j},
\end{equation}

\noindent where the factors of~$j$ account for the fact that mutations can occur in any of the~$j$ lineages during the time interval~$T_{j}$.

For a well-mixed population, we compute~$T_{j}$ by noticing that any of~$j(j-1)/2$ distinct pairs of lineages can coalesce next, and, from Eq.~(\ref{EOMWellMixedU}), each pair has a constant coalescence rate of~$\mathfrak{D}_{g}$. Hence,

\begin{equation}
\label{ETjWellMixed}
T_{j}=\frac{2}{j(j-1)\mathfrak{D}_{g}},
\end{equation}

\noindent and from Eq.~(\ref{ESTj})

\begin{equation}
\label{ESWellMixed}
S_{\rm{well-mixed}}=\frac{2\mu\mathcal{L}_{g}}{\mathfrak{D}_{g}}\sum_{j_{1}=1}^{n-1}\frac{1}{j}\approx\frac{2\mu\mathcal{L}_{g}}{\mathfrak{D}_{g}}\left[\ln(n)+\gamma-\frac{1}{2n}\right],
\end{equation}

\noindent where the approximation is valid for large sample sizes~$n$~\cite{Gradshteyn:Tables}, and~$\gamma$ is the Euler constant.

For a one-dimensional population, one could try to generalize the approach used to calculate~$T_{2}$ for multiple lineages, but this method seems prohibitively difficult for large~$n$. However, we can qualitatively understand the effects of spatial structure by considering~$n$ lineages sampled uniformly in~$x$ from the habitat. While analyzing a related problem of annihilating random walks~(see Appendix~\ref{AVoter}), \textcite{Doering:IDF}, and~\textcite{Zhong:FiniteReactionRates} showed that for a generic uniform spatial distribution of the samples the number of surviving lineages~$j$ at reverse time~$\tau$ decays as

\begin{equation}
\label{Ej1d}
j(\tau)\sim\frac{1}{\sqrt{2\pi D_{s}\tau}}
\end{equation}

\noindent for intermediate times, when, on one hand, the time is sufficiently small for any lineage to diffuse across the whole habitat~($\tau\ll L^{2}/D_{s}$), but, on the other hand, the time is sufficiently large for neighboring lineages to coalesce~($\tau\ll Ln^{-1}D_{g}^{-1}+L^{2}n^{-2}D_{s}^{-1}$). 

Since~$T_{j}$ is the time during which the number of lineages changes from~$j$ to~$j-1$, it follows from Eq.~(\ref{Ej1d}) that

\begin{equation}
\label{ETj1d}
T_{j}\approx \tau(j-1)-\tau(j)\approx - \frac{d\tau(j)}{dj}=\frac{1}{\pi D_{s} j^{3}},
\end{equation}

\noindent where~$\tau(j)$ is the inverse function of~$j(\tau)$ used in Eq.~(\ref{Ej1d}). Equation~(\ref{ETj1d}) is only valid for intermediate $j$~values:~$1\ll j\ll n$ because of the similar restrictions on Eq.~(\ref{Ej1d}). Upon comparing this one-dimensional result for~$T_{j}$ to Eq.~(\ref{ETjWellMixed}), we see that the well-mixed model overestimates the contribution to the number of segregating sites from the recent part of genealogy with a large number of lineages. This should also be true for the initial stage~$j\approx n$, where Eq.~(\ref{Ej1d}) is not valid, because faster coalescence results from the fact that a lineages has to travel only about~$L/n$ to meet its neighbor. Other statistics that rely on the relative duration of periods with~$j$ lineages should be affected in a similar way. This is particularly important when the deviations of the observed genealogical data from the predictions based on Eq.~(\ref{ETjWellMixed}) are used to infer past evolutionary events, such as a selective pressure or geographic isolation~\cite{Wakeley:Coalescent}, because some of these deviations could be due to the spatial structure of the population rather than external or internal perturbations. 

In summary, the classic theory of genetic inference can be extended to spatial populations. These extensions are not only more accurate and realistic than assuming the well-mixed-population dynamics, but also can be used to obtain information about the migration within the habitat~\cite{Wilkins:Linear}. As spatially resolved genetic data sets become more readily available, better statistical tools based on spatial population genetics models will be needed.


\section{Conclusions}
\label{SConclusions}

\begin{figure}
\includegraphics[width=\columnwidth]{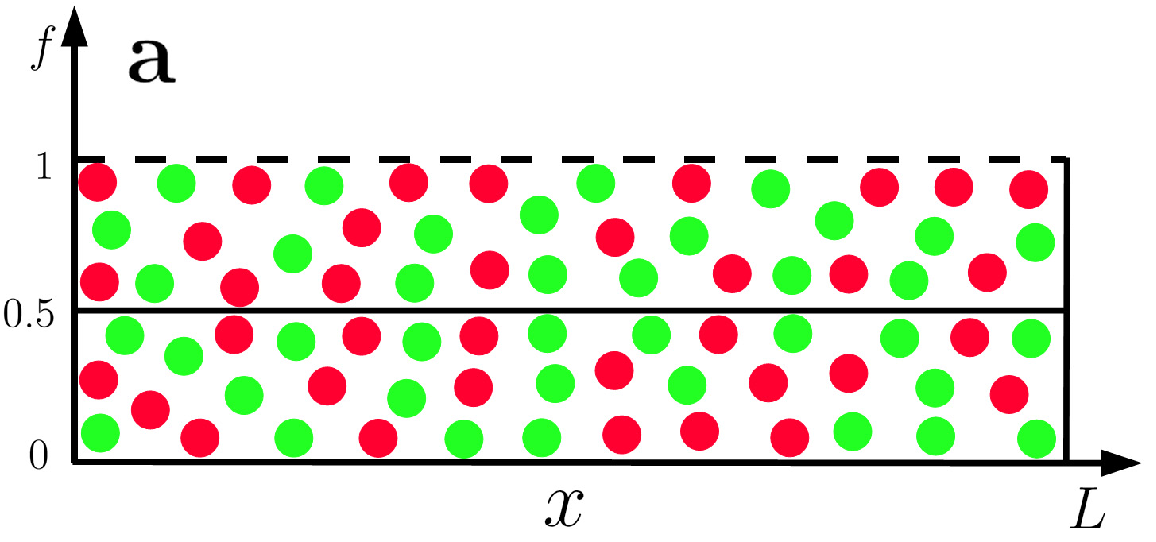}
\includegraphics[width=\columnwidth]{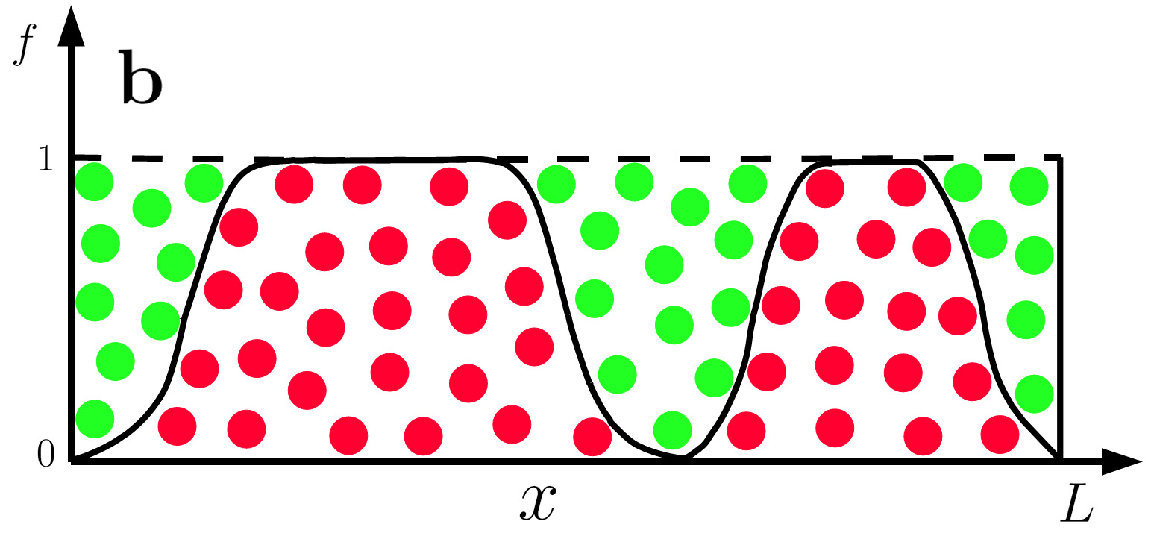}
\caption{(Color online) Illustration of spinodal-decomposition-like genetic demixing in a one-dimensional population. (a) Initially well-mixed population with red and green colors labeling different genotypes. (b) The same population several generations later. The frequency of one of the alleles is now oscillating between~$0$ and~$1$ because the population segregates into monoallelic domains.}
\label{FGeneticDriftSketch}
\end{figure}

Fluctuations due to sampling error during reproduction significantly affect the evolutionary dynamics of quasi one-dimensional populations, e.g. two-dimensional populations undergoing range expansions. These fluctuations lead to the genetic demixing illustrated in Fig.~\ref{FGeneticDriftSketch}, where an initially well-mixed population of alleles ``phase separates'' into monoallelic domains. The transition is somewhat analogous to spinodal decomposition in physics and material science~\cite{Scheucher:SpinodalDecomposition}, but is also markedly different. In particular, unlike conventional demixing phase transitions in finite-temperature statistical mechanics, genetic demixing occurs only in a \textit{low} number of spatial dimensions~$d$~($d\le2$)~\cite{Duty:Thesis,Scheucher:SpinodalDecomposition}. The dependence of genetic demixing on the number of spatial dimensions~$d$ is illustrated by the decay of local heterozygosity in the absence of selection and mutation. For long times, the functional form of the decay is given by~\cite{Duty:Thesis},

\begin{equation}
H(t,0)\sim\left\{
\begin{aligned}
&e^{-2t/(N\tau_{g})} \quad &d=0,\\
&(\tau_{f}/t)^{1/2}\quad &d=1,\\
&1/\ln(t) \quad &d=2,\\
&{\rm const} \quad &d>2.\\
\end{aligned}
\right.
\label{EHd}
\end{equation}

\noindent Note that~$d=2$ is the critical dimension. 

Here, we have shown that the one-dimensional stepping stone model has very different dynamics compared to the standard well-mixed-population models used in population genetics. Most of the differences arise because, in the spatial model, populations segregate into monoallelic domains. As a result, genetic drift and selection can only act at the boundaries of the domains, which slows down the dynamics of the model. In particular, we found that, in the neutral model without mutation, fixation occurs exponentially fast in a well-mixed population, but the decay of heterozygosity is algebraic in the spatial model. Genetic drift in the population as a whole becomes weaker with time, as spatial diffusion causes the effective population size to increase. For a linear one-dimensional model, we also found that the standard deviation of the total fraction of one of the alleles~(in the absence of selection and mutation) increases subdiffusively as~$t^{1/4}$. Selective sweeps also occur more slowly in the spatial model: for weak selection,~$s\ll D_{g}^{2}/D_{s}$, we found that the time constant of the selective sweep is quadratic in~$s$ in the linear spatial model, but it is only linear in~$s$ in the well-mixed-population model. The effects of mutation do not differ as dramatically in spatial and nonspatial models, but the stepping stone model reveals nontrivial spatial correlations and predicts a different value for the local steady state heterozygosity, proportional to~$\sqrt{\mu_{12}+\mu_{21}}$ for small mutation rates, compared to~$\mu_{12}+\mu_{21}$ in the well-mixed-population model. The evolutionary dynamics of spatial models also depends on the geometry of the expansion. For radial expansions, we found that the number of domains approaches a finite limit, which is, up to an additive constant, proportional to the square root of the initial radius of the colony~$R_{0}$. 

Our main conclusion is that the data from natural populations may not always conform to the predictions of the well-mixed-population model and, even when it does, the estimated parameters from the model may not be biologically meaningful. The spatial model contains an important additional parameter, the spatial diffusion constant parameter~$D_{s}$, which enters into many of the predictions. For example, the timescale for local fixation is given by~$D_{s}/D_{g}^{2}$ rather than~$N\tau_{g}$~[see Eq.~(\ref{EH0LimitNoMutation})] and, for small selective advantage,~$s$ is sometimes replaced by~$s^{2}D_{s}/D_{g}^{2}$, see Sec.~\ref{SSelection}. Moreover, as we saw in Sec.~\ref{SSimulations}, the timescale of fixation depends on the partitioning  of the population by the experimenter into measurement sites. Thus care must be taken when interpreting the data from the natural populations. Finally, well-mixed-population models and experiments without spatial resolution do not account for spatial correlations, which contain important information about the population~(see Secs.~\ref{SNeutralMutation} and~\ref{SGeneticInference}).


\begin{acknowledgments}
M. A. is grateful for financial support from the Danish National Research Foundation through Center for Models of Life. Overall support for this project was provided by the National Science Foundation, through Grant DMR-0654191, National Institute of General Medical Sciences Grant GM068763 of the National Centers for Systems Biology, and by the Harvard Materials Research Science and Engineering Center through DMR-0820484.
\end{acknowledgments}


\appendix

\section{\protect{The It\^o} calculus}
\label{AIto}
In this appendix, we briefly discuss the It\^o calculus. Our presentation relies on~\textcite{Risken:FPE} and~\textcite{Gardiner:Handbook}, which can be consulted for a more extensive presentation. For simplicity, we only consider nonspatial stochastic differential equations, but the results can be extended to spatial problems straightforwardly.

Let us analyze the following stochastic differential equation, which includes~Eq.~(\ref{EMoranStochastic}) as a special case,

\begin{equation}
\label{EGeneralSDE}
\frac{d\psi}{dt}=\omega(\psi)+g(\psi)\Gamma(t),
\end{equation}

\noindent where~$\Gamma(t)$ satisfies Eq.~(\ref{EMoranStochasticGamma}), and~$\omega(\psi)$ and~$g(\psi)$ are arbitrary continuously differentiable functions. From the point of view of ordinary calculus, Eq.~(\ref{EGeneralSDE}) is not well-defined because~$\Gamma(t)$ is discontinuous at every point. One way to circumvent this problem is to use discrete time steps of infinitesimal length~$\delta t$ rather than continuous time. Then, Eq.~(\ref{EGeneralSDE}) takes the following form:

\begin{equation}
\label{EDiscreteIto}
\frac{\psi(t+\delta t)-\psi(t)}{\delta t}=\omega[\psi(t)]+g[\psi(t)]\Gamma(t).
\end{equation}

\noindent However, this is not the only way to interpret Eq.~(\ref{EGeneralSDE}). For example, an alternative way to go from the continuous to a discrete description is to write Eq.~(\ref{EGeneralSDE}) as,

\begin{equation}
\label{EDiscreteStratonovich}
\begin{split}
\frac{\psi(t+\delta t)-\psi(t)}{\delta t}=&\omega\left[\frac{\psi(t)+\psi(t+\delta t)}{2}\right]\\ &+g\left[\frac{\psi(t)+\psi(t+\delta t)}{2}\right]\Gamma(t).
\end{split}
\end{equation}

\noindent In fact, there is an infinite number of ways to interpret Eq.~(\ref{EGeneralSDE}), depending on the relative weight of~$\psi(t)$ and~$\psi(t+\delta t)$ inside the arguments of the functions on the right hand side of the equation. The two most commonly used interpretations are  It\^{o}'s and Stratonovich's prescriptions. The former corresponds to Eq.~(\ref{EDiscreteIto}), and the latter to Eq.~(\ref{EDiscreteStratonovich}).

In physics, Stratonovich's prescription is commonly used because~$\Gamma(t)$ is usually an approximation to a thermal noise with small but finite correlation time; therefore, the argument of~$g(\cdot)$ should be an average value of~$\psi$ over the time that the correlations persist. In population genetics, on the other hand, Ito's prescription is appropriate because a random change of the allele frequencies depends only on the genetic composition of the population \textit{prior} to the change.

Without the stochastic term, Eqs.~(\ref{EDiscreteIto}) and~(\ref{EDiscreteStratonovich}) would yield the same results provided~$\delta t$ is sufficiently small, but the stochastic terms remain different even in the limit~$\delta t\rightarrow 0$. An easy way to see this difference is to average Eqs.~(\ref{EDiscreteIto}) and~(\ref{EDiscreteStratonovich}) with respect to the nondifferentiable noise function~$\Gamma(t)$. It\^{o}'s prescription gives~$\langle \psi(t+\delta t)\rangle-\langle\psi(t)\rangle=\langle \omega[\psi(t)]\rangle\delta t$ because~$\langle g[\psi(t)]\Gamma(t)\rangle=\langle g[\psi(t)]\rangle\langle\Gamma(t)\rangle=0$ due to the independence of~$\psi(t)$ and~$\Gamma(t)$. A similar simplification, however, cannot be applied to Stratonovich's prescription because, generically, the stochastic term depends on~$\psi(t+\delta t)$, which is not independent of~$\Gamma(t)$.

Because of the aforementioned ambiguity in interpreting stochastic differential equations with multiplicative noise, care must be taken while differentiating stochastic variables. While the rules of ordinary calculus apply to Stratonovich's prescription, special rules of the It\^{o} calculus are required for It\^{o}'s prescription when tracking the evolution of a composite function~$u[\psi(t)]$ of the stochastic variable obeying Eq.~(\ref{EGeneralSDE}). In this paper, we use It\^{o}'s formula, namely~\cite{Risken:FPE,Gardiner:Handbook}

\begin{equation}
\label{EItoFormula}
\begin{split}
\frac{d}{dt}u[\psi(t)]=&u'[\psi(t)]\omega[\psi(t)]+u'[\psi(t)]g[\psi(t)]\Gamma(t)\\ &+\frac{1}{2}u''[\psi(t)]g^{2}[\psi(t)],
\end{split}
\end{equation}

\noindent where~$u(\psi)$ is a twice continuously differentiable function, and the primes now indicate differentiation with respect to~$\psi$. The last term is the crucial addition due to the It\^{o} calculus.

We conclude this discussion with an illustration of how Eq.~(\ref{EItoFormula}) can be used by deriving Eqs.~(\ref{EMoranGeneticDriftF}) and~(\ref{EMoranGeneticDriftH}) from Eq.~(\ref{EMoranStochastic}) assuming~$s=0$ and~$\mu_{12}=\mu_{21}=0$. Thus, we start from the following equation of motion for~$f(t)$

\begin{equation}
\label{EOMIllustration}
\frac{df(t)}{dt}=\sqrt{\mathfrak{D}_{g}f(t)[1-f(t)]}\Gamma(t) \quad \mbox{(It\^{o})}.
\end{equation}

\noindent Thus,~$\psi(t)=f(t)$, $\omega[\psi(t)]=0$, and~$g[\psi(t)]=\sqrt{\mathfrak{D}_{g}\psi(t)[1-\psi(t)]}$. Since~$F(t)=\langle f(t)\rangle$, we obtain Eq.~(\ref{EMoranGeneticDriftF}) by averaging Eq.~(\ref{EOMIllustration}). For~$H(t)=\langle h(t)\rangle=\langle 2 f(t)[1-f(t)]\rangle$, we use Eq.~(\ref{EItoFormula}) with~$u[\psi(t)]=2\psi(t)[1-\psi(t)]$ to obtain the equation of motion for~$h(t)$

\begin{equation}
\label{EOMIllustrationh}
\begin{split}
\frac{dh(t)}{dt}=&0+2[1-2f(t)]\sqrt{\mathfrak{D}_{g}f(t)[1-f(t)]}\Gamma(t)\\&+\frac{1}{2}(-4)\mathfrak{D}_{g}f(t)[1-f(t)] \quad \mbox{(It\^{o})}.
\end{split}
\end{equation}

\noindent Upon averaging Eq.~(\ref{EOMIllustrationh}) with the rules described above, we obtain Eq.~(\ref{EMoranGeneticDriftH}).


\section{Solution of the Neutral Model Without Mutations}
\label{ANeutral}

In this appendix, we solve Eq.~(\ref{EHNoMutationEquation}) subject to the initial condition~$H(0,x)=H_{0}$. It is advantageous to first solve a simpler equation:

\begin{equation}
\label{ESimplifiedRD}
\frac{\partial}{\partial t}H=2D_{s}\frac{\partial^{2}}{\partial x^{2}}H-b(t)\delta(x),
\end{equation}

\noindent where~$b(t)$ is an arbitrary function of time. Equation~(\ref{ESimplifiedRD}) is a standard diffusion equation with a sink term, and it can be readily solved in the Fourier domain. The result is

\begin{equation}
\label{ESimplifiedRDSolution}
H(t,x)=H_{0}-\int_{0}^{t}dt'\frac{b(t')e^{-\frac{x^{2}}{8D_{s}(t-t')}}}{\sqrt{8\pi D_{s}(t-t')}}.
\end{equation}

\noindent Note the convolution of~$b(t')$ with the diffusion propagator. Now, we impose a self-consistency condition~$b(t)=D_{g}H(t,0)$, which leads to

\begin{equation}
\label{ESelfConsistentCondition}
H(t,0)=H_{0}-D_{g}\int_{0}^{t}dt'\frac{H(t',0)}{\sqrt{8\pi D_{s}(t-t')}}.
\end{equation}

\noindent This is Abel's integral equation of the second kind, canonically written as

\begin{equation}
\label{EAbelSecondKind}
y(x)+\lambda\int_{a}^{x}\frac{y(t)dt}{\sqrt{x-t}}=g(x),
\end{equation}

\noindent where~$g(x)$ is a known function. The general solution of Eq.~(\ref{EAbelSecondKind}) given in~\textcite{Polyanin:IntegralEquations} reads

\begin{equation}
\label{EAbelSecondKindSolution}
\begin{split}
&y(x)=G(x)+\pi\lambda^2\int_{a}^{x}e^{\pi\lambda^2(x-t)}G(t)dt, \quad \mbox{where}\\
&G(x)=g(x)-\lambda\int_{a}^{x}\frac{g(t)dt}{\sqrt{x-t}}.
\end{split}
\end{equation}

Equations~(\ref{EHSolutionNoMutation})~and~(\ref{EH0SolutionNoMutation}) follow from Eqs.~(\ref{ESimplifiedRDSolution}),~(\ref{ESelfConsistentCondition}),~(\ref{EAbelSecondKind}), and~(\ref{EAbelSecondKindSolution}).

For radial expansions considered in Sec.~\ref{SInflation}, one can solve the equation of motion for~$H(\mathsf{t},\phi)$ by following the same set of steps.

\section{Average domain density from the spatial heterozygosity~$H(t,x)$}
\label{ADomainSize}

In this appendix, we derive the relationship between the spatial heterozygosity,~$H(t,x)$, and the average domain density~$n_{d}(t)$. From~$n_{d}$, we can get a domain size by defining~$\ell \equiv n_{d}^{-1}$. The result for the domain density is valid for an arbitrary number of alleles, so in this appendix we use a broader definition of~$H(x,t)$ as the average probability of sampling at time~$t$ two different alleles from two demes distance~$x$ apart. We assume that the domains have formed, and they are on average much larger than the boundary regions.

Let~$h(t,x_{1},x_{2})$ equal to~$1$ if both~$x_{1}$ and~$x_{2}$ are occupied by organisms in different allelic state and~$0$ otherwise. To compute~$\ell$, we use an alternative definition of~$H(t,x)$ with ensemble average replaced by space average:

\begin{equation}
\label{ESpaceAverageH}
H(t,x)=\lim_{L\rightarrow\infty}\frac{1}{L}\int_{0}^{L}h(t,\xi,\xi+x)d\xi,
\end{equation}

\noindent where we assume periodic boundary conditions. Let us compute~$H(t,x+\delta x)-H(t,x)=\lim_{L\rightarrow\infty}\frac{1}{L}\int_{0}^{L}[h(t,\xi,\xi+\delta x)-h(t,\xi,\xi)]d\xi$ for~$\delta x$ small compared to typical domain size, but large compared to the deme spacing~$a$. To do so, we expand both sides in~$\delta x$. At the lowest order in~$\delta x$, each domain boundary contributes~$\delta x$ to the right hand side; therefore,~$\frac{\partial}{\partial x}H(t,+0)$ equals the density of the domain boundaries. Upon defining the average domain size~$\ell(t)$ as the inverse of the domain boundary density, we obtain the following relationship:

\begin{equation}
\label{EDomainSizedHdx}
\ell(t)=\left(\frac{\partial H(t,+0)}{\partial x}\right)^{-1}.
\end{equation}

\noindent This relation is analogous to the one derived in~\textcite{Benavraham:ExactCoalescence}.

We can further simplify Eq.~(\ref{EDomainSizedHdx}) by observing that

\begin{equation}
\label{EH0dHdx}
\frac{\partial H(t,+0)}{\partial x}=\frac{D_{g}H(t,0)}{4D_{s}},
\end{equation}

\noindent which follows from integrating Eq.~(\ref{EContinuous1dSSMHHomogeneous}) or Eq.~(\ref{EHInfiniteAlleleEquation}) with respect to~$x$ from~$-\epsilon$ to~$\epsilon$,~$0<\epsilon\ll 1$, and noticing that~$H(t,x)$ is an even function of~$x$. The final result then reads

\begin{equation}
\label{EDomainSizeH0}
\ell(t)=\frac{4D_{s}}{D_{g}H(t,0)}.
\end{equation}

\noindent It should be emphasized that this result is only valid in the limit of very large domain sizes compared to the boundary regions, which means~$H(t,0)\ll1$. Therefore the leading term in~$H(t,0)$ is sufficient at this level of approximation. Note that Eqs.~\ref{EDomainSizedHdx} and~\ref{EDomainSizeH0} are valid in the presence of genetic drift, migration, selection, and mutation. For radially expanding populations subject to inflation, Eq.~\ref{EDomainSizedHdx} remains valid, but Eq.~(~\ref{EDomainSizeH0}) is replaced by Eq.~(\ref{EDomainSizedHdphi}).


\section{Infinite Alleles Model}
\label{AInfiniteAlleleModel}

In this appendix, we extend the analysis of the stepping stone model with mutations presented in Sec.~\ref{SNeutralMutation} to the infinite alleles model. The infinite alleles model assumes that every new mutation creates a new allele, which is a good approximation for genes encoded by a large number of nucleotides because the number of all possible mutations is much larger than the number of all possible back mutations~\cite{Hartl:PopulationGenetics}. The equation of motion for~$H(t,x)$, which we interpret as the average probability of sampling two different alleles from demes $x$~apart, can be derived by following two lineages backward in time, as done in Sec.~\ref{SNeutralNoMutation}. In the presence of mutation, the right hand side of Eq.~(\ref{EHNoMutationEquation}) should contain an additional term describing the rate of increase of~$H(t,x)$ due to mutations in both of the lineages. Because, in the infinite alleles model, a mutation changes the probability that the organisms have different alleles from~$H$ to~$1$, that is by~$1-H$, the new term is $2\mu(1-H)$, where~$\mu$ is the mutation rate that is assumed to be the same for all types of mutations. Thus, Eq.~(\ref{EHNoMutationEquation}) becomes

\begin{equation}
\label{EHInfiniteAlleleEquation}
\frac{\partial}{\partial t}H=2D_{s}\frac{\partial^{2}}{\partial x^{2}}H+2\mu (1-H)-D_{g}H\delta(x)
\end{equation}

\noindent for the infinite alleles model~[compare Eq.~\ref{EMutationHEquation}].
The stationary solution of Eq.~(\ref{EHInfiniteAlleleEquation}) is given  by

\begin{equation}
\label{EHInfiniteAlleleStationarySolution}
H(\infty,x)=1-\frac{e^{-\sqrt{\frac{\mu}{D_{s}}}|x|}}{1+\frac{1}{4}\frac{D_{g}}{\sqrt{D_{s}\mu}}}.
\end{equation}

\noindent At large separations,~$H(\infty,x)$ approaches one, which is consistent with the infinite number of alleles. Locally,~$H(\infty,0)=(1+\frac{1}{4}\frac{D_{g}}{\sqrt{D_{s}\mu}})^{-1}$, and if~$H(\infty,0)\ll 1$ the population is segregated into domains containing only one allelic type. The average size of such domains follows from Eq.~(\ref{EDomainSizeH0}):
\begin{equation}
\label{EAverageDomainSizeInfiniteAlleleModel}
\ell=\frac{4D_{s}}{D_{g}}\left(1+\frac{1}{4}\frac{D_{g}}{\sqrt{D_{s}\mu}}\right)\approx\sqrt{\frac{D_{s}}{\mu}},
\end{equation}

\noindent where the last equality follows from the assumption that~$H(\infty,0)\ll1$.
The approach to the stationary state can be either obtained by methods of Appendix~\ref{ANeutral} or by the change of variables~$H(t,x)=H(\infty,t)+e^{-2\mu t}\hat{H}(t,x)$, which reduces Eq.~(\ref{EHInfiniteAlleleEquation}) to Eq.~(\ref{EHNoMutationEquation}). The result is that the slowest decaying mode vanishes as~$\tilde{C}t^{-1/2}e^{-2\mu t}$, where~$\tilde{C}$ is a constant.

The infinite allele model and Eq.~(\ref{EHInfiniteAlleleEquation}) has been analyzed before by~\textcite{Malecot:Dynamics} and~\textcite{Nagylaki:DecayGeneticVariability}, who calculated the stationary solution and the long time approach to the equilibrium. Our results are consistent with their findings.


\section{A model with several neutral alleles}
\label{AMultipleColors}

A model with~$q$ neutral alleles is an intermediate case between the two-alleles model that we focus on in this paper and the infinite alleles model discussed in Appendix~\ref{AInfiniteAlleleModel}. The $q$-alleles model is also analogous to nonequilibrium $q$-state Potts models. In this appendix, we briefly outline how the~$q$-alleles model can be formulated and solved in the language of one and two-point correlation functions, compare our analytical predictions to simulations, and extend Eq.~(\ref{ETotalFractionVarianceSolutionLimit}) to the undulating-front model.

To specify the~$q$-alleles model, we let~$f_{i}(t,x)$ be the frequency of allele~$i$ at time~$t$ and position~$x$; these quantities satisfy~$\sum_{i=1}^{q}f_{i}(t,x)=1$. The spatial diffusion and coalescence probability of two lineages are still characterized by~$D_{s}$ and~$D_{g}$ respectively. Intra-allelic mutations are described by the mutation matrix~$\mu_{ij}$, which is the probability of allele~$i$ mutating into allele~$j$ if~$i\ne j$. When~$i=j$, we let~$\mu_{ii}=-\sum_{j=1,\; j\ne i}^{q}\mu_{ij}$ to describe the outflow of alleles from allelic state~$i$ due to mutations. 

The dynamics of the~$q$-alleles model can be analyzed by considering one-point correlation functions~$F_{i}(t,x)=\langle f_{i}(t,x)\rangle$ and two-point correlation functions~$F_{ij}(t,x_{1},x_{2})=\langle f_{i}(t,x_{1})f_{j}(t,x_{2})\rangle$. $F_{i}(t,x)$~is the probability to find allele~$i$ at position~$x$ at time~$t$, and~$F_{ij}(t,x_{1},x_{2})$ is the probability to simultaneously find at time~$t$ allele~$i$ at position~$x_{1}$ and allele~$j$ at position~$x_{2}$. The evolution equations for these correlation functions are obtained by tracing one and two lineages backward in time; the results are

\begin{equation}
\label{EMultipleColorsEquationOne}
\frac{\partial F_{i}(t,x)}{\partial t}=D_{s}\frac{\partial^{2}F_{i}(t,x)}{\partial x^{2}}+\sum_{j=1}^{q}\mu_{ji}F_{j}(t,x),
\end{equation}

\begin{equation}
\label{EMultipleColorsEquationTwo}
\begin{split}
\frac{\partial F_{ij}(t,x_{1},x_{2})}{\partial t}=&D_{s}\left(\frac{\partial^{2}}{\partial x_{1}^{2}}+\frac{\partial^{2}}{\partial x_{2}^{2}}\right)F_{ij}(t,x_{1},x_{2})\\ &+D_{g}\delta(x_{1}-x_{2})[\delta_{ij}F_{i}(t,x_{1})\\ &-F_{ij}(t,x_{1},x_{2})]\\ &+ \sum_{i'=1}^{q}\sum_{j'=1}^{q}[\mu_{i'i}F_{i'j}(t,x_{1},x_{2})\\ &+\mu_{j'j}F_{ij'}(t,x_{1},x_{2})],
\end{split}
\end{equation}

\noindent where~$\delta_{ij}$ is Kronecker's delta, which is zero if~$i\ne j$ and one otherwise. Thus, for a generic mutation matrix~$\mu_{ij}$ one has to solve a system of coupled linear partial differential equations. 

For simplicity and the ease of comparison with the other results in this paper, let us assume spatial homogeneity and identical mutation rates between any two alleles,~$\mu_{i\ne j}=\mu/q$. Under these assumptions, Eq.~(\ref{EMultipleColorsEquationTwo}) can be simplified by introducing averaged spatial heterozygosity

\begin{equation}
\label{EManyAlleleHeterozygosityDefinition}
H(t,x)=\sum_{i=1}^{q}\sum_{\substack{j=1 \\ j\ne i}}^{q}F_{ij}(t,0,x)=\sum_{i=1}^{q}\langle f_{i}(t,0)[1-f_{i}(t,x)]\rangle,
\end{equation}

\noindent which is the probability to sample two different alleles at time~$t$ distance~$x$ apart. The equation of motion for~$H(t,x)$ can be derived both from~Eq.~(\ref{EMultipleColorsEquationTwo}) and, more simply, by tracing two lineages backward in time:

\begin{equation}
\label{EHMultipleColorsEquation}
\frac{\partial}{\partial t}H=2D_{s}\frac{\partial^{2}}{\partial x^{2}}H+2\mu \left(\frac{q-1}{q}-H\right)-D_{g}H\delta(x).
\end{equation}

\noindent Note that Eq.~(\ref{EHMultipleColorsEquation}) agrees with Eq.~(\ref{EHInfiniteAlleleEquation}) in the limit~$q\rightarrow\infty$ and with Eq.~(\ref{EMutationHEquation}) for~$\mu_{12}=\mu_{21}=\mu/2$. Since Eq.~(\ref{EHMultipleColorsEquation}) has the same functional form as Eq.~(\ref{EHInfiniteAlleleEquation}), the methods of Appendix~\ref{AInfiniteAlleleModel} can be used to solve for~$H(t,x)$.

\begin{figure}
\includegraphics[width=\columnwidth]{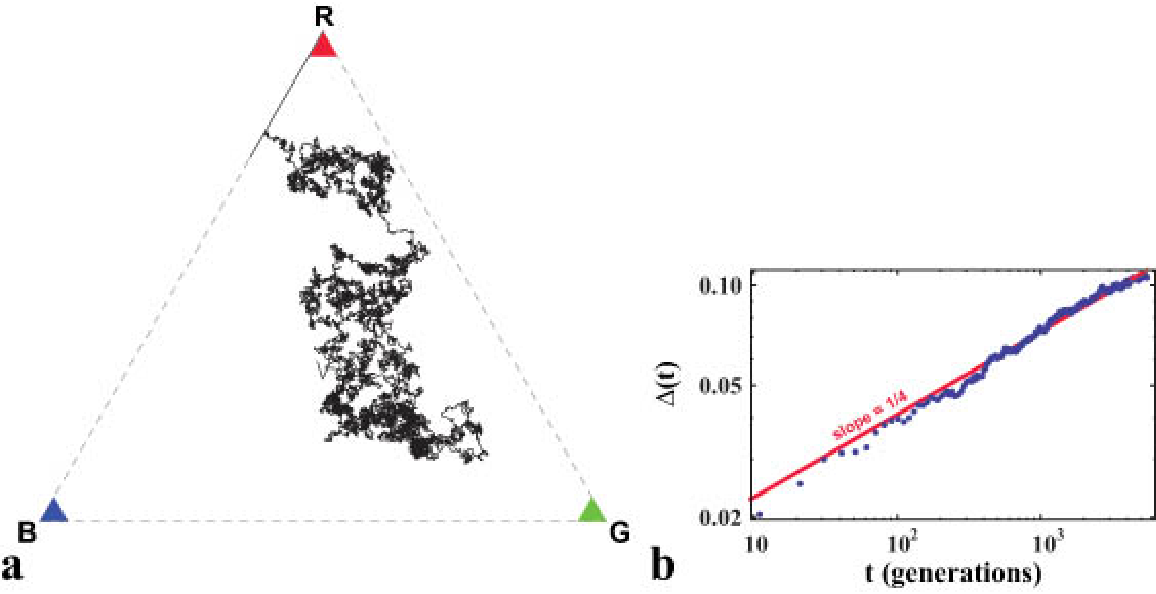}
\caption{(Color online)  Genetic drift during a linear range expansion in the flat-front model with three alleles. (a)~The genetic composition of the population~$[\mathfrak{f}_{1}(t),\mathfrak{f}_{2}(t),\mathfrak{f}_{3}(t)]$ projected on the plane~$\sum_{i=1}^{3}\mathfrak{f}_{i}(t)=1$ in a single run of the neutral $3$-alleles model with a flat front. The population is finite,~$L=1000$, and there are no mutations. 
	(b)~The average standard deviation of the frequency of allele one~$\Delta(t)$, shown in blue, is obtained from~$200$ realizations of the simulations described in~a. The red solid line shows the best power-law fit, and the slope is close to the exponent expected from Eq.~(\protect{\ref{ETotalFractionVarianceSolutionLimit}}). The gray area encloses the points within one standard deviation from the mean. At~$t=0$, each site is assigned either allele one or allele two with equal probability, which corresponds to the center of the triangle in (a).}
	\label{FThreeColors}
	\end{figure}

In the absence of mutations, Eq.~(\ref{EHMultipleColorsEquation}) is identical to Eq.~(\ref{EHNoMutationEquation}), as we briefly mentioned in Sec.~\ref{SNeutralNoMutation}. However, $q$-alleles models with different~$q$ may have slightly different dynamics due to $q$-dependent initial conditions: for example, an initially well-mixed population is represented by~$H(0,x)=H_{0}=1-1/q$. Thus the results of Sec.~\ref{SNeutralNoMutation} apply to the $q$-alleles model, provided appropriate initial conditions are used. In particular, we expect the standard deviation of~$\mathfrak{f}_{i}(t)$, the total frequency of allele~$i$, in a finite population to grow as~$t^{1/4}$. This is indeed confirmed by our simulations shown in Fig.~\ref{FThreeColors}. Spatial correlations in the nonequilibrium $q$-state Potts model have recently been analyzed by Masser and ben-Avraham~\cite{Masser:Potts}, who also found that two-point correlation functions obey the same $q$-independent equation of motion.

	Finally, one can obtain the behavior of the standard deviation of the total frequency of allele one,~$\Delta(t)$, in the undulating-front model by the following scaling argument. We consider~$\Delta(t)$ at large times after monoallelic domains have formed. Let~$N_{d}(t)$ be the number of domains consisting of allele one and~$d_{k}(t), \; k=1,2,...,N_{d}(t)$ be lengths of these domains. Then,~$\Delta(t)$ is given by

	\begin{equation}
	\label{EDeltaLengthDefinition}
	\Delta(t)=\sqrt{\frac{1}{L^{2}}\left\langle\left[\sum_{k=1}^{N_{d}(t)}d_{k}(t)-\langle\sum_{k=1}^{N_{d}(t)}d_{k}(t)\rangle\right]^{2}\right\rangle}.
	\end{equation}

	\noindent We simplify Eq.~(\ref{EDeltaLengthDefinition}) by making an approximation that~$N_{d}(t)$ and~$d_{k}(t)$ for~$k=1,2,...,N_{d}(t)$ are independent random variables, which gives

	\begin{equation}
	\label{EDeltaLengthSimplificationOne}
	\begin{split}
\Delta^{2}(t) \approx & \frac{1}{L^{2}}\left\{\langle N_{d}(t)\rangle [\langle d_{1}^{2}(t)\rangle-\langle d_{1}(t)\rangle^{2}] \right. \\ & \left.+\langle d_{1}(t)\rangle^{2}[\langle N^{2}_{d}(t)\rangle-\langle N_{d}(t)\rangle^{2}]\right\},
\end{split}
\end{equation}

\noindent where we used the fact that~$d_{i}(t)$ are identically distributed. 

By using first passage time analysis discussed in~\textcite{Redner:FirstPassageTime}, one can show that~$\langle N^{2}_{d}(t)\rangle-\langle N_{d}(t)\rangle^{2}\propto \langle N_{d}(t)\rangle [\langle d_{1}^{2}(t)\rangle-\langle d_{1}(t)\rangle^{2}]/\langle d_{1}(t)\rangle^{2}\propto L[\langle d_{1}^{2}(t)\rangle-\langle d_{1}(t)\rangle^{2}]/\langle d_{1}(t)\rangle^{3}$. Thus 

\begin{equation}
\label{EDeltaLengthSimplificationTwo}
\Delta^{2}(t)\propto\frac{1}{L\langle d_{1}(t)\rangle} [\langle d_{1}^{2}(t)\rangle-\langle d_{1}(t)\rangle^{2}].
\end{equation}

\begin{figure}
\includegraphics[width=\columnwidth]{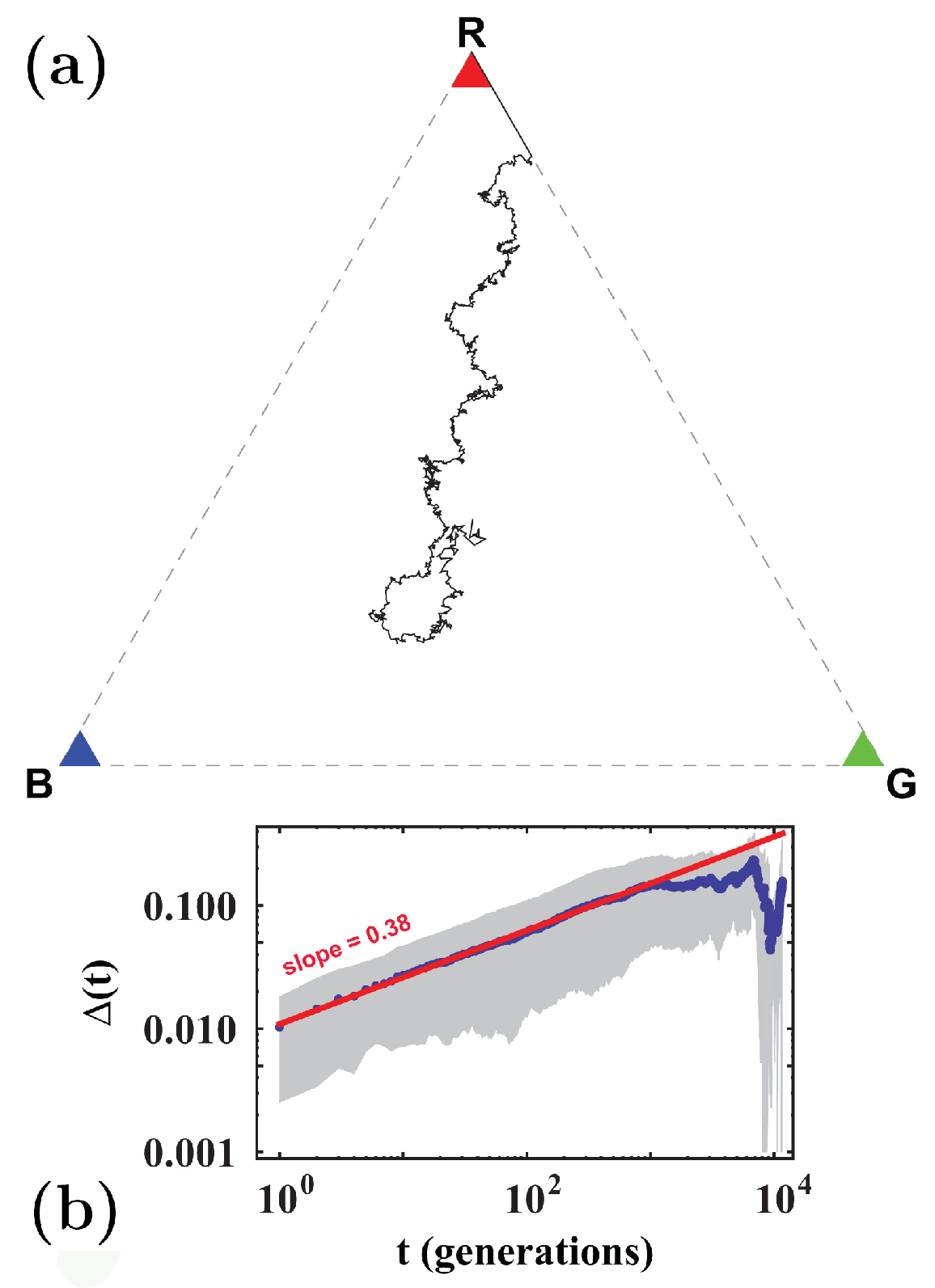}
\caption{(Color online)  Genetic drift during a linear range expansion in the undulating-front model with three alleles. (a)~The genetic composition of the population~$[\mathfrak{f}_{1}(t),\mathfrak{f}_{2}(t),\mathfrak{f}_{3}(t)]$ projected on the plane~$\sum_{i=1}^{3}\mathfrak{f}_{i}(t)=1$ in a single run of the neutral $3$-alleles model with an undulating front. The population is finite,~$L=1000$, and there are no mutations. 
(b)~The average standard deviation of the frequency of allele one~$\Delta(t)$, shown in blue, is obtained from~$200$ realizations of the simulations described in a. The red solid line shows the best power-law fit, and the slope is close to the exponent expected from Eq.~(\protect{\ref{EDeltaLengthSimplificationFinal}}). The gray area encloses the points within one standard deviation from the mean. At~$t=0$, each site is assigned either allele one or allele two with equal probability, which corresponds to the center of the triangle in (a).}
\label{FUndulatingFrontThreeColors}
\end{figure}

\noindent Upon recalling, that, in the undulating-front model,~$\langle d_{1}(t)\rangle\propto t^{\zeta}$, and $\langle d_{1}(t)\rangle^{2}\propto t^{2\zeta}$, we conclude that

\begin{equation}
\label{EDeltaLengthSimplificationFinal}
\Delta(t)\propto\frac{t^{\zeta/2}}{\sqrt{L}}\propto\frac{t^{1/3}}{\sqrt{L}},
\end{equation}

\noindent where, in the last proportionality, we used~$\zeta=2/3$ from~\textcite{Saito:EdenModel}. Equation~(\ref{EDeltaLengthSimplificationFinal}) is in good agreement with the simulations of the undulating-front model shown in Fig.~\ref{FUndulatingFrontThreeColors}.

\section{Connection with the voter model and one-dimensional reaction kinetics}
\label{AVoter}

The stepping stone model with only one organism per island or ``deme,''~$N=1$, has been extensively studied in probability theory~\cite{Liggett:Book, Durrett:LectureNotes} and nonequilibrium statistical mechanics~\cite{Odor:NonEquilibrium}, where it is known as the voter model. The model typically considers a set of voters on a hypercubic lattice in $d$-dimensions. Each voter holds one of the~$q$ possible opinions about an issue~(corresponding to~$q$ alleles in population genetics), and, at a certain rate, each voter reconsiders the issue, and adopts the opinion of a randomly chosen nearest neighbor. The voter model can be mapped onto the dynamics of the $q$-state Potts model at zero temperature. In one and two dimensions, opinions in the voter model coarsen spatially with time, and the model approaches one of the~$q$ absorbing states, in which all the voters have the same opinion~\cite{CoxGriffeath:VoterModel,Duty:Thesis}. In higher dimensions, the voters still form cluster of opinions, but these clusters stop growing after reaching a certain limiting size. Selection and mutation are typically not considered in voter models.

The voter model can be solved exactly by tracing the history of opinion adoptions backward in time~\cite{CoxGriffeath:VoterModel,Scheucher:SpinodalDecomposition,Duty:Thesis}. The opinion of a given individual performs a random walk as we follow the opinion from its current holder to its ultimate ancestor. With this observation, we can easily understand how the behavior of the voter model depends on the number of spatial dimensions. In one and two dimensions, a pair of random walks always meet~\cite{Redner:FirstPassageTime}, so the histories of opinion adoptions starting from two different voters will eventually converge to a single voter as we trace them backward in time. Therefore, any two voters should have the same opinion after a sufficiently long time has elapsed. In higher dimensions, however, there is a finite probability that two random walkers never meet~\cite{Redner:FirstPassageTime}; therefore, the voters never agree, and an absorbing state is never reached.

Another important property of the voter model is that the dynamics occurs only at the boundaries between the opinion clusters; inside a cluster the opinions cannot change because every voter has the same opinion as its nearest neighbors. This property is particularly useful in one spatial dimension, where it allows us to map the dynamics of the voter model to the one-dimensional diffusion-limited chemical kinetics of point particles. We identify each domain wall with a particle performing a random walk due to opinion changes at the boundary. When two particles meet, they react with two possible outcomes. They annihilate~($A+A\rightarrow 0$) if the flanking domains have the same opinion or coalesce~($A+A\rightarrow A$) otherwise. If the initial state is uncorrelated, the annihilation occurs with probability~$1/(q-1)$, and the coalescence with probability~$(q-2)/(q-1)$. In one-dimension, this reaction diffusion system has been analyzed by~\textcite{Masser:Potts}, who found that the density of the domain walls decays as~$t^{-1/2}$ in agreement with Eq.~(\ref{EDomainSizeNoMutation}). A related model of annihilating random walks for radial and linear range expansions was solved by~\textcite{Hallatschek:LifeFront}.

It is not surprising that the voter model and the stepping stone model have the same long time behavior in one dimension. At long times, most of the voters belong to large domains; therefore, we do not affect the system by combining neighboring sites into larger coarse-grained demes, as in Sec.~\ref{SSimulations}. For these large demes, the equations of motion of the stepping stone model are valid, so the two models are equivalent in the long time limit. The voter and the stepping stone models are also equivalent in the small-$D_{s}$ limit of very slow migration. In this case, each deme reaches fixation much faster than it sends out or accepts new migrants; hence, the stepping stone model reduces to the voter model with one voter representing an entire deme.

We can further illustrate the connection between the stepping stone model and the voter model by calculating the probability that two voters $l$~sites apart have different opinions. This probability is analogous to the average spatial heterozygosity, so we call it~$H(t,l)$. The equation of motion for~$H(t,l)$ is obtained by following the histories of opinion adoptions backward in time. Since~$H(t,l)$ changes only due to the diffusion of the history traces, the equation of motion reads

\begin{equation}
\frac{d}{dt}H(t,l)=[H(t,l-1)+H(t,l+1)-2H(t,l)],
\label{AEHDiscrete}
\end{equation}

\noindent where we measure time in such units that the rate of opinion adoption is set to unity. While Eq.~(\ref{AEHDiscrete}) can be solved exactly~\cite{Houchmandzadeh:NeutralEcology}, it is more instructive to go to the continuum limit, in which the equation of motion for~$H(t,x)$ takes the following form

\begin{equation}
\label{AEHVoter}
\frac{\partial}{\partial t}H(t,x)=2D_{s}\frac{\partial^{2}}{\partial x^{2}}H(t,x),
\end{equation}

\noindent where~$D_{s}$ denotes the spatial diffusion constant as in Eq.~(\ref{EHNoMutationEquation}).

Upon comparing Eqs.~(\ref{EHNoMutationEquation}) and~(\ref{AEHVoter}), one might naively conclude that the voter model corresponds to~$D_{g}=0$ limit of the stepping stone model; in fact, the opposite is true: the voter model corresponds to the limit~$D_{g}=\infty$. Qualitatively, one can see this from the fact that~$D_{g}\propto N^{-1}$, so, as the deme size~$N$ approaches its lowest value of~$1$, we expect~$D_{g}$ to increase. On more rigorous grounds, we should note that the role of the delta function in Eq.~(\ref{EHNoMutationEquation}) is to enforce a boundary condition at~$x=0$, provided one considers~$H(t,x)$ only for~$x>0$. This boundary condition is derived in Appendix~\ref{ADomainSize} and is given by Eq.~(\ref{EH0dHdx}). The corresponding boundary condition for Eq.~(\ref{AEHVoter}) is~$H(t,0)=0$ because the probability of one voter having two different opinions is zero. We indeed recover~$H(t,0)=0$ by letting~$D_{g}\rightarrow\infty$ in Eq.~(\ref{EH0dHdx}).

One can solve Eq.~(\ref{AEHVoter}) for the initial condition~$H(0,x)=H_{0}$ by the Laplace transform in time or a self-similar ansatz; the solution reads

\begin{equation}
\label{AEVoterSolution}
H(t,x)=H_{0}\erf\left(\frac{|x|}{\sqrt{8D_{s}t}}\right).
\end{equation}

\noindent We can now compute the average size of the domains with the help of Eq.~(\ref{EDomainSizedHdx}). As we expect, the result is given by Eq.~(\ref{EDomainSizeNoMutation}) because the long time limits of the stepping stone model and the voter model agree.

The~$D_{g}=\infty$ approximation is particularly valuable for circular fronts undergoing inflation because the exact solution of the stepping stone model in this case~[Eqs.~(\ref{EHCircularSolution}) and~(\ref{EH0CircularSolution})] is rather unwieldy. The equation of motion for~$H(t,\varphi)$ in the voter model with inflation is given by

\begin{equation}
\label{AEHCircularVoter}
\frac{\partial}{\partial t}H(t,\varphi)=\frac{2D_{s}}{(R_{0}+\mathsf{v}t)^{2}}\frac{\partial^{2}}{\partial \varphi^{2}}H(t,\varphi).
\end{equation}

One can solve Eq.~(\ref{AEHCircularVoter}) in the Fourier domain, and compute the nontrivial limit-shape as~$t\rightarrow\infty$. The result reads

\begin{equation}
\label{AEVoterSolutionInflation}
H(\infty,\varphi)=H_{0}\erf\left(|\varphi|\sqrt{\frac{R_{0}\mathsf{v}}{8D_{s}}}\right).
\end{equation}

\noindent With the help of the angular version of Eq.~(\ref{EDomainSizedHdx}) [see Eq.~(\ref{EDomainSizedHdphi})], we calculate the final number of sectors:

\begin{equation}
\mathcal{N}(\infty)=H_{0}\sqrt{\frac{2\pi R_{0}\mathsf{v}}{D_{s}}},
\label{AENRVoter}
\end{equation}

\noindent which agrees with Eq.~(\ref{ENR}) in the limit~$D_{g}=\infty$. In the same limit, we can also obtain an analytical expression for the long time variance~$\nu(\infty)$:

\begin{equation}
\label{EANuRVoter}
\nu(\infty)=\frac{1}{4\pi}\int_{-\pi}^{\pi}[H_{0}-H(\infty,\varphi)]d\varphi=H_{0}\sqrt{\frac{2D_{s}}{\pi^{3}R_{0}\mathsf{v}}},
\end{equation}

\noindent where we used the relationship between the variance~$\nu(t)$ and the global heterozygosity~$\mathcal{H}(t,\varphi)$ given by the spatial generalization of Eq.~(\ref{EHVarianceWellMixed}).

Finally, we note that the mapping to a one-dimensional reaction-diffusion system of particles could be generalized to account for super-diffusive boundaries in the undulating-front model, for example, by considering continuous time L\'{e}vy flights instead of random walks; see~\cite{Hinrichsen:LevyFlights}. In the chemical kinetics picture, one can also account for mutations by introducing a birth process~$0\rightarrow 2A$ and for natural selection by imposing an attraction between the particles flanking domains of the deleterious allele. 


\begin{thebibliography}{61}
\expandafter\ifx\csname natexlab\endcsname\relax\def\natexlab#1{#1}\fi
\expandafter\ifx\csname bibnamefont\endcsname\relax
  \def\bibnamefont#1{#1}\fi
\expandafter\ifx\csname bibfnamefont\endcsname\relax
  \def\bibfnamefont#1{#1}\fi
\expandafter\ifx\csname citenamefont\endcsname\relax
  \def\citenamefont#1{#1}\fi
\expandafter\ifx\csname url\endcsname\relax
  \def\url#1{\texttt{#1}}\fi
\expandafter\ifx\csname urlprefix\endcsname\relax\def\urlprefix{URL }\fi
\providecommand{\bibinfo}[2]{#2}
\providecommand{\eprint}[2][]{\url{#2}}

\bibitem[{\citenamefont{Araten} \emph{et~al.}(2005)\citenamefont{Araten, Golde,
  Zhang, Thaler, Gargiulo, Notaro, and Luzzatto}}]{Araten:MutationRates}
\bibinfo{author}{\bibnamefont{Araten}, \bibfnamefont{D.}},
  \bibinfo{author}{\bibfnamefont{D.}~\bibnamefont{Golde}},
  \bibinfo{author}{\bibfnamefont{R.}~\bibnamefont{Zhang}},
  \bibinfo{author}{\bibfnamefont{H.}~\bibnamefont{Thaler}},
  \bibinfo{author}{\bibfnamefont{L.}~\bibnamefont{Gargiulo}},
  \bibinfo{author}{\bibfnamefont{R.}~\bibnamefont{Notaro}}, and
  \bibinfo{author}{\bibfnamefont{L.}~\bibnamefont{Luzzatto}},
  \bibinfo{year}{2005}, \bibinfo{journal}{Cancer Res.}
  \textbf{\bibinfo{volume}{65}}, \bibinfo{pages}{8111}.

\bibitem[{\citenamefont{Bar-Yam}(2005)}]{MakingThingsWork}
\bibinfo{author}{\bibnamefont{Bar-Yam}, \bibfnamefont{Y.}},
  \bibinfo{year}{2005}, \emph{\bibinfo{title}{{Making Things Work: Solving
  Complex Problems in a Complex World}}} (\bibinfo{publisher}{Knowledge
  Press}).

\bibitem[{\citenamefont{Barton} \emph{et~al.}(2002)\citenamefont{Barton,
  Depaulis, and Etheridge}}]{Barton:NeutralEvolution}
\bibinfo{author}{\bibnamefont{Barton}, \bibfnamefont{N.}},
  \bibinfo{author}{\bibfnamefont{F.}~\bibnamefont{Depaulis}}, and
  \bibinfo{author}{\bibfnamefont{A.}~\bibnamefont{Etheridge}},
  \bibinfo{year}{2002}, \bibinfo{journal}{Theor. Popul. Biol.}
  \textbf{\bibinfo{volume}{61}}, \bibinfo{pages}{31}.

\bibitem[{\citenamefont{{ben-Avraham}}(1998)}]{Benavraham:ExactCoalescence}
\bibinfo{author}{\bibnamefont{{ben-Avraham}}, \bibfnamefont{D.}},
  \bibinfo{year}{1998}, \bibinfo{journal}{Phys. Rev. Lett.}
  \textbf{\bibinfo{volume}{81}}, \bibinfo{pages}{4756}.

\bibitem[{\citenamefont{Blythe and McKane}(2007)}]{Blythe:Review}
\bibinfo{author}{\bibnamefont{Blythe}, \bibfnamefont{R.}}, and
  \bibinfo{author}{\bibfnamefont{A.}~\bibnamefont{McKane}},
  \bibinfo{year}{2007}, \bibinfo{journal}{J. Stat. Mech}
  \textbf{\bibinfo{volume}{7018}}, \bibinfo{pages}{1}.

\bibitem[{\citenamefont{Bramson and
  Griffeath}(1980)}]{Bramson:DomainSizeDistributionARW}
\bibinfo{author}{\bibnamefont{Bramson}, \bibfnamefont{M.}}, and
  \bibinfo{author}{\bibfnamefont{D.}~\bibnamefont{Griffeath}},
  \bibinfo{year}{1980}, \bibinfo{journal}{Ann. Probab.}
  \textbf{\bibinfo{volume}{8}}, \bibinfo{pages}{183}.

\bibitem[{\citenamefont{Bramson and Lebowitz}(1991)}]{Bramson:FiniteSize}
\bibinfo{author}{\bibnamefont{Bramson}, \bibfnamefont{M.}}, and
  \bibinfo{author}{\bibfnamefont{J.}~\bibnamefont{Lebowitz}},
  \bibinfo{year}{1991}, \bibinfo{journal}{J. Stat. Phys.}
  \textbf{\bibinfo{volume}{62}}, \bibinfo{pages}{297}.

\bibitem[{\citenamefont{Charlesworth}
  \emph{et~al.}(2003)\citenamefont{Charlesworth, Charlesworth, and
  Barton}}]{Charlesworth:NeutralVariation}
\bibinfo{author}{\bibnamefont{Charlesworth}, \bibfnamefont{B.}},
  \bibinfo{author}{\bibfnamefont{D.}~\bibnamefont{Charlesworth}}, and
  \bibinfo{author}{\bibfnamefont{N.}~\bibnamefont{Barton}},
  \bibinfo{year}{2003}, \bibinfo{journal}{Annu. Rev. Ecol. Evol. Syst.}
  \textbf{\bibinfo{volume}{34}}, \bibinfo{pages}{99}.

\bibitem[{\citenamefont{Cox and Griffeath}(1986)}]{CoxGriffeath:VoterModel}
\bibinfo{author}{\bibnamefont{Cox}, \bibfnamefont{J.}}, and
  \bibinfo{author}{\bibfnamefont{D.}~\bibnamefont{Griffeath}},
  \bibinfo{year}{1986}, \bibinfo{journal}{Ann. Probab.}
  \textbf{\bibinfo{volume}{14}}, \bibinfo{pages}{347}.

\bibitem[{\citenamefont{Crow and Kimura}(1970)}]{Crow:PopulationGenetics}
\bibinfo{author}{\bibnamefont{Crow}, \bibfnamefont{J.}}, and
  \bibinfo{author}{\bibfnamefont{M.}~\bibnamefont{Kimura}},
  \bibinfo{year}{1970}, \emph{\bibinfo{title}{{An Introduction to Population
  Genetics Theory}}} (\bibinfo{publisher}{Harper \& Row, New York}).

\bibitem[{\citenamefont{Doering and ben Avraham}(1988)}]{Doering:IDF}
\bibinfo{author}{\bibnamefont{Doering}, \bibfnamefont{C.}}, and
  \bibinfo{author}{\bibfnamefont{D.}~\bibnamefont{ben Avraham}},
  \bibinfo{year}{1988}, \bibinfo{journal}{Phys. Rev. A}
  \textbf{\bibinfo{volume}{38}}, \bibinfo{pages}{3035}.

\bibitem[{\citenamefont{Doering} \emph{et~al.}(2003)\citenamefont{Doering,
  Mueller, and Smereka}}]{Doering:FisherWaveWeakSelection}
\bibinfo{author}{\bibnamefont{Doering}, \bibfnamefont{C.}},
  \bibinfo{author}{\bibfnamefont{C.}~\bibnamefont{Mueller}}, and
  \bibinfo{author}{\bibfnamefont{P.}~\bibnamefont{Smereka}},
  \bibinfo{year}{2003}, \bibinfo{journal}{Physica A}
  \textbf{\bibinfo{volume}{325}}, \bibinfo{pages}{243}.

\bibitem[{\citenamefont{Drake}(1991)}]{Drake:MutationRates}
\bibinfo{author}{\bibnamefont{Drake}, \bibfnamefont{J.}}, \bibinfo{year}{1991},
  \bibinfo{journal}{Proc. Natl. Acad. Sci. USA} \textbf{\bibinfo{volume}{88}},
  \bibinfo{pages}{7160}.

\bibitem[{\citenamefont{Durrett}(1988)}]{Durrett:LectureNotes}
\bibinfo{author}{\bibnamefont{Durrett}, \bibfnamefont{R.}},
  \bibinfo{year}{1988}, \emph{\bibinfo{title}{{Lecture Notes on Particle
  systems and Percolation}}} (\bibinfo{publisher}{Wadsworth Publishing
  Company}).

\bibitem[{\citenamefont{Duty}(2000)}]{Duty:Thesis}
\bibinfo{author}{\bibnamefont{Duty}, \bibfnamefont{T.~L.}},
  \bibinfo{year}{2000}, \bibinfo{type}{{PhD} thesis}, \bibinfo{school}{(The
  University of British Columbia)}.

\bibitem[{\citenamefont{Fisher}(1937)}]{Fisher:FisherWave}
\bibinfo{author}{\bibnamefont{Fisher}, \bibfnamefont{R.}},
  \bibinfo{year}{1937}, \bibinfo{journal}{Ann. Eugenics}
  \textbf{\bibinfo{volume}{7}}, \bibinfo{pages}{353}.

\bibitem[{\citenamefont{Gardiner}(1985)}]{Gardiner:Handbook}
\bibinfo{author}{\bibnamefont{Gardiner}, \bibfnamefont{C.}},
  \bibinfo{year}{1985}, \emph{\bibinfo{title}{{Handbook of Stochastic
  Methods}}} (\bibinfo{publisher}{Springer, New York}).

\bibitem[{\citenamefont{Gibson}(1965)}]{Gibson:Oklahoma}
\bibinfo{author}{\bibnamefont{Gibson}, \bibfnamefont{A.}},
  \bibinfo{year}{1965}, \emph{\bibinfo{title}{{Oklahoma: A History of Five
  Centuries}}} (\bibinfo{publisher}{Harlow, Norman}).

\bibitem[{\citenamefont{Goldenfeld}(1992)}]{Goldenfeld:Lectures}
\bibinfo{author}{\bibnamefont{Goldenfeld}, \bibfnamefont{N.}},
  \bibinfo{year}{1992}, \emph{\bibinfo{title}{{Lectures on Phase Transitions
  and the Renormalization Group}}} (\bibinfo{publisher}{Westview Press}).

\bibitem[{\citenamefont{Gradshteyn and Ryzhik}(1980)}]{Gradshteyn:Tables}
\bibinfo{author}{\bibnamefont{Gradshteyn}, \bibfnamefont{I.}}, and
  \bibinfo{author}{\bibfnamefont{I.}~\bibnamefont{Ryzhik}},
  \bibinfo{year}{1980}, \emph{\bibinfo{title}{{Table of Integrals, Series, and
  Products}}} (\bibinfo{publisher}{Academic Press, New York}).

\bibitem[{\citenamefont{Guth}(1981)}]{UniverseInflation}
\bibinfo{author}{\bibnamefont{Guth}, \bibfnamefont{A.}}, \bibinfo{year}{1981},
  \bibinfo{journal}{Phys. Rev. D} \textbf{\bibinfo{volume}{23}},
  \bibinfo{pages}{347}.

\bibitem[{\citenamefont{Hallatschek}
  \emph{et~al.}(2007)\citenamefont{Hallatschek, Hersen, Ramanathan, and {D. R.
  Nelson}}}]{HallatschekNelson:ExperimentalSegregation}
\bibinfo{author}{\bibnamefont{Hallatschek}, \bibfnamefont{O.}},
  \bibinfo{author}{\bibfnamefont{P.}~\bibnamefont{Hersen}},
  \bibinfo{author}{\bibfnamefont{S.}~\bibnamefont{Ramanathan}}, and
  \bibinfo{author}{\bibnamefont{{D. R. Nelson}}}, \bibinfo{year}{2007},
  \bibinfo{journal}{Proc. Natl. Acad. Sci. USA} \textbf{\bibinfo{volume}{104}},
  \bibinfo{pages}{19926}.

\bibitem[{\citenamefont{Hallatschek and {K. S.
  Korolev}}(2009)}]{Hallatschek:FisherWave}
\bibinfo{author}{\bibnamefont{Hallatschek}, \bibfnamefont{O.}}, and
  \bibinfo{author}{\bibnamefont{{K. S. Korolev}}}, \bibinfo{year}{2009},
  \bibinfo{journal}{Phys. Rev. Lett.} \textbf{\bibinfo{volume}{103}},
  \bibinfo{pages}{108103}.

\bibitem[{\citenamefont{Hallatschek and Nelson}(2010)}]{Hallatschek:LifeFront}
\bibinfo{author}{\bibnamefont{Hallatschek}, \bibfnamefont{O.}}, and
  \bibinfo{author}{\bibfnamefont{D.~R.} \bibnamefont{Nelson}},
  \bibinfo{year}{2010}, \bibinfo{journal}{Evolution}
  \textbf{\bibinfo{volume}{64}}, \bibinfo{pages}{193}.

\bibitem[{\citenamefont{Hartl and Clark}(1989)}]{Hartl:PopulationGenetics}
\bibinfo{author}{\bibnamefont{Hartl}, \bibfnamefont{D.}}, and
  \bibinfo{author}{\bibfnamefont{A.}~\bibnamefont{Clark}},
  \bibinfo{year}{1989}, \emph{\bibinfo{title}{{Principles of population
  genetics}}} (\bibinfo{publisher}{Sinauer Associates, Sunderland}).

\bibitem[{\citenamefont{Hewitt}(1996)}]{Hewitt:SouthNorthGradient}
\bibinfo{author}{\bibnamefont{Hewitt}, \bibfnamefont{G.}},
  \bibinfo{year}{1996}, \bibinfo{journal}{Biol. J. Linn. Soc.}
  \textbf{\bibinfo{volume}{58}}, \bibinfo{pages}{247}.

\bibitem[{\citenamefont{Hinrichsen and Howard}(1999)}]{Hinrichsen:LevyFlights}
\bibinfo{author}{\bibnamefont{Hinrichsen}, \bibfnamefont{H.}}, and
  \bibinfo{author}{\bibfnamefont{M.}~\bibnamefont{Howard}},
  \bibinfo{year}{1999}, \bibinfo{journal}{Eur. Phys. J. B}
  \textbf{\bibinfo{volume}{7}}(\bibinfo{number}{4}), \bibinfo{pages}{635}.

\bibitem[{\citenamefont{Houchmandzadeh and
  Vallade}(2003)}]{Houchmandzadeh:NeutralEcology}
\bibinfo{author}{\bibnamefont{Houchmandzadeh}, \bibfnamefont{B.}}, and
  \bibinfo{author}{\bibfnamefont{M.}~\bibnamefont{Vallade}},
  \bibinfo{year}{2003}, \bibinfo{journal}{Phys. Rev. E}
  \textbf{\bibinfo{volume}{68}}, \bibinfo{pages}{61912}.

\bibitem[{\citenamefont{Kimura}(1955)}]{Kimura:GeneticDrift}
\bibinfo{author}{\bibnamefont{Kimura}, \bibfnamefont{M.}},
  \bibinfo{year}{1955}, \bibinfo{journal}{Proc. Natl. Acad. Sci. USA}
  \textbf{\bibinfo{volume}{41}}, \bibinfo{pages}{144}.

\bibitem[{\citenamefont{Kimura}(1969)}]{Kimura:InfiniteSiteModel}
\bibinfo{author}{\bibnamefont{Kimura}, \bibfnamefont{M.}},
  \bibinfo{year}{1969}, \bibinfo{journal}{Genetics}
  \textbf{\bibinfo{volume}{61}}, \bibinfo{pages}{893}.

\bibitem[{\citenamefont{Kimura}(1983)}]{Kimura:NeutralEvolution}
\bibinfo{author}{\bibnamefont{Kimura}, \bibfnamefont{M.}},
  \bibinfo{year}{1983}, \emph{\bibinfo{title}{{The Neutral Theory of Molecular
  Evolution}}} (\bibinfo{publisher}{Cambridge University Press}).

\bibitem[{\citenamefont{Kimura and Weiss}(1964)}]{KimuraWeiss:SSM}
\bibinfo{author}{\bibnamefont{Kimura}, \bibfnamefont{M.}}, and
  \bibinfo{author}{\bibfnamefont{G.}~\bibnamefont{Weiss}},
  \bibinfo{year}{1964}, \bibinfo{journal}{Genetics}
  \textbf{\bibinfo{volume}{49}}, \bibinfo{pages}{561}.

\bibitem[{\citenamefont{Kingman}(1982)}]{Kingman:Coalescent}
\bibinfo{author}{\bibnamefont{Kingman}, \bibfnamefont{J.}},
  \bibinfo{year}{1982}, \bibinfo{journal}{J. Appl. Prob.} ,
  \bibinfo{pages}{27}.

\bibitem[{\citenamefont{Kolmogorov}
  \emph{et~al.}(1937)\citenamefont{Kolmogorov, Petrovsky, and
  Piscounov}}]{Kolmogorov:FKPPEquation}
\bibinfo{author}{\bibnamefont{Kolmogorov}, \bibfnamefont{A.}},
  \bibinfo{author}{\bibfnamefont{N.}~\bibnamefont{Petrovsky}}, and
  \bibinfo{author}{\bibfnamefont{N.}~\bibnamefont{Piscounov}},
  \bibinfo{year}{1937}, \bibinfo{journal}{Moscow University Bulletin of
  Mathematics} \textbf{\bibinfo{volume}{1}}, \bibinfo{pages}{1}.

\bibitem[{\citenamefont{Liggett}(2004)}]{Liggett:Book}
\bibinfo{author}{\bibnamefont{Liggett}, \bibfnamefont{T.}},
  \bibinfo{year}{2004}, \emph{\bibinfo{title}{{Interacting Particle Systems}}}
  (\bibinfo{publisher}{Springer Verlag}).

\bibitem[{\citenamefont{Lin}(1991)}]{Lin:ClosureSchemes}
\bibinfo{author}{\bibnamefont{Lin}, \bibfnamefont{J.}}, \bibinfo{year}{1991},
  \bibinfo{journal}{Phys. Rev. A} \textbf{\bibinfo{volume}{44}},
  \bibinfo{pages}{6706}.

\bibitem[{\citenamefont{Mal{\'e}cot}(1955)}]{Malecot:DecreaseReltationshipDist%
ance}
\bibinfo{author}{\bibnamefont{Mal{\'e}cot}, \bibfnamefont{G.}},
  \bibinfo{year}{1955}, \bibinfo{journal}{Cold Springs Harbor Symp. Quant.
  Biol} \textbf{\bibinfo{volume}{20}}, \bibinfo{pages}{52}.

\bibitem[{\citenamefont{Mal{\'e}cot}(1975)}]{Malecot:Dynamics}
\bibinfo{author}{\bibnamefont{Mal{\'e}cot}, \bibfnamefont{G.}},
  \bibinfo{year}{1975}, \bibinfo{journal}{Theor. Popul. Biol.}
  \textbf{\bibinfo{volume}{8}}, \bibinfo{pages}{212}.

\bibitem[{\citenamefont{Masser and {ben-Avraham}}(2000)}]{Masser:Potts}
\bibinfo{author}{\bibnamefont{Masser}, \bibfnamefont{T.}}, and
  \bibinfo{author}{\bibfnamefont{D.}~\bibnamefont{{ben-Avraham}}},
  \bibinfo{year}{2000}, \bibinfo{journal}{Phys. Lett. A}
  \textbf{\bibinfo{volume}{275}}, \bibinfo{pages}{382}.

\bibitem[{\citenamefont{Mayr}(1942)}]{Mayr:FounderEffect}
\bibinfo{author}{\bibnamefont{Mayr}, \bibfnamefont{E.}}, \bibinfo{year}{1942},
  \emph{\bibinfo{title}{{Systematics and the Origin of Species from the
  Viewpoint of a Zoologist}}} (\bibinfo{publisher}{Columbia University, New
  York}).

\bibitem[{\citenamefont{Murray}(2003)}]{Murray:MathematicalBiology}
\bibinfo{author}{\bibnamefont{Murray}, \bibfnamefont{J.}},
  \bibinfo{year}{2003}, \emph{\bibinfo{title}{{Mathematical Biology}}}
  (\bibinfo{publisher}{Springer}).

\bibitem[{\citenamefont{Nagylaki}(1974)}]{Nagylaki:DecayGeneticVariability}
\bibinfo{author}{\bibnamefont{Nagylaki}, \bibfnamefont{T.}},
  \bibinfo{year}{1974}, \bibinfo{journal}{Proc. Natl. Acad. Sci. USA}
  \textbf{\bibinfo{volume}{71}}, \bibinfo{pages}{2932}.

\bibitem[{\citenamefont{Nagylaki}(1978)}]{Nagylaki:Hierarchy}
\bibinfo{author}{\bibnamefont{Nagylaki}, \bibfnamefont{T.}},
  \bibinfo{year}{1978}, \bibinfo{journal}{Proc. Natl. Acad. Sci. USA}
  \textbf{\bibinfo{volume}{75}}, \bibinfo{pages}{423}.

\bibitem[{\citenamefont{Nordborg}(1998)}]{Nordborg:Neanderthal}
\bibinfo{author}{\bibnamefont{Nordborg}, \bibfnamefont{M.}},
  \bibinfo{year}{1998}, \bibinfo{journal}{The Am. J. of Hum. Genet.}
  \textbf{\bibinfo{volume}{63}}, \bibinfo{pages}{1237}.

\bibitem[{\citenamefont{Nowak}(2006)}]{Nowak:EvolutionaryDynamics}
\bibinfo{author}{\bibnamefont{Nowak}, \bibfnamefont{M.}}, \bibinfo{year}{2006},
  \emph{\bibinfo{title}{{Evolutionary Dynamics: Exploring the Equations of
  Life}}} (\bibinfo{publisher}{Harvard University, Cambridge MA}).

\bibitem[{\citenamefont{\'{O}dor}(2004)}]{Odor:NonEquilibrium}
\bibinfo{author}{\bibnamefont{\'{O}dor}, \bibfnamefont{G.}},
  \bibinfo{year}{2004}, \bibinfo{journal}{Rev. Mod. Phys.}
  \textbf{\bibinfo{volume}{76}}, \bibinfo{pages}{633}.

\bibitem[{\citenamefont{Poli} \emph{et~al.}(2008)\citenamefont{Poli, Langdon,
  and McPhee}}]{Poli:GeneticProgramming}
\bibinfo{author}{\bibnamefont{Poli}, \bibfnamefont{R.}},
  \bibinfo{author}{\bibfnamefont{W.}~\bibnamefont{Langdon}}, and
  \bibinfo{author}{\bibfnamefont{N.}~\bibnamefont{McPhee}},
  \bibinfo{year}{2008}, \emph{\bibinfo{title}{{A Field Guide to Genetic
  Programming}}} (\bibinfo{publisher}{Lulu Enterprises, UK Ltd}).

\bibitem[{\citenamefont{Polyanin and
  Manzhirov}(1998)}]{Polyanin:IntegralEquations}
\bibinfo{author}{\bibnamefont{Polyanin}, \bibfnamefont{A.}}, and
  \bibinfo{author}{\bibfnamefont{A.}~\bibnamefont{Manzhirov}},
  \bibinfo{year}{1998}, \emph{\bibinfo{title}{{Handbook of Integral
  Equations}}} (\bibinfo{publisher}{CRC Press, Boca Raton}).

\bibitem[{\citenamefont{Ramachandran}
  \emph{et~al.}(2005)\citenamefont{Ramachandran, Deshpande, Roseman, Rosenberg,
  Feldman, and Cavalli-Sforza}}]{Ramachandran:MigrationFromAfrica}
\bibinfo{author}{\bibnamefont{Ramachandran}, \bibfnamefont{S.}},
  \bibinfo{author}{\bibfnamefont{O.}~\bibnamefont{Deshpande}},
  \bibinfo{author}{\bibfnamefont{C.}~\bibnamefont{Roseman}},
  \bibinfo{author}{\bibfnamefont{N.}~\bibnamefont{Rosenberg}},
  \bibinfo{author}{\bibfnamefont{M.}~\bibnamefont{Feldman}}, and
  \bibinfo{author}{\bibfnamefont{L.}~\bibnamefont{Cavalli-Sforza}},
  \bibinfo{year}{2005}, \bibinfo{journal}{Proc. Natl. Acad. Sci. USA}
  \textbf{\bibinfo{volume}{102}}, \bibinfo{pages}{15942}.

\bibitem[{\citenamefont{Redner}(2001)}]{Redner:FirstPassageTime}
\bibinfo{author}{\bibnamefont{Redner}, \bibfnamefont{S.}},
  \bibinfo{year}{2001}, \emph{\bibinfo{title}{{A Guide to First Passage
  Processes}}} (\bibinfo{publisher}{Cambridge University Press}).

\bibitem[{\citenamefont{Risken}(1989)}]{Risken:FPE}
\bibinfo{author}{\bibnamefont{Risken}, \bibfnamefont{H.}},
  \bibinfo{year}{1989}, \emph{\bibinfo{title}{{The Fokker-Planck equation:
  Methods of Solution and Applications}}} (\bibinfo{publisher}{Springer, Berlin
  and Heidelberg}).

\bibitem[{\citenamefont{Saito and
  M{\"u}ller-Krumbhaar}(1995)}]{Saito:EdenModel}
\bibinfo{author}{\bibnamefont{Saito}, \bibfnamefont{Y.}}, and
  \bibinfo{author}{\bibfnamefont{H.}~\bibnamefont{M{\"u}ller-Krumbhaar}},
  \bibinfo{year}{1995}, \bibinfo{journal}{Phys. Rev. Lett.}
  \textbf{\bibinfo{volume}{74}}, \bibinfo{pages}{4325}.

\bibitem[{\citenamefont{Scheucher and
  Spohn}(1988)}]{Scheucher:SpinodalDecomposition}
\bibinfo{author}{\bibnamefont{Scheucher}, \bibfnamefont{M.}}, and
  \bibinfo{author}{\bibfnamefont{H.}~\bibnamefont{Spohn}},
  \bibinfo{year}{1988}, \bibinfo{journal}{J. Stat. Phys.}
  \textbf{\bibinfo{volume}{53}}, \bibinfo{pages}{279}.

\bibitem[{\citenamefont{Templeton}(2002)}]{Templeton:MigrationFromAfrica}
\bibinfo{author}{\bibnamefont{Templeton}, \bibfnamefont{A.}},
  \bibinfo{year}{2002}, \bibinfo{journal}{Nature}
  \textbf{\bibinfo{volume}{416}}, \bibinfo{pages}{45}.

\bibitem[{\citenamefont{Wakeley}(2008)}]{Wakeley:Coalescent}
\bibinfo{author}{\bibnamefont{Wakeley}, \bibfnamefont{J.}},
  \bibinfo{year}{2008}, \emph{\bibinfo{title}{{Coalescent Theory: An
  Introduction}}} (\bibinfo{publisher}{Roberts \& Company Publishers}).

\bibitem[{\citenamefont{Wilkins and Wakeley}(2002)}]{Wilkins:Linear}
\bibinfo{author}{\bibnamefont{Wilkins}, \bibfnamefont{J.}}, and
  \bibinfo{author}{\bibfnamefont{J.}~\bibnamefont{Wakeley}},
  \bibinfo{year}{2002}, \bibinfo{journal}{Genetics}
  \textbf{\bibinfo{volume}{161}}, \bibinfo{pages}{873}.

\bibitem[{\citenamefont{Wilkinson-Herbots}(1998)}]{Wilkinson:1dCoalescent}
\bibinfo{author}{\bibnamefont{Wilkinson-Herbots}, \bibfnamefont{H.}},
  \bibinfo{year}{1998}, \bibinfo{journal}{J. Math. Biol.}
  \textbf{\bibinfo{volume}{37}}, \bibinfo{pages}{535}.

\bibitem[{\citenamefont{Wilson and Kogut}(1974)}]{Wilson:Renormalization}
\bibinfo{author}{\bibnamefont{Wilson}, \bibfnamefont{K.}}, and
  \bibinfo{author}{\bibfnamefont{J.}~\bibnamefont{Kogut}},
  \bibinfo{year}{1974}, \bibinfo{journal}{Phys. Rep.}
  \textbf{\bibinfo{volume}{12}}, \bibinfo{pages}{75}.

\bibitem[{\citenamefont{Wright}(1931)}]{Wright:IslandModel}
\bibinfo{author}{\bibnamefont{Wright}, \bibfnamefont{S.}},
  \bibinfo{year}{1931}, \bibinfo{journal}{Genetics}
  \textbf{\bibinfo{volume}{16}}, \bibinfo{pages}{97}.

\bibitem[{\citenamefont{Wright}(1943)}]{Wright:IsolationByDistance}
\bibinfo{author}{\bibnamefont{Wright}, \bibfnamefont{S.}},
  \bibinfo{year}{1943}, \bibinfo{journal}{Genetics}
  \textbf{\bibinfo{volume}{28}}, \bibinfo{pages}{114}.

\bibitem[{\citenamefont{Zhong and {D.
  ben-Avraham}}(1995)}]{Zhong:FiniteReactionRates}
\bibinfo{author}{\bibnamefont{Zhong}, \bibfnamefont{D.}}, and
  \bibinfo{author}{\bibnamefont{{D. ben-Avraham}}}, \bibinfo{year}{1995},
  \bibinfo{journal}{Phys. Lett. A} \textbf{\bibinfo{volume}{209}},
  \bibinfo{pages}{333}.

\end{thebibliography}

\end{document}